**Dr. Petar Radanliev**
Parks Road,
Oxford OX1 3PJ
United Kingdom
Email: petar.radanliev@cs.ox.ac.uk

BA Hons., MSc., Ph.D. Post-Doctorate

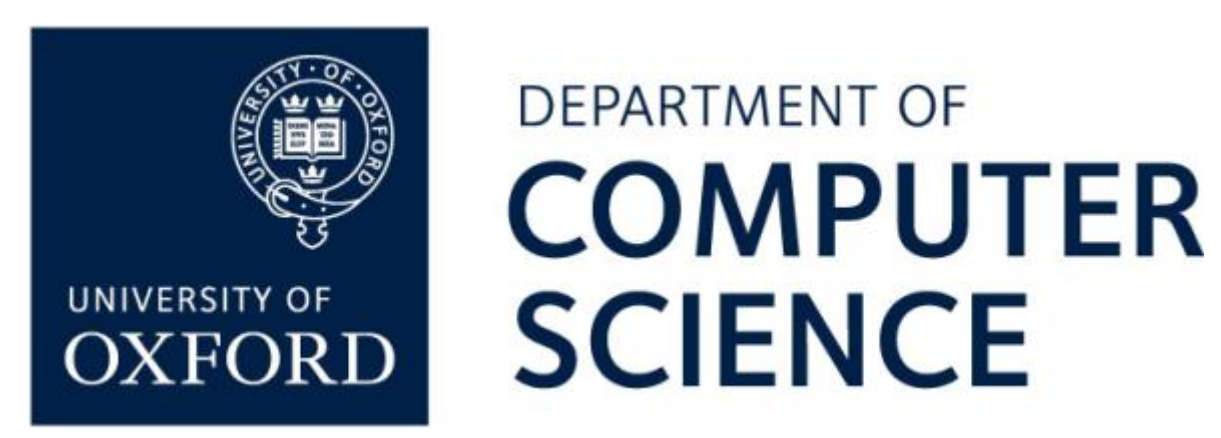

# AI-Driven Adaptive Adversaries and the Erosion of Cryptographic Trust in Public Key Systems

**Corresponding author email*: petar.radanliev@cs.ox.ac.uk

Petar Radanliev*

Department of Computer Sciences, University of Oxford, Wolfson Building, Parks Rd, Oxford OX1 3QG; The Alan Turing Institute, British Library, 96 Euston Rd., London NW1 2DB; Email: * petar.radanliev@cs.ox.ac.uk

**Other IDs**

ORCID: https://orcid.org/0000-0001-5629-6857

ResearcherID: L-7509-2015

ResearcherID: M-2176-2017

Scopus Author ID: 57003734400

Loop profile: 839254

AI-Driven Adaptive Adversaries vs. Traditional Cryptography

| Characteristic | Adversary Type | Attack Focus | Security Definition | Threat Actors | Attack Process | Attack Surfaces | Attack Mechanisms |
|---|---|---|---|---|---|---|---|
| Traditional Cryptography | Static | Cryptographic Algorithms | Strong Algorithms + Large Keys | Classical Cryptanalysis | Static | Entropy, Private Key Storage, Side-channel Leakage, Certificate Validation, Key Exchange Protocols, Cloud/Virtualised Environments | Brute Force, Static Malware |
| AI-Driven Adaptive Adversaries | Adaptive | Implementation Weaknesses | Adaptive System Resilience | AI-enabled Malware, Quantum-assisted Adversaries | Observe, Learn, Optimize, Exploit, Adapt | Entropy, Private Key Storage, Side-channel Leakage, Certificate Validation, Key Exchange Protocols, Cloud/Virtualised Environments | Machine Learning, Reinforcement Learning, Side-channel Analysis, Entropy Prediction, Adaptive MITM, Certificate Abuse, Runtime Behavioural Modeling |

| Characteristic | Key Failure Modes | Critical Insight | Research Focus | Key Size | Post-Quantum Cryptography | Evolution |
|---|---|---|---|---|---|---|
| Traditional Cryptography | Private Key Compromise, Signature Forgery, Trust Infrastructure Collapse, Identity Impersonation, Session Hijacking, Cryptographic Trust Erosion | Break Cryptographic Mathematics Directly | Algorithm Robustness | Larger Key Sizes Alone Insufficient | Does Not Solve Implementation-Layer Vulnerabilities | Phase 1: Classical Cryptanalysis |
| AI-Driven Adaptive Adversaries | Private Key Compromise, Signature Forgery, Trust Infrastructure Collapse, Identity Impersonation, Session Hijacking, Cryptographic Trust Erosion | Exploit Observability, Implementation Leakage, Trust Dependencies, Adaptive Optimization | AI-enabled Cryptographic Attacks | Larger Key Sizes Alone Insufficient | Does Not Solve Implementation-Layer Vulnerabilities | Phase 3: Adaptive AI-driven Optimization |

| Characteristic | Systemic Consequences | Defensive Shift | Required Controls | Final Conclusion |
|---|---|---|---|---|
| Traditional Cryptography | Confidentiality Loss, Integrity Failure, Authentication Collapse, Non-repudiation Failure, Breakdown of PKI Trust | Static Security → Adaptive Cryptographic Resilience | AI-aware monitoring, Continuous entropy validation, Adaptive trust management, Runtime anomaly detection, Dynamic key governance, Hardware-backed isolation, AI-enhanced defensive systems, PQC integrated with adaptive controls | Security Determined by Algorithm Strength |
| AI-Driven Adaptive Adversaries | Confidentiality Loss, Integrity Failure, Authentication Collapse, Non-repudiation Failure, Breakdown of PKI Trust | Static Security → Adaptive Cryptographic Resilience | AI-aware monitoring, Continuous entropy validation, Adaptive trust management, Runtime anomaly detection, Dynamic key governance, Hardware-backed isolation, AI-enhanced defensive systems, PQC integrated with adaptive controls | Security Determined by System Resilience |

**Dr. Petar Radanliev**
Parks Road,
Oxford OX1 3PJ
United Kingdom
Email: petar.radanliev@cs.ox.ac.uk

BA Hons., MSc., Ph.D. Post-Doctorate

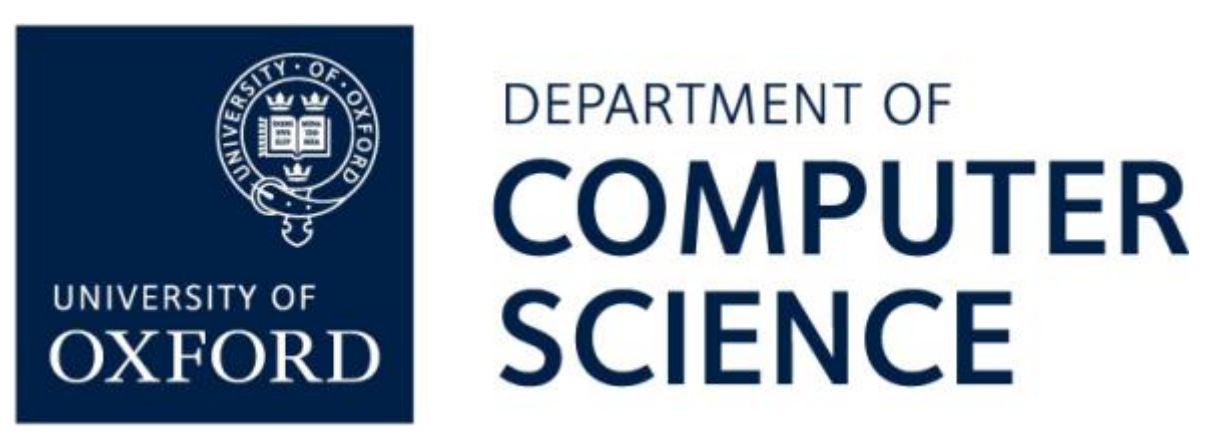

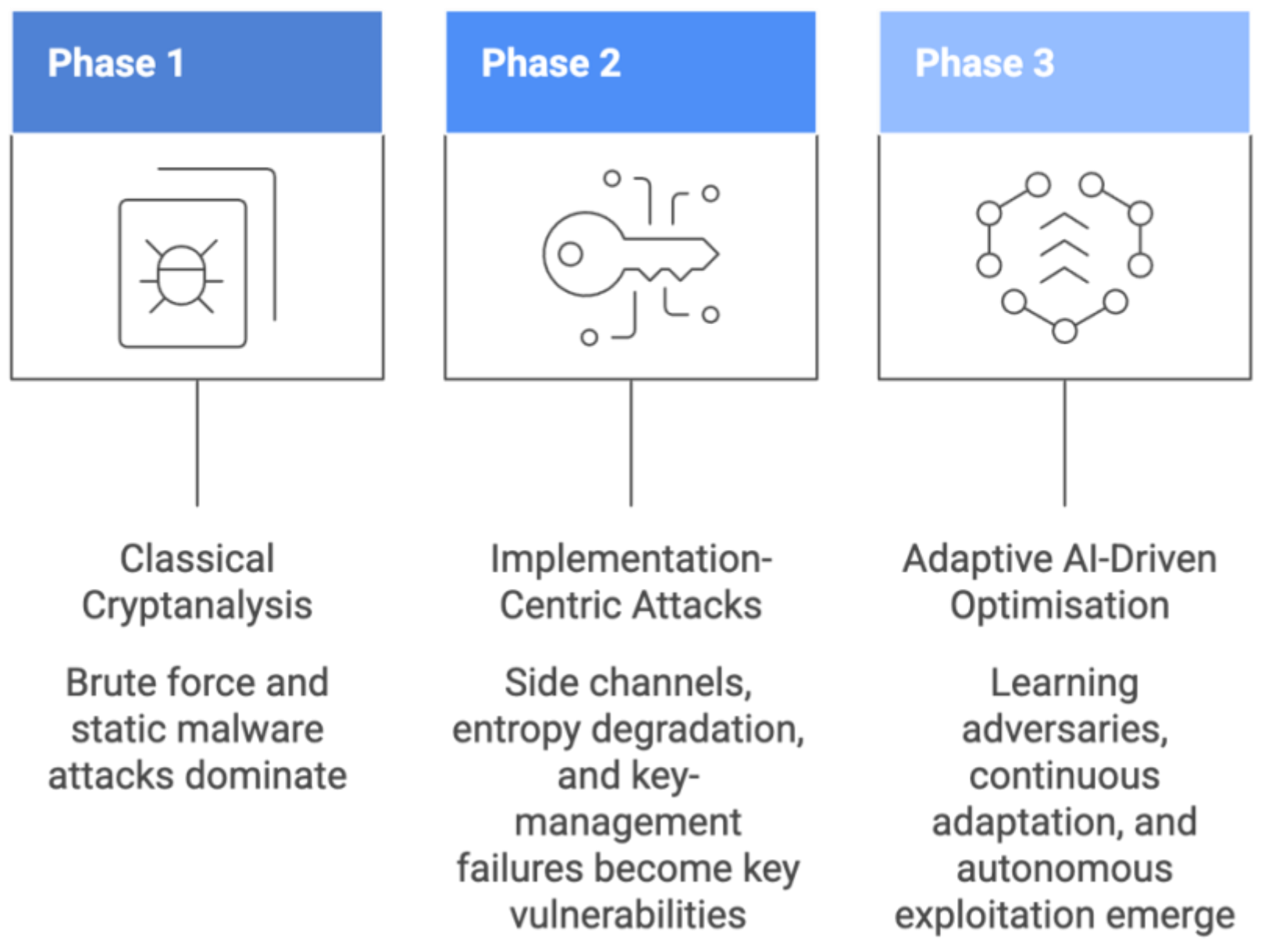

Abstract:

This paper examines the erosion of Public Key Cryptography (PKC) security under adaptive adversarial optimisation driven by artificial intelligence. The problem addressed is the growing mismatch between algorithm-centric cryptographic security models and operational attack realities, where adversaries exploit implementation-level observability rather than breaking cryptographic primitives. The methodology integrates a reproducible bibliometric analysis of Web of Science records, qualitative evidence from twenty expert interviews and three industry workshops, and a technical synthesis of AI-enabled attack mechanisms across the cryptographic lifecycle. Results show that existing research is structurally concentrated on algorithmic robustness, with no significant focus on AI-driven attack vectors, while 82% of practitioners attribute private key compromise to AI-augmented optimisation and side-channel inference. The paper's contribution is fourfold: (1) identification of a systemic research gap in AI-enabled cryptographic attacks; (2) development of an adaptive adversarial threat model spanning key generation to validation; (3) empirical validation of implementation-layer compromise mechanisms; and (4) formulation of AI-aware cryptographic resilience requirements extending beyond post-quantum approaches. The findings demonstrate that cryptographic security must be reconceptualised as an adaptive, system-level property rather than a function of algorithm strength alone.

***Keywords**:* Public Key Cryptography; Adaptive Adversarial Optimisation; AI-Enabled Malware; Private Key Compromise; Side-Channel Inference; Post-Quantum Cryptography

# 1. Problem Context and Research Motivation

Public Key Cryptography (PKC) is deployed under the assumption that security is determined by computational hardness and formally verified protocol correctness. This

**Dr. Petar Radanliev**
Parks Road,
Oxford OX1 3PJ
United Kingdom
Email: petar.radanliev@cs.ox.ac.uk
BA Hons., MSc., Ph.D. Post-Doctorate

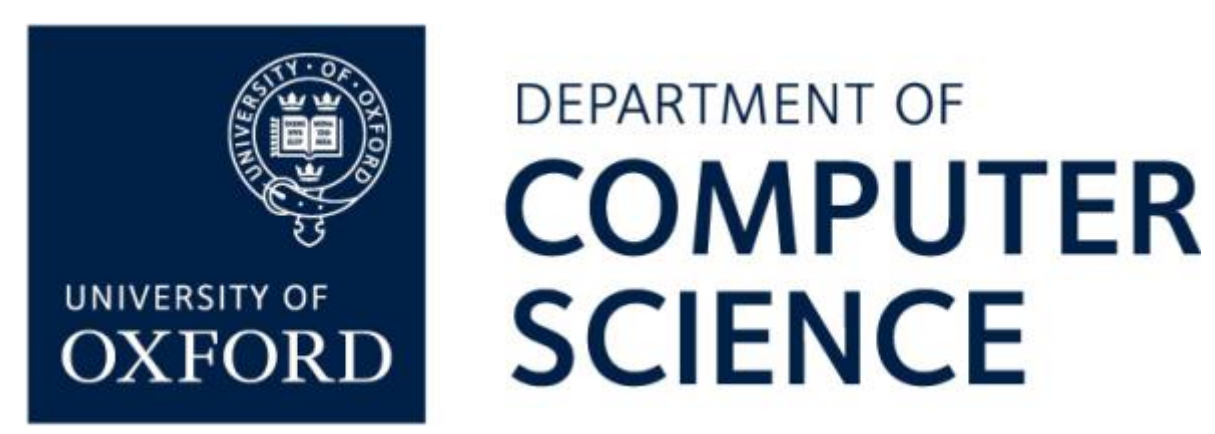

assumption implies that adversaries are bounded by infeasible computation and that compromise occurs only through cryptanalytic breakthroughs or implementation error. However, operational evidence contradicts this model. Contemporary adversaries do not primarily target cryptographic primitives; they target the conditions under which those primitives are implemented and executed.

The emergence of artificial intelligence–enabled attack systems introduces a qualitatively different threat model. Polymorphic and fully morphing malware systems apply machine learning and reinforcement learning to observe, infer, and optimise attacks against cryptographic processes in real time. These systems exploit entropy weaknesses, side-channel leakage, and key-management dependencies without violating the formal properties of cryptographic algorithms. As a result, security degradation occurs through iterative optimisation rather than brute-force computation.

This shift creates a structural misalignment between cryptographic theory and operational security. Existing research remains concentrated on algorithm design, key length, and protocol verification, while under-representing adversaries capable of learning from system behaviour. Empirical evidence indicates that private key compromise increasingly arises from adaptive inference and optimisation techniques, particularly in cloud-based and virtualised environments where observability is high.

The consequence is that cryptographic assurance, as currently defined, does not correspond to real-world resilience. Systems that satisfy formal security definitions may still be vulnerable to sustained, low-noise exploitation driven by adaptive adversarial strategies. Addressing this discrepancy requires reframing cryptographic security from a static property of algorithms to a dynamic property of systems interacting with adversarial learning processes. This paper is motivated by the need to formalise this shift and to derive security requirements that reflect the operational realities of AI-enabled adversaries.

## Introduction to Cryptographic Resilience in AI-Enabled Adversarial Environments

Cryptography [1], or cryptology, has its roots in the Ancient Greek words ‘kryptós’, meaning hidden or secret, ‘graphein’, signifying to write, and ‘logia’, denoting study [2]. It is a key element in modern-day security [3], grounded in cryptographic algorithms constructed around the ‘computational hardness assumption’ [4].

Public Key Cryptography underpins the confidentiality, integrity, authentication, and non-repudiation guarantees of modern digital systems. Its security rests on formally defined computational hardness assumptions, including the infeasibility of integer factorisation, discrete logarithms, and related one-way functions under classical computation. These assumptions enable secure key exchange, digital signatures, encrypted communications, and trusted identity verification across financial systems, critical infrastructure, cloud platforms, and distributed ledgers.

In operational environments, cryptographic mechanisms support chip-and-PIN payment systems, secure authentication protocols, military and diplomatic communications, public blockchain infrastructures, and regulatory compliance regimes governing data protection and transactional integrity [5]. Transport Layer Security (TLS), public key infrastructures (PKI), and digital signature schemes form the backbone of secure web services, email systems, and software update mechanisms. Cryptographic primitives such as hash functions, asymmetric key pairs, and zero-knowledge proofs further enable privacy-preserving computation and decentralised trust models.

**Dr. Petar Radanliev**
Parks Road,
Oxford OX1 3PJ
United Kingdom
Email: petar.radanliev@cs.ox.ac.uk

BA Hons., MSc., Ph.D. Post-Doctorate

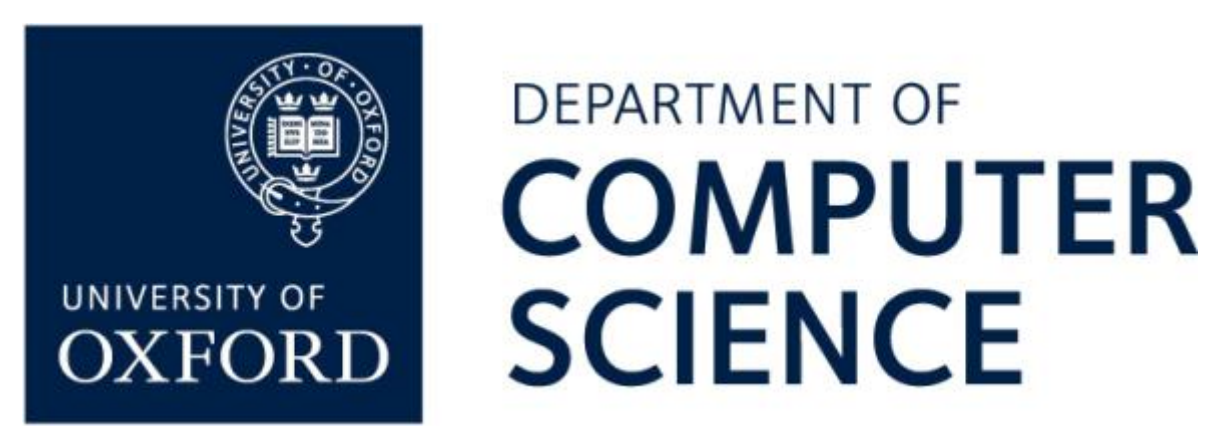

Contemporary cryptographic deployments rely on a combination of symmetric and asymmetric algorithms. The Advanced Encryption Standard (AES) remains the dominant symmetric cipher for bulk data protection due to its efficiency and resistance to known cryptanalytic attacks. Legacy algorithms such as Triple DES persist in regulated environments despite well-documented security limitations. Triple DES is distinctive for its thrice encryption methodology using the Data Encryption Standard (DES) cipher, a technique known as the Data Encryption Algorithm (DEA), originally derived from the Lucifer cipher [6].

Asymmetric cryptography is dominated by RSA and elliptic curve schemes, which enable scalable key distribution and digital signature verification [7]. These algorithms operate within regulatory frameworks such as PCI-DSS and GDPR, which mandate cryptographic safeguards for personal, financial, and operational data [8], [9].

The security model underlying these systems assumes that cryptographic algorithms fail only through infeasible computation, not through adaptive adversarial optimisation. That assumption no longer holds. Artificial intelligence has reshaped the cryptographic threat landscape by enabling adversaries to optimise attacks against cryptographic implementations rather than attempting to defeat cryptographic primitives directly. Polymorphic and fully morphing malware constitute a substantive escalation in adversarial capability because they combine machine learning, reinforcement learning, and evolutionary optimisation to adapt continuously to defensive controls, execution environments, and system responses. Unlike conventional malware, which depends on static signatures or predetermined execution paths, these systems modify their observable and internal characteristics in response to feedback, rendering detection, attribution, and mitigation increasingly unreliable.

Polymorphic malware alters its external representation across executions while preserving semantic functionality, enabling systematic evasion of signature-based detection and static analysis. Fully morphing malware extends this behaviour by evolving execution logic, control flow, and exploitation strategies in real time, guided by environmental observations and defensive reactions. When directed at cryptographic targets, these capabilities enable adversaries to exploit weaknesses in entropy generation, side-channel leakage, key-management workflows, and certificate validation processes with a degree of precision and persistence that exceeds the limits of traditional attack models. The resulting attacks do not rely on exhaustive brute-force enumeration but instead reduce effective security margins incrementally through adaptive optimisation.

Public Key Cryptography is particularly exposed to these attack dynamics because its operational security depends on assumptions that extend beyond mathematical hardness. Secure key generation requires high-quality entropy; private-key protection depends on strict isolation guarantees; trust relationships rely on the integrity of certificate authorities and key-management systems. AI-enabled adaptive malware systematically targets these dependencies. By exploiting entropy degradation in virtualised environments, optimising side-channel extraction through iterative learning, and identifying transient trust failures in PKI and cloud-based cryptographic services, private keys can be compromised without violating the formal security properties of the underlying algorithms.

Operational evidence from industrial incident response and threat intelligence indicates that private-key compromise increasingly results from AI-augmented optimisation of attack surfaces rather than from procedural error, misconfiguration, or insider activity. This shift undermines security models that treat cryptographic compromise as a rare or exceptional

**Dr. Petar Radanliev**
Parks Road,
Oxford OX1 3PJ
United Kingdom
Email: petar.radanliev@cs.ox.ac.uk

BA Hons., MSc., Ph.D. Post-Doctorate

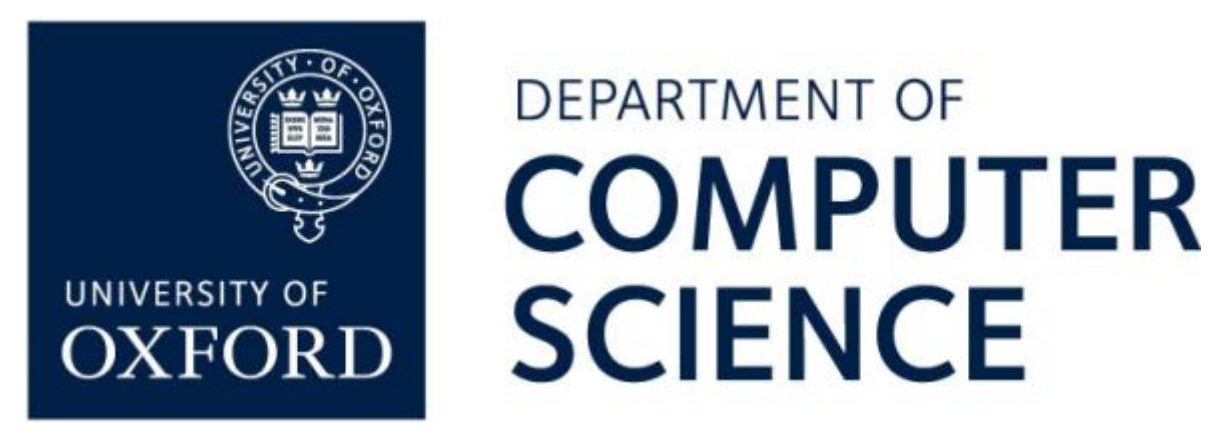

event. Instead, adaptive adversarial systems transform key compromise into a scalable, repeatable process driven by continuous learning and environmental adaptation.

A structural mismatch therefore exists between prevailing cryptographic threat models and the operational reality of AI-enabled adversaries. Much of the existing literature continues to prioritise algorithmic strength, key length, and formal proofs of security, while under-analysing the exploitation of cryptographic implementations by adversaries capable of learning, adapting, and optimising in real time. Post-quantum cryptography mitigates future quantum computational threats but does not inherently address AI-driven exploitation of entropy sources, side-channel emissions, key-management processes, or certificate infrastructures. Cryptographic resilience increasingly depends on the interaction between algorithmic hardness, implementation security, adversarial adaptability, and defensive observability rather than on any single control dimension.

The analytical focus therefore centres on identifying cryptographic failure modes exploited by AI-driven polymorphic and fully morphing malware across key generation, exchange, storage, and validation; characterising the security, legal, and trust implications of private-key compromise under adaptive adversarial conditions; evaluating the limitations of key size and algorithm selection as primary defensive controls when attacks are optimised through machine learning; and deriving cryptographic design requirements that integrate post-quantum resilience with AI-aware defensive mechanisms.

The analysis integrates bibliometric evidence and industrial qualitative data to examine cryptographic vulnerability under AI-enabled attack models. Laboratory implementation of cryptographic algorithms or malware systems is deliberately excluded. Emphasis remains on threat characterisation, failure-mode analysis, and defensible security requirements rather than tool-level performance benchmarking.

This work advances the state of knowledge beyond existing cryptographic and cybersecurity literature in four specific ways. First, it introduces an adaptive adversarial optimisation model that formally reframes cryptographic compromise as a function of adversarial learning over system observability rather than computational infeasibility. Second, it provides the first empirically grounded mapping between AI-enabled malware behaviour and cryptographic lifecycle vulnerabilities, integrating bibliometric evidence with practitioner-derived data. Third, it develops a cross-layer threat taxonomy that unifies entropy degradation, side-channel inference, and trust infrastructure exploitation under a single analytical framework, which is absent in current literature. Fourth, it proposes a design paradigm for AI-aware cryptographic resilience, extending beyond post-quantum cryptography by incorporating continuous monitoring, adaptive key governance, and dynamic trust validation. Collectively, these contributions differentiate this study from prior work that remains constrained to algorithm-centric or protocol-level security analysis.

The remainder of the paper is structured as follows. Section 2 presents the methodological framework. Section 3 analyses bibliometric evidence of research gaps. Section 4 examines limitations of existing approaches. Sections 5–8 develop the adaptive adversarial model across cryptographic mechanisms and attack vectors. Section 9 presents empirical results, followed by a discussion of implications for cryptographic resilience.

The core problem addressed in this study is that contemporary cryptographic security models assume static, computationally bounded adversaries, whereas real-world attackers increasingly employ adaptive, AI-driven optimisation that targets implementation-level observability and trust dependencies.

**Dr. Petar Radanliev**
Parks Road,
Oxford OX1 3PJ
United Kingdom
Email: petar.radanliev@cs.ox.ac.uk

BA Hons., MSc., Ph.D. Post-Doctorate

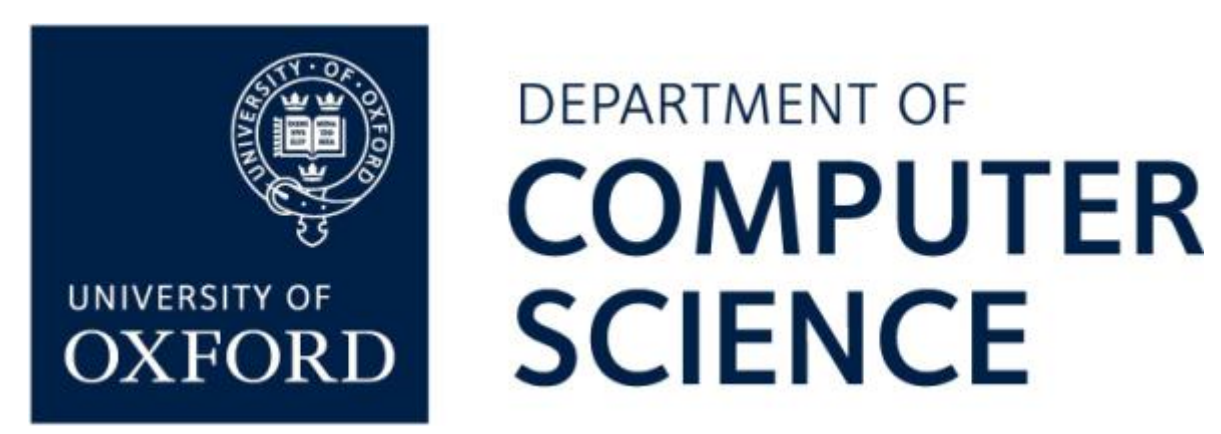

## 2. Methodological Framework for Evaluating Cryptographic Resilience under Adaptive Adversaries

The analytical design combines bibliometric analysis, qualitative empirical evidence, and cryptographic threat modelling to examine how AI-enabled polymorphic and fully morphing malware exploits Public Key Cryptography (PKC) [10] at the implementation and operational layers. The methodological structure is deliberately analytical–empirical, prioritising reproducibility, triangulation, and technical interpretability over narrative synthesis. Literature review serves as contextual grounding, while primary analytical weight is placed on structured data extraction, empirical validation, and failure-mode analysis.

The analytical process applies thematic analysis to identify recurring cryptographic failure patterns associated with adaptive malware behaviour. Thematic extraction focuses on attack mechanisms affecting key generation entropy, private-key isolation, side-channel exposure, key exchange integrity, and certificate validation reliability. These themes are derived through systematic coding of bibliometric metadata and qualitative interview material, enabling comparison between academic research emphasis and observed industrial threat exposure. Attention is directed toward mechanisms by which polymorphic malware evades static detection through structural mutation, and by which fully morphing malware evolves exploitation strategies in response to defensive countermeasures and runtime conditions.

To improve methodological transparency and reproducibility, the analytical workflow is formalised as a structured multi-stage process:

**Step 1: Data Acquisition** – Retrieval of bibliometric records using a deterministic query and extraction of structured metadata fields.

**Step 2: Bibliometric Processing** – Application of co-occurrence analysis, clustering, and factorial methods using the bibliometrix R package.

**Step 3: Qualitative Coding** – Thematic coding of interview and workshop data using a predefined schema aligned with cryptographic lifecycle stages.

**Step 4: Cross-Dataset Triangulation** – Mapping bibliometric themes against empirical observations to identify convergence and divergence.

**Step 5: Threat Modelling** – Construction of an adaptive adversarial model linking AI-enabled attack mechanisms to specific cryptographic dependencies.

**Step 6: Synthesis and Validation** – Derivation of cryptographic failure modes and validation through consistency across datasets.

This structured workflow ensures that analytical conclusions are not inferred heuristically but emerge from reproducible and systematically integrated evidence streams.

Triangulation is employed to enhance validity by cross-referencing three independent evidence streams: peer-reviewed scientific literature, industrial qualitative data, and documented real-world incident reports relating to AI-enabled malware activity. Convergence across these sources is treated as an indicator of analytical robustness, while divergence is examined as evidence of research gaps or misaligned threat models. Ethical and methodological safeguards are applied throughout data handling and analysis to ensure integrity, confidentiality, and analytical consistency.

The selection of a mixed analytical–empirical design is justified by the need to capture structural research gaps and operational threat realities. Bibliometric analysis alone would identify publication trends but not attack mechanisms, while qualitative evidence without

**Dr. Petar Radanliev**
Parks Road,
Oxford OX1 3PJ
United Kingdom
Email: petar.radanliev@cs.ox.ac.uk

BA Hons., MSc., Ph.D. Post-Doctorate

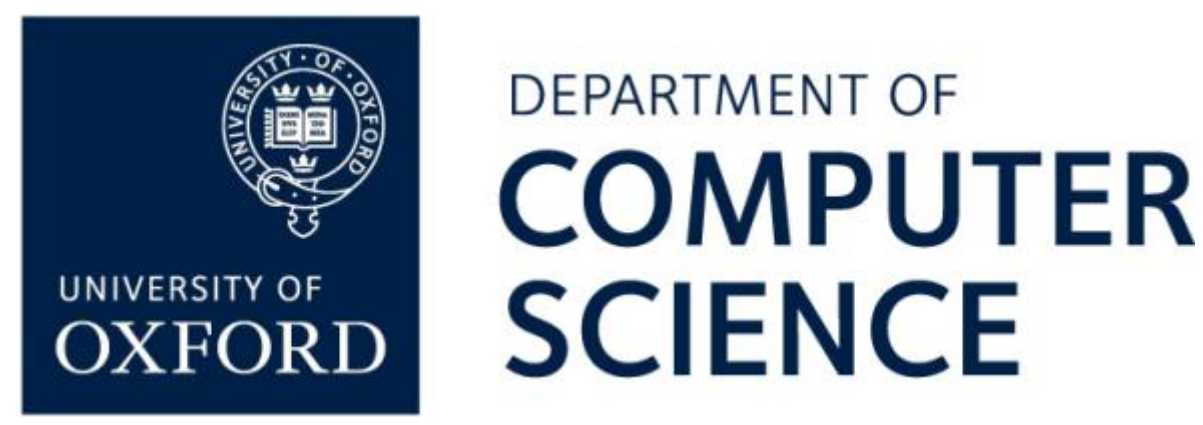

structural context would lack generalisability. The integration of both enables identification of systemic misalignment between research focus and adversarial behaviour, which is central to the study's objective.

The analytical scope excludes laboratory implementation of cryptographic algorithms or malware artefacts. Instead, cryptographic security implications are derived from observed attack patterns, empirical evidence, and implementation-level failure analysis. AI-aware countermeasures are specified as design requirements grounded in identified vulnerabilities rather than as deployed or benchmarked systems. This approach reflects the objective of evaluating cryptographic resilience under adaptive adversarial pressure rather than measuring tool-level performance.

A structured visual analytical framework supports the methodology by contextualising cryptographic deployment characteristics. The analytical perspective illustrated in Figure 1 integrates four dimensions relevant to cryptographic exposure: algorithm usage distribution, regulatory compliance prevalence, application-domain concentration, and comparative performance characteristics.

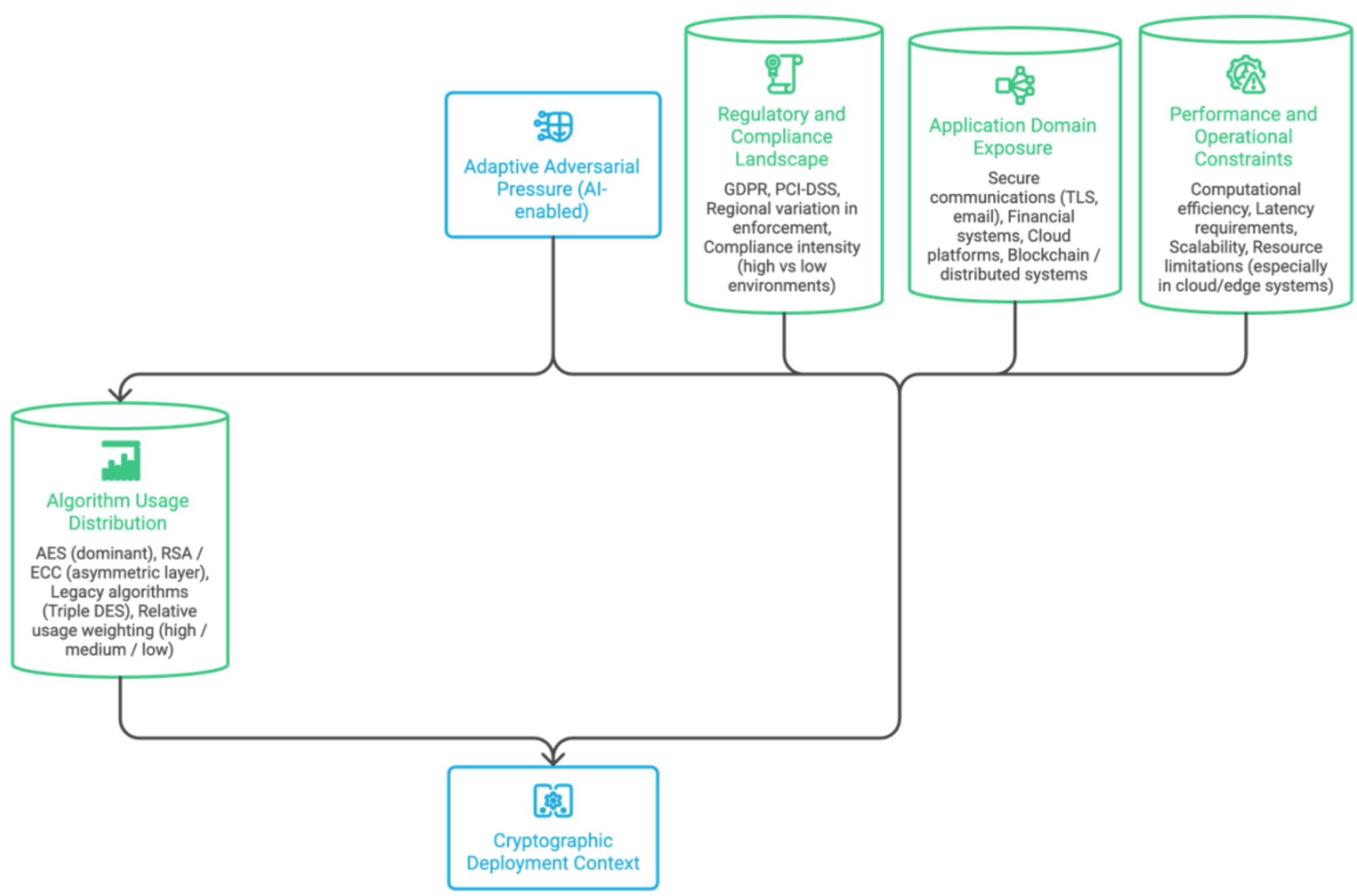

*Figure 1: Analytical Framework for Cryptographic Deployment and Exposure*

Figure 1 is included to provide a visual abstraction of the analytical dimensions used in the study and to support interpretation of subsequent empirical findings. Figure 1 presents a four-dimensional framework in which cryptographic deployment is shaped by the interaction between algorithm usage distribution, regulatory and compliance requirements, application-domain concentration, and performance constraints. These dimensions collectively define the operational exposure surface of cryptographic systems, illustrating that security is not determined solely by algorithm selection but by the broader system context in which

**Dr. Petar Radanliev**
Parks Road,
Oxford OX1 3PJ
United Kingdom
Email: petar.radanliev@cs.ox.ac.uk

BA Hons., MSc., Ph.D. Post-Doctorate

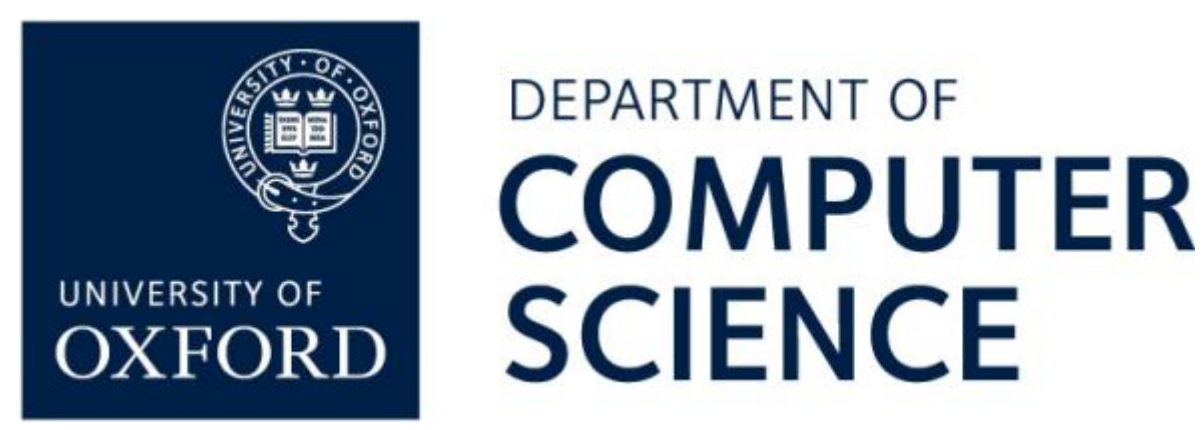

cryptography is implemented. The framework further highlights how adaptive adversarial pressure, particularly from AI-enabled attack mechanisms, intersects with these dimensions to amplify implementation-level vulnerabilities.

The encryption algorithm distribution highlights industry reliance on AES, legacy persistence of Triple DES, and continued use of RSA-based asymmetric schemes. Regulatory compliance analysis reflects differential adherence across jurisdictions, indicating varying enforcement and risk exposure. Application-domain frequency illustrates the concentration of cryptographic usage in secure transactions and encrypted communications, with lower but non-trivial deployment in crypto-economic systems. Comparative performance analysis provides a high-level indication of algorithm efficiency and operational suitability, informing later discussion on why performance-centric selection alone is insufficient under adaptive attack conditions.

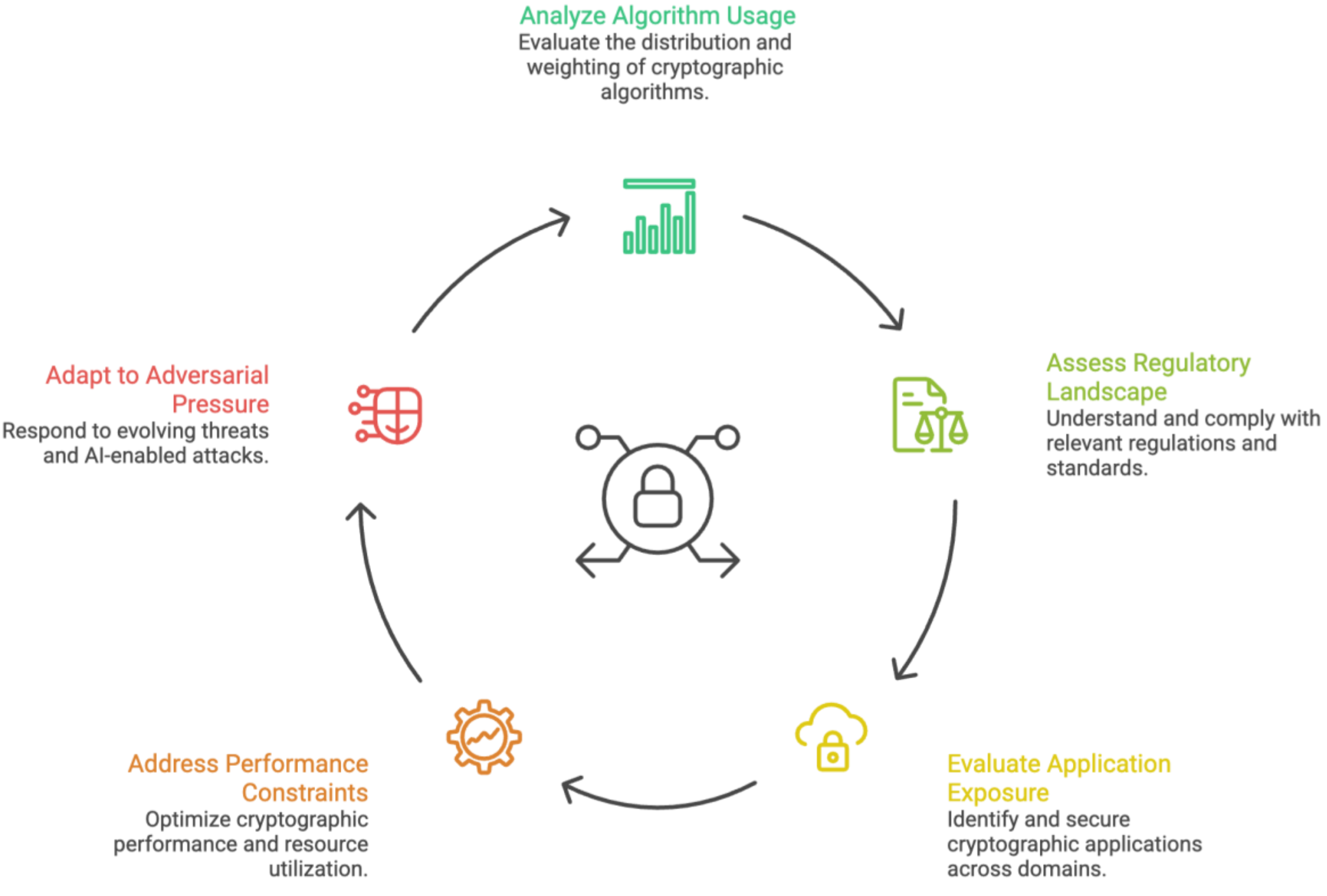

*Figure 2: Cryptographic Deployment Cycle*

Figure 2 details the ‘Cryptographic Deployment Cycle’, where two datasets underpin the empirical analysis. The first consists of a bibliometric corpus of twenty-three peer-reviewed publications retrieved from the Web of Science Core Collection using the exact search string “Cyber-attacks on Public Key Cryptography” across SCIE, SSCI, CPCI, and ESCI indices. Extracted metadata includes authorship, institutional affiliation, geographic origin, publication year, keywords, and citation metrics. The dataset is fully reproducible through identical query parameters and database scope.

The second dataset comprises qualitative empirical evidence drawn from twenty semi-structured expert interviews and three industry workshops conducted with cybersecurity practitioners from Cisco between 2022 and 2023. Interview protocols focus on observed

**Dr. Petar Radanliev**
Parks Road,
Oxford OX1 3PJ
United Kingdom
Email: petar.radanliev@cs.ox.ac.uk

BA Hons., MSc., Ph.D. Post-Doctorate

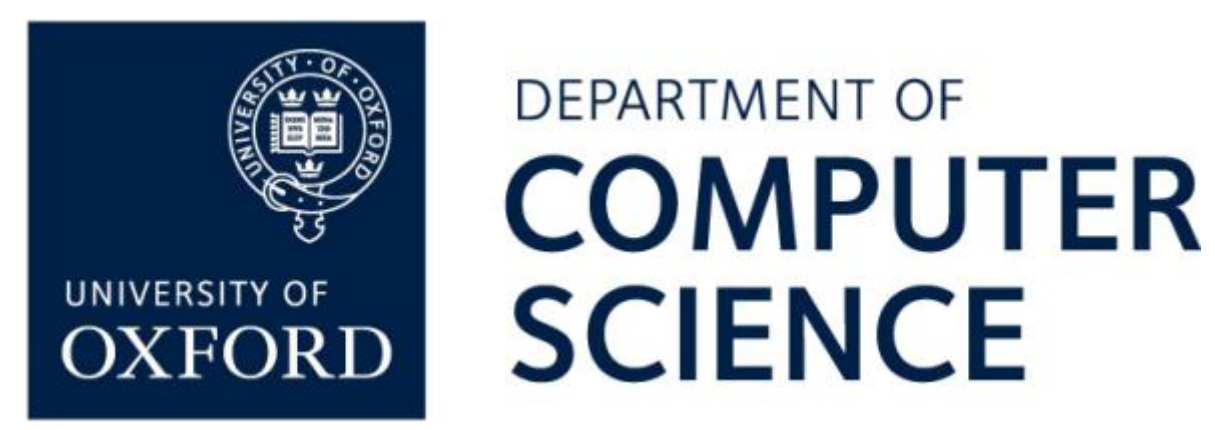

cryptographic failure incidents, mechanisms of private-key compromise, adaptive malware behaviour, and the operational impact of AI-enabled attacks on cryptographic infrastructures. Data are anonymised and thematically coded following established qualitative research practices.

Analytical reproducibility is ensured through deterministic bibliometric processing using the *bibliometrix* package within the R statistical environment, including co-occurrence analysis, hierarchical clustering, factorial analysis, and correspondence analysis. Qualitative coding reliability is validated through inter-coder agreement assessment, achieving a Cohen's κ coefficient of 0.86, indicating high consistency in theme identification. Together, these procedures establish a transparent and replicable methodological foundation for analysing cryptographic vulnerability under AI-enabled adaptive adversarial conditions.

Figure 3 introduces the conceptual structure of the paper by mapping the transition from algorithm-centric cryptographic assumptions to adaptive adversarial optimisation, and the resulting pathways to systemic cryptographic failure.

**Dr. Petar Radanliev**
Parks Road,
Oxford OX1 3PJ
United Kingdom
Email: petar.radanliev@cs.ox.ac.uk

BA Hons., MSc., Ph.D. Post-Doctorate

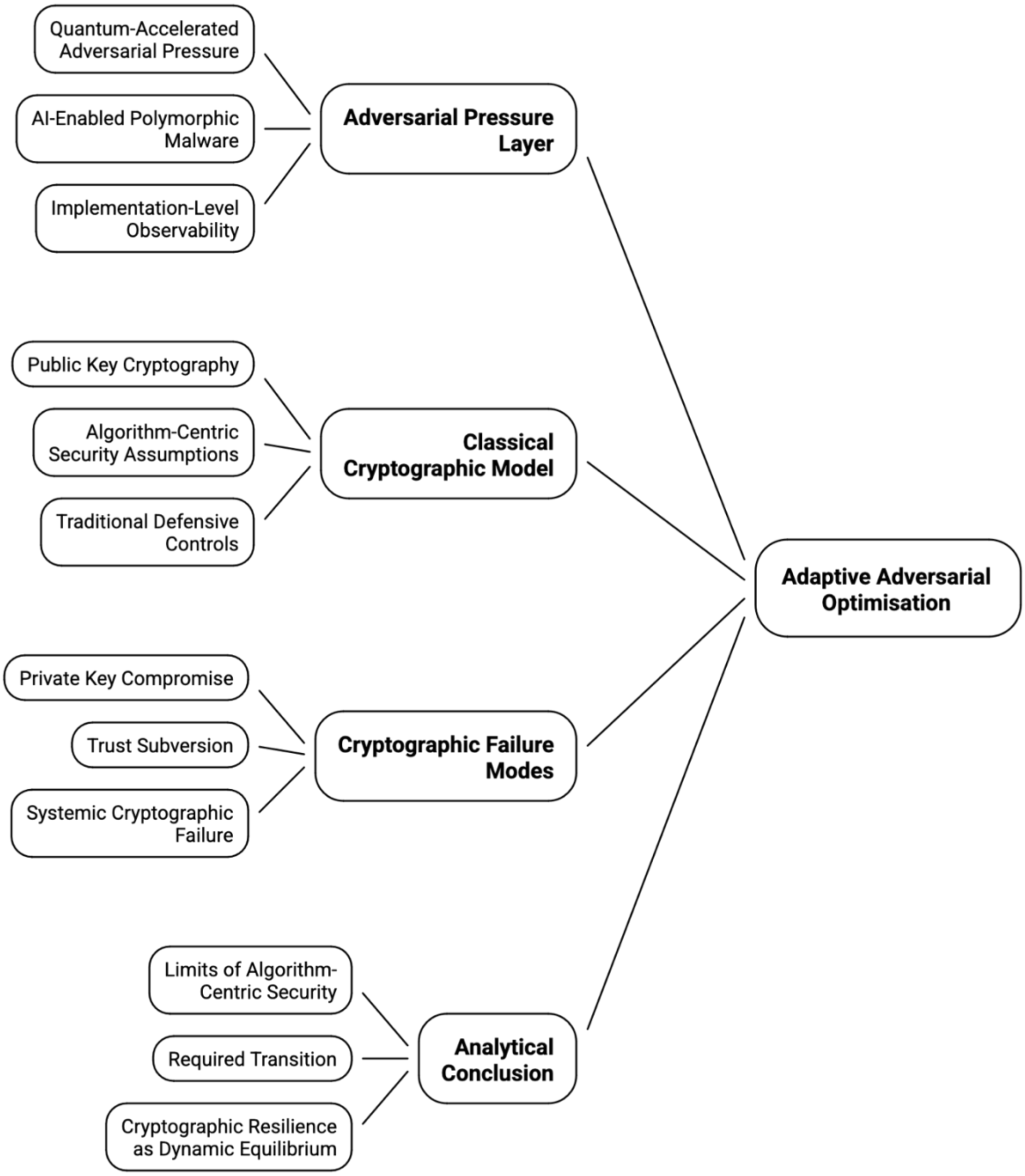

*Figure 3: Conceptual Structure of Adaptive Adversarial Optimisation and Cryptographic Failure*

Figure 3 presents a layered analytical model illustrating the transition from classical, algorithm-centric cryptographic assumptions to adaptive adversarial optimisation. The upper layer captures adversarial pressure driven by AI-enabled polymorphic and fully morphing malware, reinforced by quantum acceleration and implementation-level observability. This pressure introduces feedback-driven optimisation loops that destabilise static security assumptions within the classical cryptographic model. The intermediate layer shows how

**Dr. Petar Radanliev**
Parks Road,
Oxford OX1 3PJ
United Kingdom
Email: petar.radanliev@cs.ox.ac.uk

BA Hons., MSc., Ph.D. Post-Doctorate

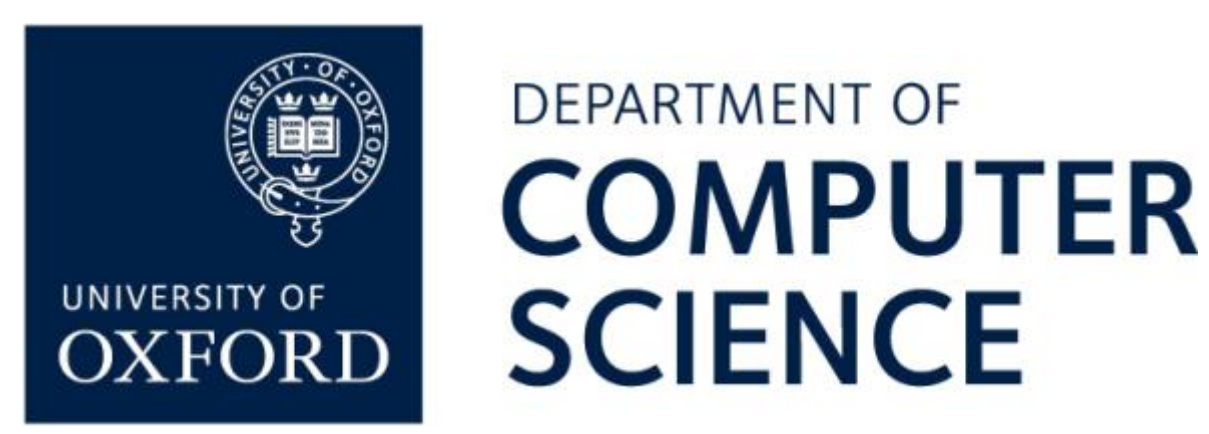

these assumptions collapse, leading to private key compromise, trust subversion, and ultimately systemic cryptographic failure. The lower layer formalises the analytical implication: cryptographic security can no longer be understood as a function of algorithmic strength alone, but must transition towards adaptive resilience, conceptualised as a dynamic equilibrium between adversarial learning and defensive response. Figure 3 shows how AI-enabled polymorphic and fully morphing malware, amplified by quantum-accelerated pressure and implementation-level observability, collapses classical security assumptions and necessitates a shift towards cryptographic resilience as a dynamic equilibrium.

## Formalisation of the Adaptive Adversarial Threat Model

The adaptive adversarial threat model developed in this study departs from classical cryptographic assumptions by explicitly modelling the adversary as a learning system operating over observable cryptographic processes. Rather than treating the adversary as computationally bounded but behaviourally static, the model assumes access to continuous observation channels, including timing characteristics, memory access patterns, side-channel emissions, and protocol-level interactions. Adversarial capability is therefore defined by three interdependent properties: (i) learning capacity, whereby machine learning techniques infer statistical structure from repeated cryptographic executions; (ii) observability, reflecting the degree to which implementation-level artefacts expose information about cryptographic state; and (iii) adaptive strategy optimisation, through which adversaries iteratively refine attack policies in response to defensive controls and environmental feedback. Under these conditions, cryptographic systems are not treated as black boxes with fixed security margins, but as partially observable systems whose internal states can be approximated over time.

System exposure within this model is formalised as a function of the interaction between cryptographic operations and their execution environment. Key generation, signature computation, key exchange, and validation processes produce observable artefacts that, while individually non-exploitable, become statistically informative under aggregation and iterative analysis. The adversarial optimisation loop is therefore characterised by a closed feedback cycle: observation, inference, action, environmental response, updated observation. Reinforcement learning and probabilistic inference enable adversaries to prioritise attack vectors that maximise information gain while minimising detection probability. This transforms cryptographic compromise from a discrete event into a convergent process, where security degradation emerges asymptotically as adversarial models improve. Within this formalisation, cryptographic resilience is not solely determined by algorithmic hardness, but by the rate at which defensive mechanisms can disrupt or outpace adversarial learning, thereby constraining the optimisation process and limiting effective observability.

$$\pi^{*} = \arg\max_{\pi \in \Pi} \mathbb{E}\left[\sum_{t=0}^{T} \gamma^{t} r(o_t, a_t)\right], \quad \text{subject to } o_t = \mathcal{O}(s_t), a_t \sim \pi(o_t)$$

In this formulation, the adversary is modelled as an adaptive policy \pi \in \Pi that selects actions a_t based on observable outputs o_t derived from the underlying cryptographic system state s_t. The observation function \mathcal{O}(s_t) captures implementation-level leakage and execution artefacts, including timing behaviour, memory access patterns, and side-channel emissions. The adversarial objective is to maximise the expected cumulative reward over a finite horizon T, where the reward function r(o_t, a_t) encodes information gain, key inference accuracy, or successful compromise. The optimisation process is discounted by a factor \gamma \in (0,1], reflecting the relative importance of immediate versus long-term exploitation. This can also be represented as:

**Dr. Petar Radanliev**
Parks Road,
Oxford OX1 3PJ
United Kingdom
Email: petar.radanliev@cs.ox.ac.uk

BA Hons., MSc., Ph.D. Post-Doctorate

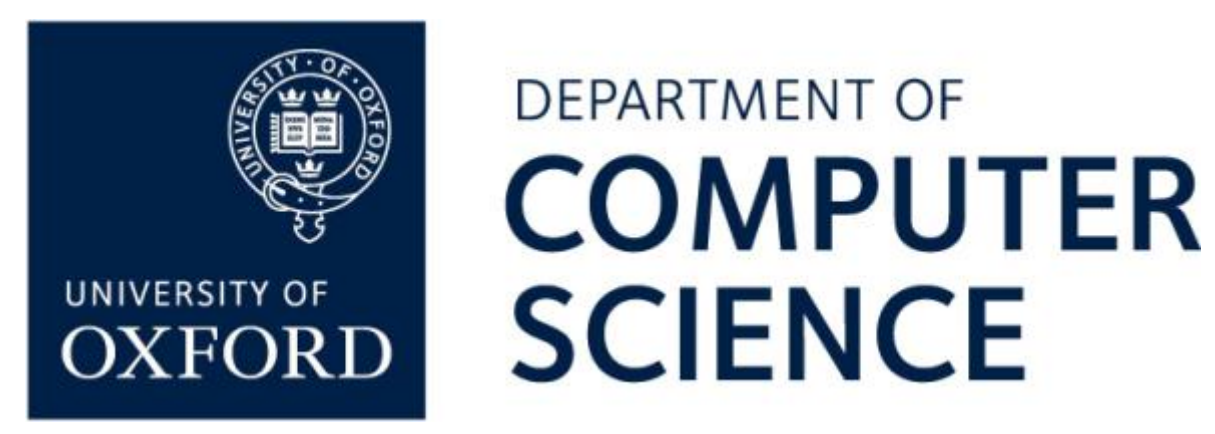

$$\pi^* = \arg\max_{\pi \in \Pi} \mathbb{E}\left[\sum_{t=0}^{T} \gamma^t\, r(o_t, a_t)\right], \quad \text{subject to } o_t = \mathcal{O}(s_t),\ a_t \sim \pi(o_t)$$

This formulation captures the adaptive adversarial optimisation loop as a sequential decision process in which observation, inference, and action are iteratively refined. Rather than relying on exhaustive search or static attack strategies, the adversary updates its policy \pi based on observed system responses, progressively improving attack efficiency while minimising detectability. Cryptographic security, under this model, is therefore not solely determined by algorithmic hardness, but by the extent to which system observability can be constrained and the rate at which defensive mechanisms can disrupt policy convergence.

# 3. Bibliometric Evidence of Research Gaps in Cyber-Attacks on Public Key Cryptography

Before we start with a deep analysis of cyber-attacks on PK cryptography, we need to review what research has been conducted in this area. We opted for bibliometric analysis to analyse large volumes of data at speed. The first step in the bibliometric analysis included the Web of Science Core Collection to identify the volume of knowledge in this area. The Web of Science Core Collection includes:

- Science Citation Index Expanded (SCIE) (Coverage:1965-present)
- Social Sciences Citation Index (SSCI) (Coverage:1965-present)
- Arts & Humanities Citation Index (AHCI) (Coverage:1975-present)
- Book Citation Index (BKCI) (Coverage: 2005-present)
- Conference Proceedings Citation Index (CPCI) (Coverage:1990-present)
- Emerging Sources Citation Index (ESCI) (Coverage: 2017-present)

We started with the Web of Science Core Collection because it represents more than 21,000 peer-reviewed, high-quality scholarly journals published worldwide in over 250 sciences, social sciences, and arts & humanities disciplines, as well as conference proceedings and book data.

The first search was conducted on ‘Cyber-attacks on PK Cryptography’, producing only 23 results. Considering this topic's importance in cybersecurity, the number of data records seems extremely low. The 23 results are analysed with the Web of Science Analyse Results tool - Figure 4.

**Dr. Petar Radanliev**
Parks Road,
Oxford OX1 3PJ
United Kingdom
Email: petar.radanliev@cs.ox.ac.uk

BA Hons., MSc., Ph.D. Post-Doctorate

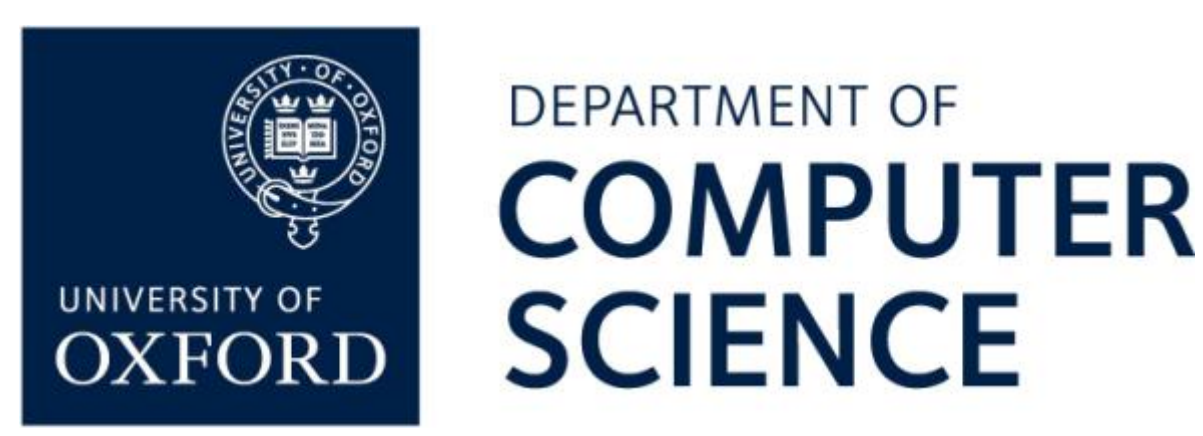

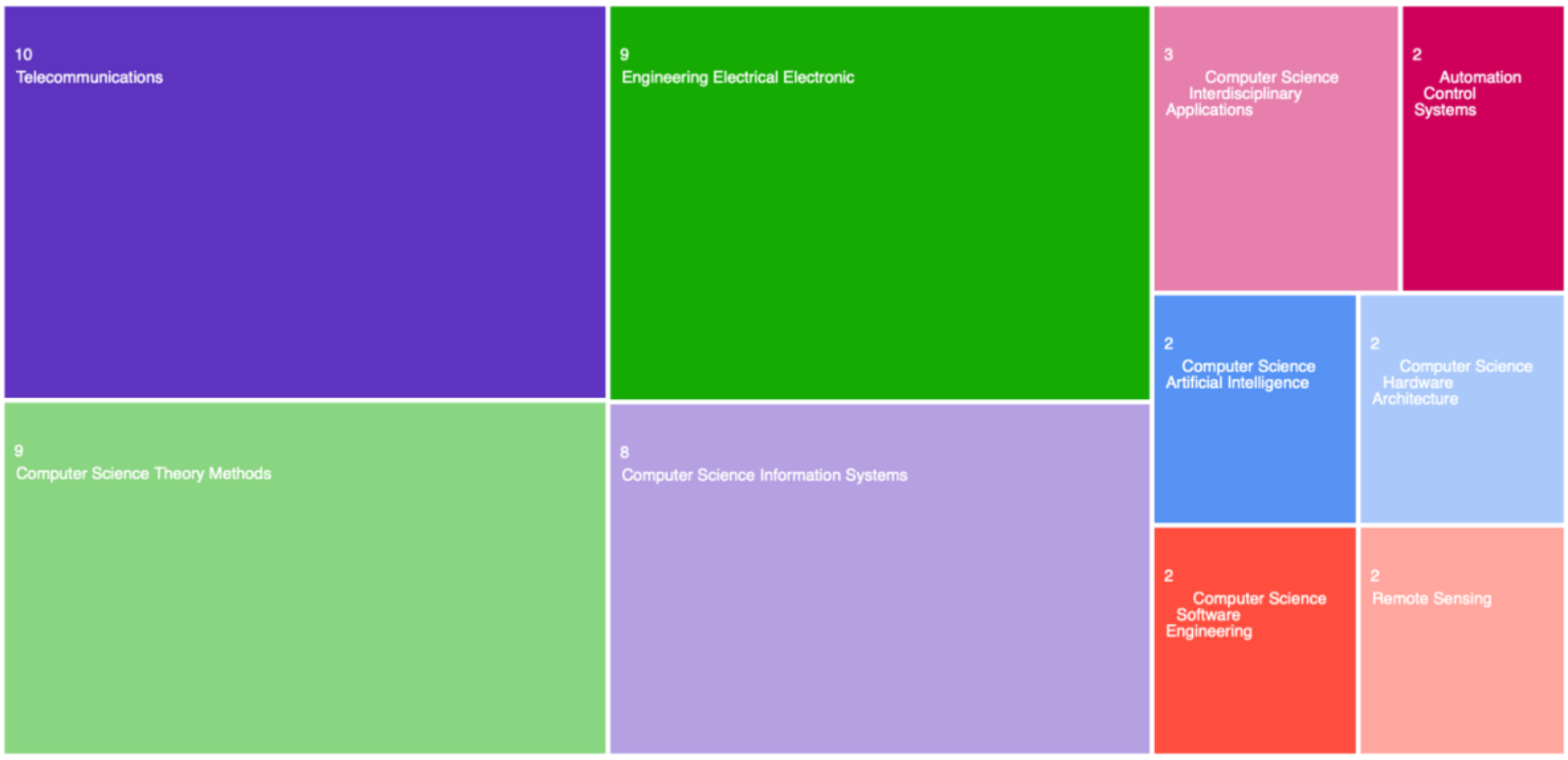

*Figure 4: Web of Science data records on cyber-assaults against Public Key (PK) Cryptography*

What we can see in Figure 4, is that the main areas of research on this topic are Telecommunications, Theory Methods, Electrical Engineering, Information Systems, and, to a lesser extent, Software Engineering and Artificial Intelligence. We wanted to analyse the data in more detail to extract key information from main areas of interest and to cluster the data records in specific categories related to where most of the research was conducted, who the leading authors, institutions, countries, and what are the correlations between different research studies. Bibliometric analysis was conducted using the bibliometrix package within R Studio, employing co-occurrence network analysis, hierarchical clustering, factorial analysis, and correspondence analysis to identify thematic structures, geographic distributions, and research gaps within the extracted dataset [11].

The first type of analysis we performed was just to get an overview of the data records. As we can see in Figure 5, the entire timespan of the Web of Science data records is from 2021 to 2023, with 23 sources, where most of the sources are from collaborative work, and only one single authored document is recorded in the Web of Science data records.

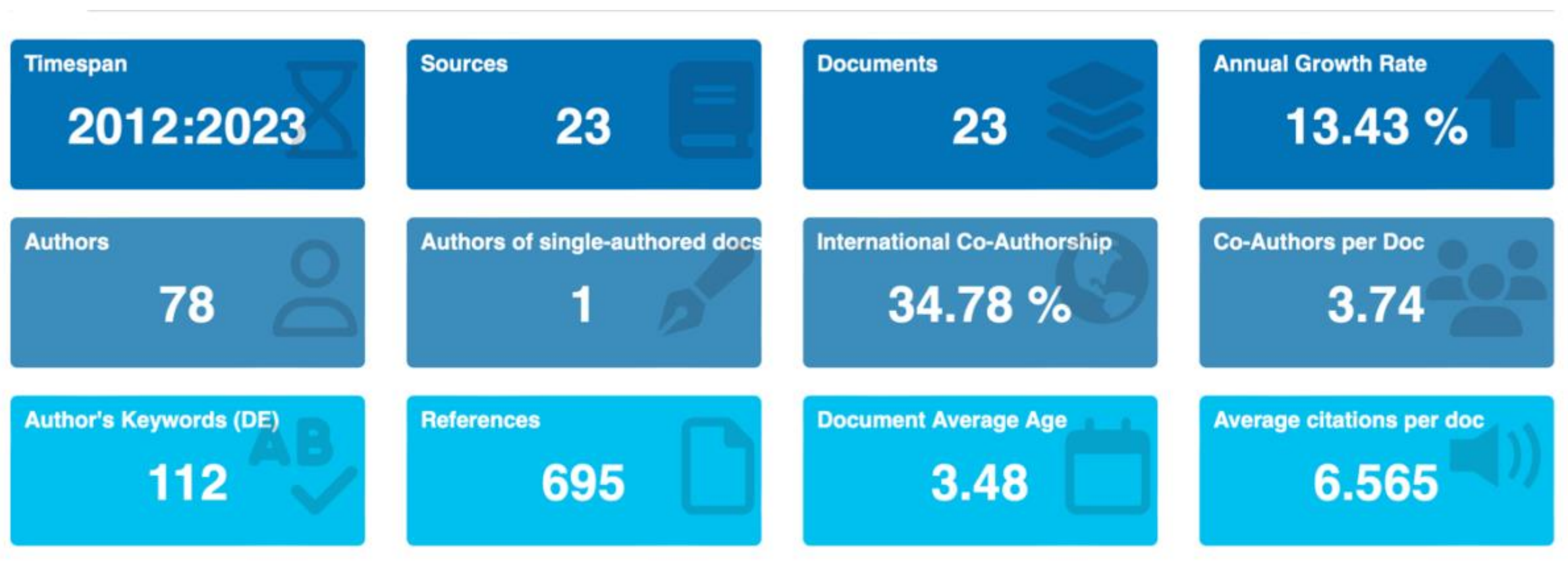

*Figure 5: Overview of the Web of Science data records*

**Dr. Petar Radanliev**
Parks Road,
Oxford OX1 3PJ
United Kingdom
Email: petar.radanliev@cs.ox.ac.uk

BA Hons., MSc., Ph.D. Post-Doctorate

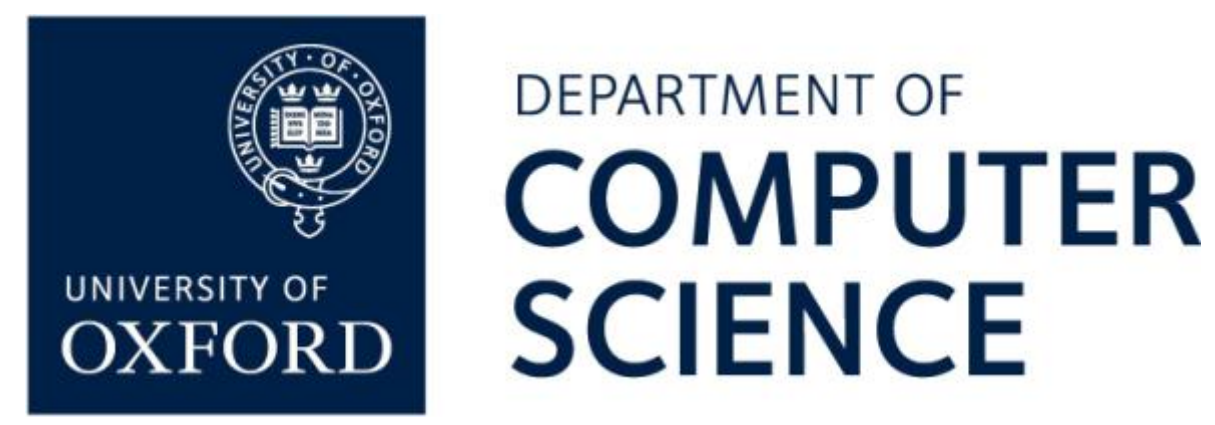

We wanted to separate the data records by keywords, countries, and affiliations in the following analysis. To visualise this, we developed a Three-Field Plot Figure 6.

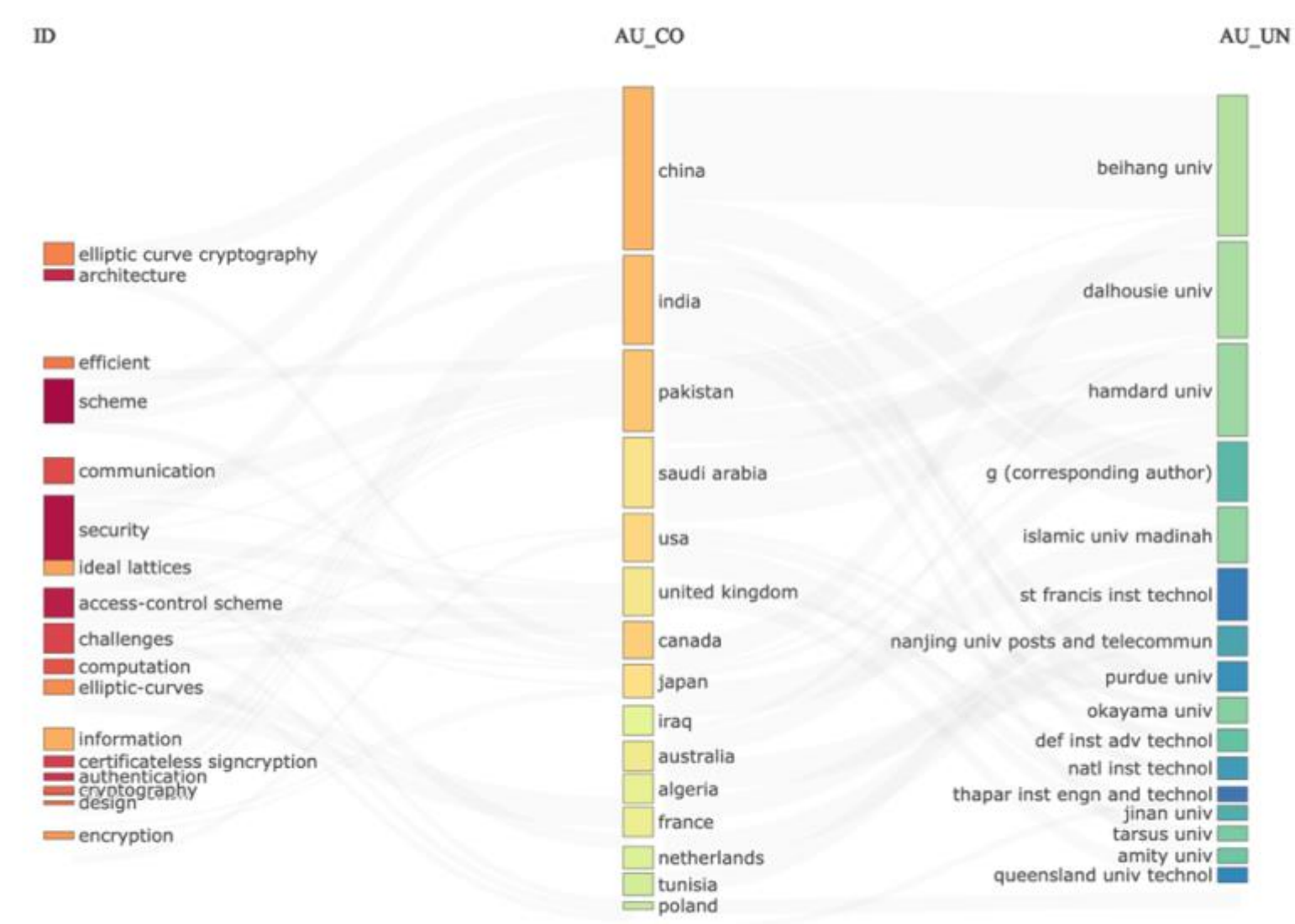

*Figure 6: Three-field plot categorising the data on keywords, countries, and affiliations.*

From analysing the data records in a Three-field plot, we noticed that China, India, Pakistan, and Saudi Arabia dominate this research area, which is somewhat unexpected because cryptography, as a subject, is traditionally associated with the United States and the United Kingdom. When we analysed the connections in the United States and research records, we discovered that most of the research in this area has been conducted in the field of encryption. To compare this with the United Kingdom, when we analysed the data records by keywords, we determined that most work conducted in the United Kingdom was related to security and access control. We intend to make this dataset publicly available for other researchers to analyse the data with different statistical methods, but we need to emphasise that we have clearly outlined the process of obtaining this dataset. The data was extracted from the Web of Science database, with the search ‘Cyber-attacks on PK Cryptography’, and we typed the exact words in the Web of Science search engine. Even without the dataset, if, for example, the dataset is deleted by some accident in the future, this same dataset can be reproduced by following these steps.

In the final analysis, we separated the data records into dendrogram subcategories - Figure 7, clustered correlated categories with Factorial Analysis - Figure 8, and performed Correspondence Analysis based on Dim1 and Dim2 - Figure 9.

**Dr. Petar Radanliev**
Parks Road,
Oxford OX1 3PJ
United Kingdom
Email: petar.radanliev@cs.ox.ac.uk

BA Hons., MSc., Ph.D. Post-Doctorate

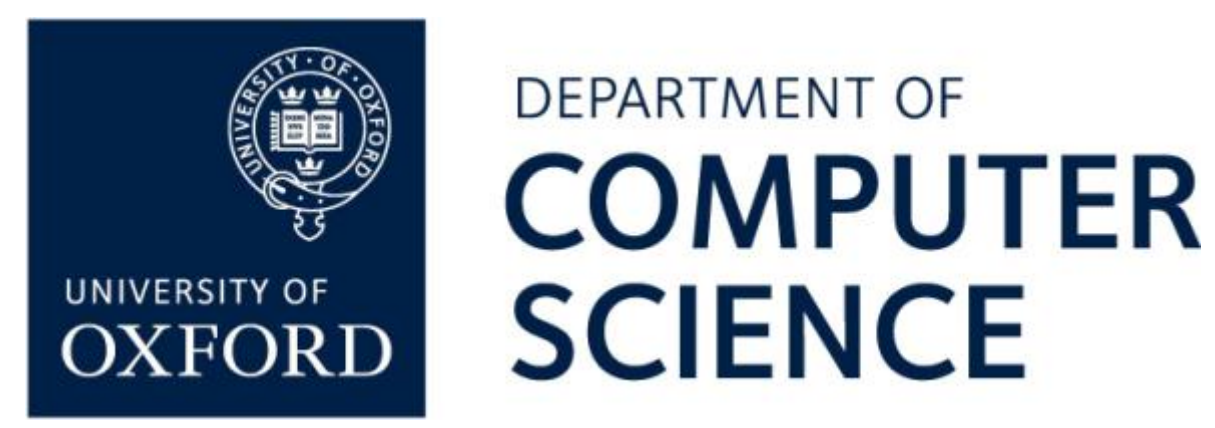

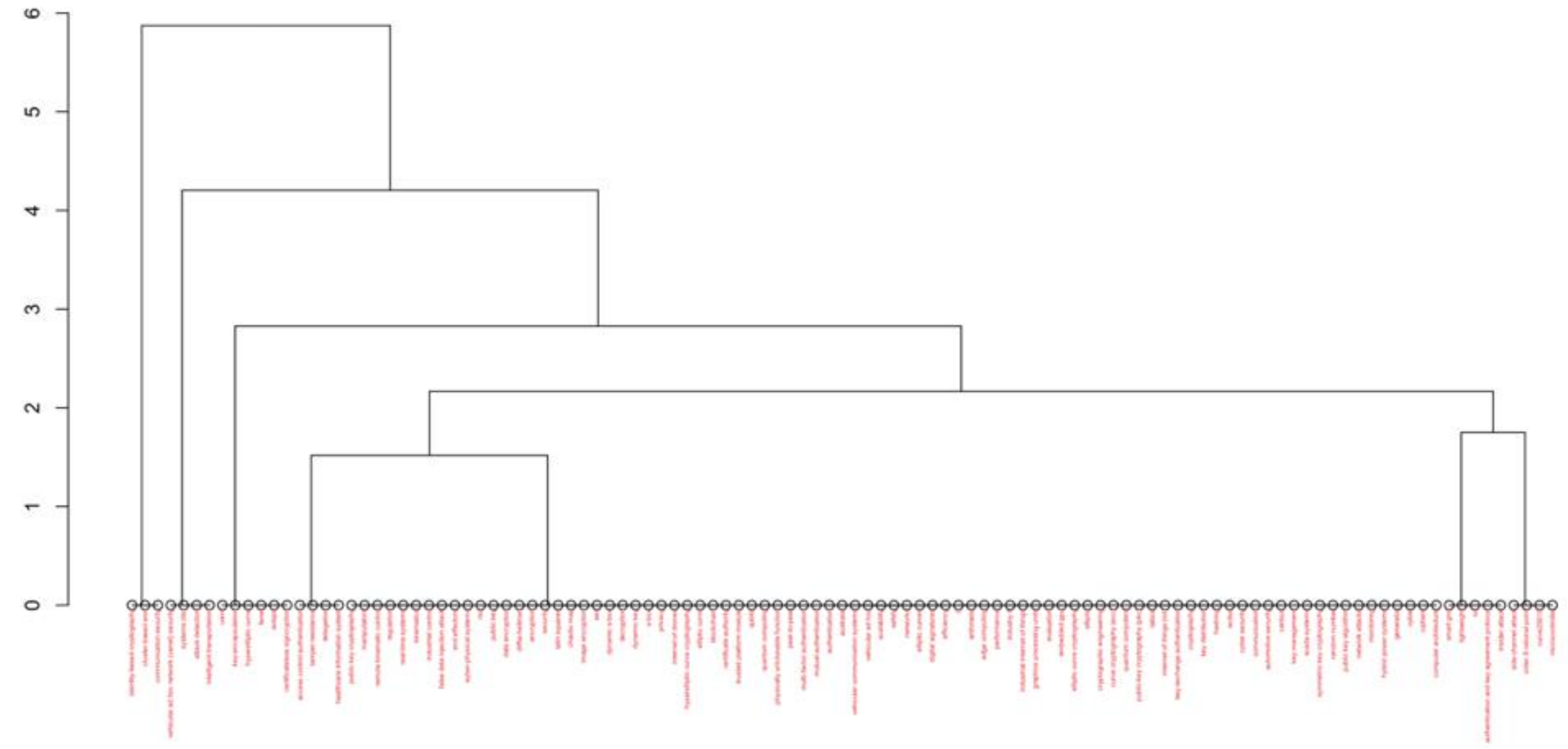

*Figure 7: Dendrogram*

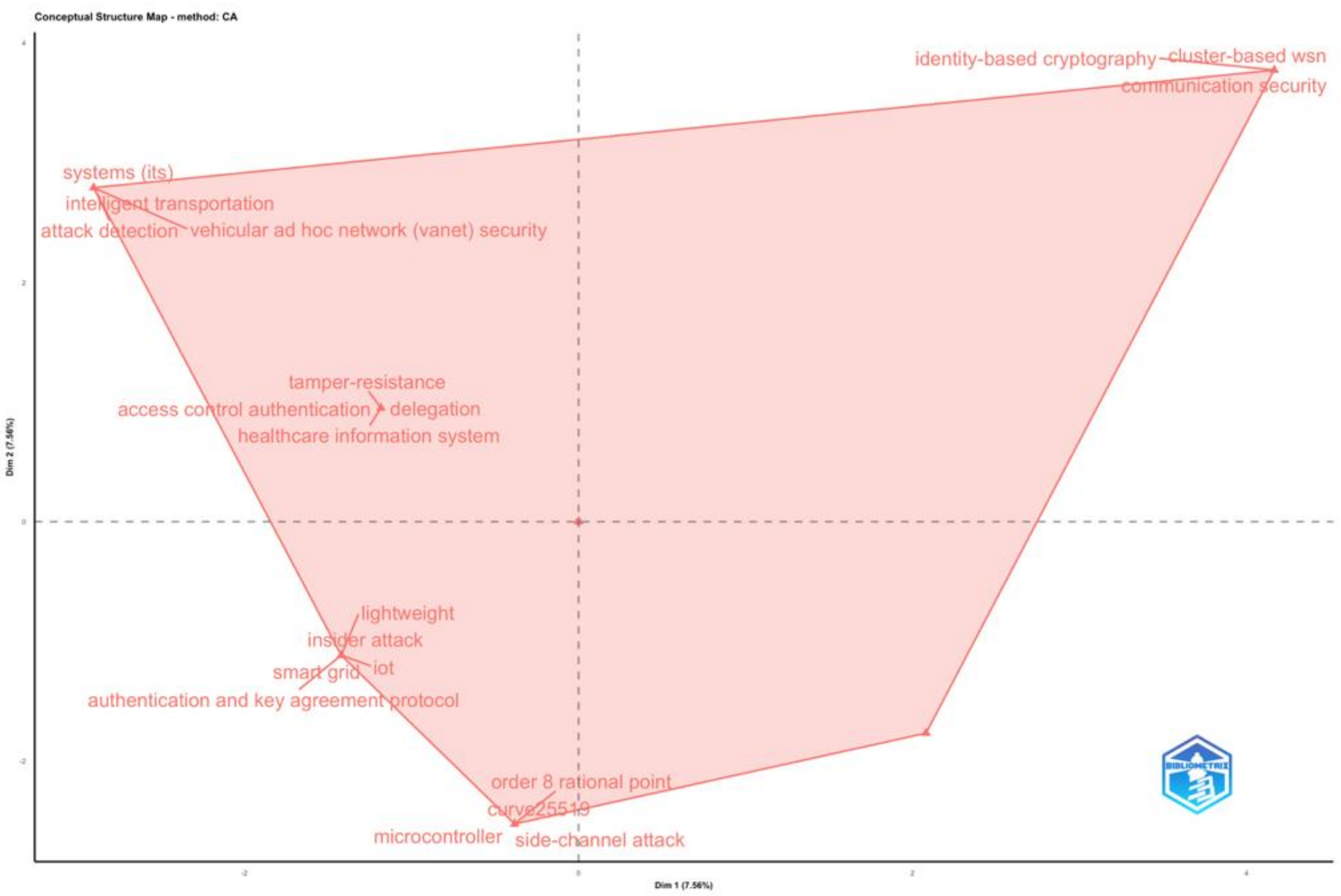

*Figure 8: Factorial Analysis*

**Dr. Petar Radanliev**
Parks Road,
Oxford OX1 3PJ
United Kingdom
Email: petar.radanliev@cs.ox.ac.uk

BA Hons., MSc., Ph.D. Post-Doctorate

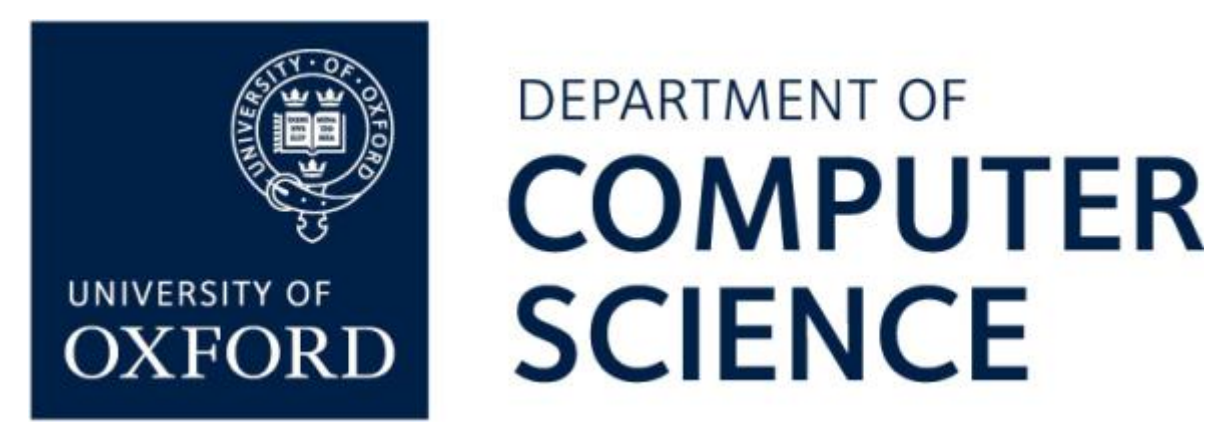

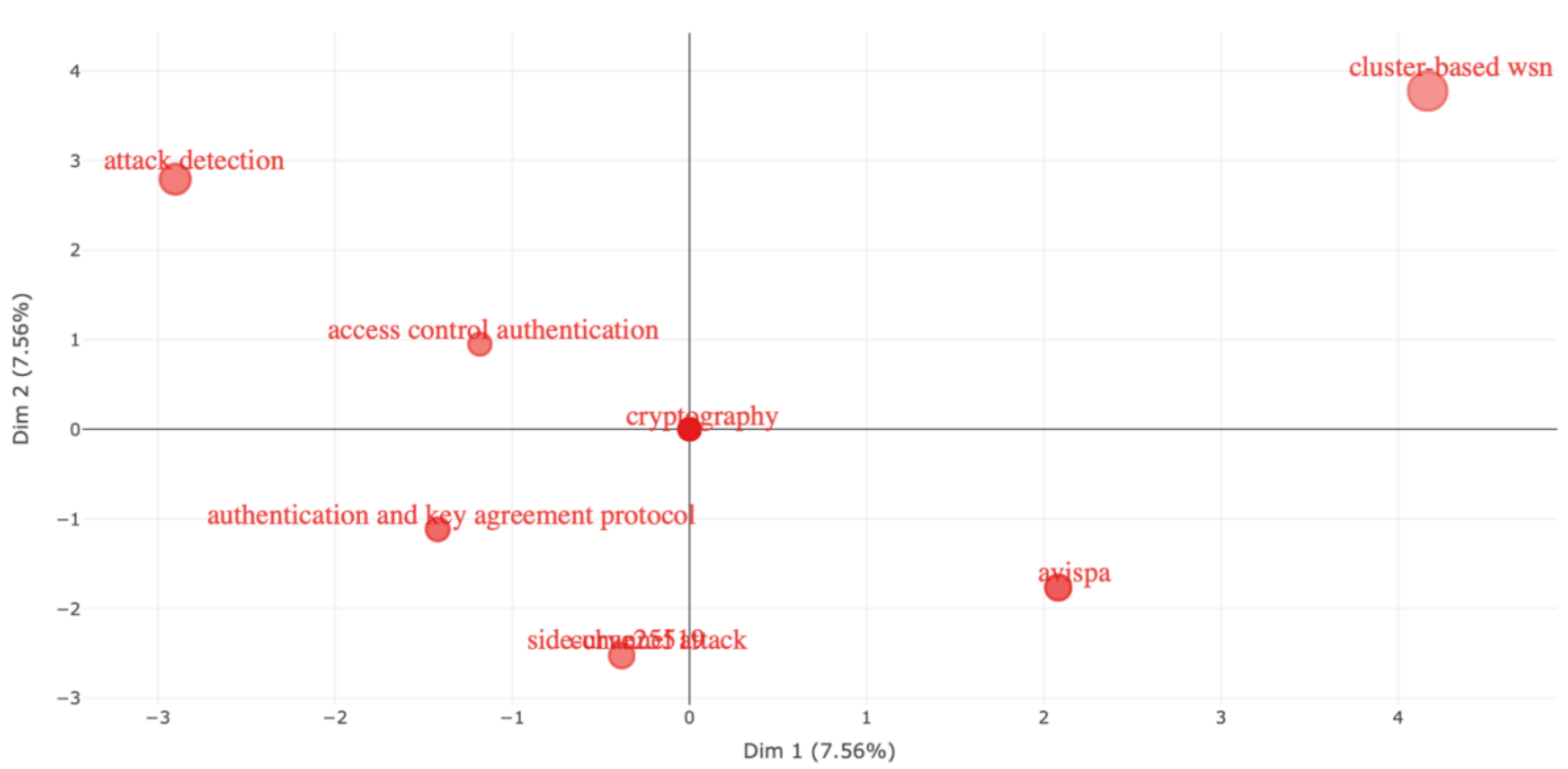

*Figure 9: Correspondence Analysis based on Dim1 and Dim2*

What becomes evident in these three figures is not what's included but what is missing. In none of these figures we see any evidence of research on the leading cyber-attacks on PK cryptography. In other words, these are the data records from the leading scientific research database, which integrates multiple databases in one single data record. This specific data record is on the specific research area of cyber-attacks on PK cryptography. Yet, we see no evidence of any significant clusters or categories emerging for any specific cyber-attack on PK cryptography. Hence, in this research, we focus on addressing this gap, and we present new knowledge on the key factors related to cyber-attacks on PK cryptography, including the specific cyber-attacks, the approaches, the methods, the prevention techniques, and many other related topics to this specific research area.

# 4. Limitations of Existing Cryptographic Research under Adaptive and AI-Enabled Threat Models

Research on cyber-attacks against Public Key Cryptography has historically concentrated on algorithmic robustness, protocol correctness, and resistance to classical cryptanalytic techniques. Foundational work in this area has focused on strengthening encryption schemes, improving key distribution mechanisms, and formally verifying cryptographic correctness under static threat assumptions. However, the rapid emergence of artificial intelligence–enabled polymorphic and fully morphing malware has exposed limitations in these approaches, particularly where security depends on implementation integrity, entropy quality, and trusted infrastructure rather than purely on mathematical hardness.

Several studies propose cryptographic enhancements without explicitly accounting for adaptive adversaries. Work on securing networked control systems through RSA-based encryption demonstrates how homomorphic properties can protect control parameters and encrypted signals under conventional attack models [12]. While effective against passive interception and basic manipulation, such approaches do not address adversaries capable of dynamically reconfiguring attack strategies in response to system feedback. AI-enabled

**Dr. Petar Radanliev**
Parks Road,
Oxford OX1 3PJ
United Kingdom
Email: petar.radanliev@cs.ox.ac.uk
BA Hons., MSc., Ph.D. Post-Doctorate

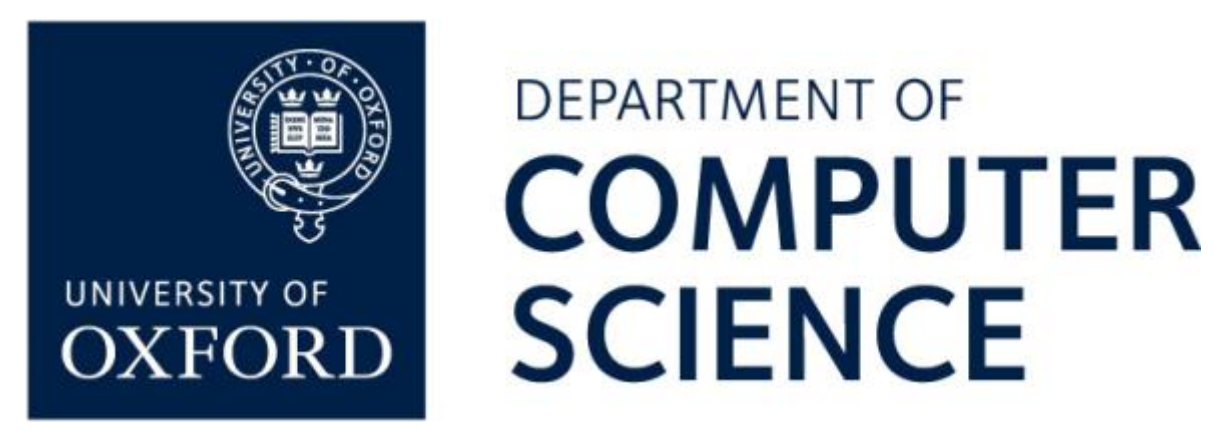

polymorphic malware can bypass static encryption assumptions by exploiting implementation-level weaknesses, rendering purely algorithm-centric protections insufficient.

Related research on securing One-Time Password mechanisms similarly prioritises encryption efficiency and transmission security [13]. These approaches improve resistance to unauthorised access in constrained environments but remain vulnerable to fully morphing malware capable of learning authentication patterns and mimicking legitimate behaviour. By adapting to timing characteristics, retry logic, and distribution workflows, AI-enabled malware can subvert OTP systems without breaking the underlying cryptographic primitives, highlighting a recurring disconnect between theoretical security and operational resilience.

Formal verification of cryptographic systems has been advanced as a means of improving assurance, particularly in cyber-physical systems [14]. Verification techniques strengthen confidence in protocol correctness and resistance to known attack classes, yet they assume stable adversarial behaviour. Fully morphing malware undermines this assumption by evolving exploitation logic in response to deployed defences, exposing a gap between formally verified security properties and runtime adversarial adaptation. This gap necessitates complementary mechanisms capable of detecting and responding to behavioural evolution rather than relying solely on static correctness guarantees.

More general treatments of cryptography in network security provide valuable overviews of encryption techniques across hardware and software contexts [15]. These contributions reinforce the importance of cryptographic primitives for confidentiality and integrity but largely omit the implications of AI-enabled adversaries. Polymorphic and fully morphing malware introduces attack vectors that operate orthogonally to traditional cryptanalysis, targeting entropy sources, key isolation boundaries, side-channel emissions, and certificate validation workflows rather than algorithmic weaknesses.

Recent proposals for novel public-key encryption schemes, including constructions inspired by near-ring criteria and physical system analogies such as the Einstein crystal model, demonstrate improved resistance to linear, differential, and noise-based attacks [16]. While these schemes enhance robustness against established attack classes, their resilience under AI-driven adversarial optimisation remains untested. Fully morphing malware equipped with predictive modelling capabilities may exploit implementation artefacts, parameter selection patterns, or operational constraints, bypassing protections designed primarily for static threat environments.

Across the reviewed literature, a consistent pattern emerges: cryptographic security is predominantly evaluated under assumptions of fixed adversarial behaviour and bounded attack strategies. Polymorphic and fully morphing malware violates these assumptions by introducing continuous learning, adaptive optimisation, and environment-specific exploitation. As a result, traditional defences that focus on algorithm strength, key size, or protocol correctness fail to capture the dominant risk vectors observed in operational environments.

Post-quantum cryptography addresses an orthogonal but equally significant challenge by mitigating vulnerabilities arising from quantum computation, particularly against factorisation and discrete logarithm problems. While these advances are essential for long-term cryptographic viability, they do not inherently address AI-enabled exploitation of entropy sources, side-channel leakage, key-management processes, or certificate infrastructures. When deployed within conventional software and cloud environments, post-quantum schemes remain exposed to adaptive adversaries capable of optimising attacks against implementation-level weaknesses.

**Dr. Petar Radanliev**
Parks Road,
Oxford OX1 3PJ
United Kingdom
Email: petar.radanliev@cs.ox.ac.uk

BA Hons., MSc., Ph.D. Post-Doctorate

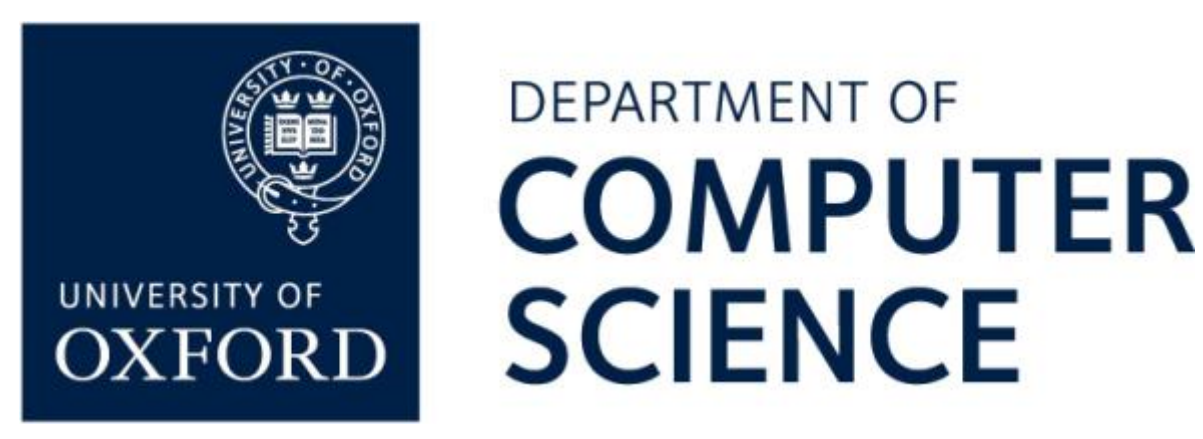

Empirical and institutional sources further reinforce this assessment. Industrial incident reports, standards documentation, and regulatory guidance from organisations such as CSAF, NTIA, CISA, and NIST document increasing exposure of cryptographic systems to adaptive threats, particularly within supply-chain, cloud, and critical infrastructure contexts CSAF [17]–[60]. Legislative and policy initiatives, including cybersecurity transparency mandates and executive orders on national cybersecurity, emphasise improved visibility and resilience but do not yet fully integrate AI-aware cryptographic threat models [61]–[63].

Taken together, the literature demonstrates a clear gap between cryptographic research priorities and the operational realities of AI-enabled adversarial behaviour. Existing studies provide strong foundations in algorithmic security and formal assurance but insufficiently address adversaries capable of learning, adapting, and optimising in real time. The intersection between adaptive artificial intelligence and cryptographic deployment therefore remains underexplored, despite its immediate relevance to the confidentiality, integrity, and trust guarantees upon which modern digital systems depend.

*Table 1: Comparison between Existing Cryptographic Approaches and Proposed Framework*

| Dimension | Existing Approaches | This Paper |
|---|---|---|
| Threat Model | Static, computationally bounded | Adaptive, learning-driven adversary |
| Focus | Algorithms & protocols | Full cryptographic lifecycle |
| Attack Surface | Cryptanalysis | Implementation + observability |
| AI Consideration | Minimal / absent | Central analytical variable |
| Validation | Formal proofs | Empirical + bibliometric + analytical |
| Defence Strategy | Key size, protocol strength | AI-aware monitoring + adaptive governance |
| PQC Integration | Algorithm substitution | System-level integration with adaptive controls |

The comparison in Table 1 demonstrates that existing methods treat cryptographic security as a static property derived from algorithm selection, whereas the proposed framework models it as a dynamic system influenced by adversarial adaptation. This distinction is critical because it explains why formally secure systems fail in operational environments despite compliance with established standards.

# 5. Evolution of Core Cryptographic Mechanisms under AI-Enabled Adversarial Adaptation

The Digital signature schemes constitute a critical trust anchor in Public Key Cryptography, providing authentication, integrity assurance, and non-repudiation across distributed digital systems. From a cryptographic perspective, the signature lifecycle comprises four tightly coupled stages: key generation, message hashing, signature computation, and verification. Classical security analysis assumes that each stage operates under stable adversarial conditions, where attacks are bounded by computational infeasibility rather than adaptive optimisation.

Figure 10 conceptualises the structural evolution of cryptographic attack surfaces (see Figure 9), providing a visual anchor for the transition from static to adaptive adversarial models and organising adversarial capabilities into three distinct but sequential threat phases. The diagram distinguishes between algorithm-centric threats grounded in classical

**Dr. Petar Radanliev**
Parks Road,
Oxford OX1 3PJ
United Kingdom
Email: petar.radanliev@cs.ox.ac.uk

BA Hons., MSc., Ph.D. Post-Doctorate

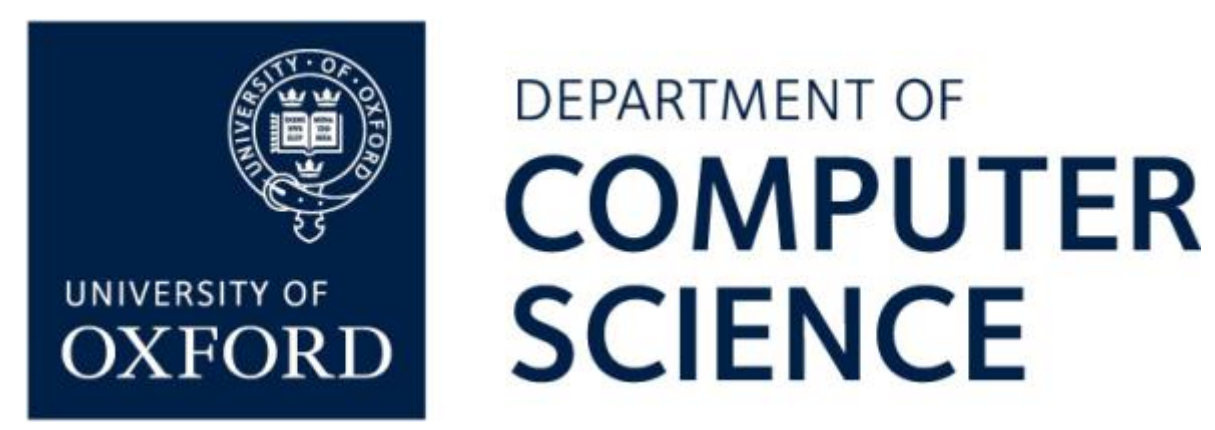

cryptanalysis, implementation-centric threats exploiting operational and system-level weaknesses, and adaptive AI-driven optimisation that enables continuous adversarial learning across the cryptographic lifecycle. This phased representation provides an analytical baseline for understanding how cryptographic compromise has shifted from mathematically bounded attacks to dynamically optimised exploitation strategies.

**Dr. Petar Radanliev**
Parks Road,
Oxford OX1 3PJ
United Kingdom
Email: petar.radanliev@cs.ox.ac.uk

BA Hons., MSc., Ph.D. Post-Doctorate

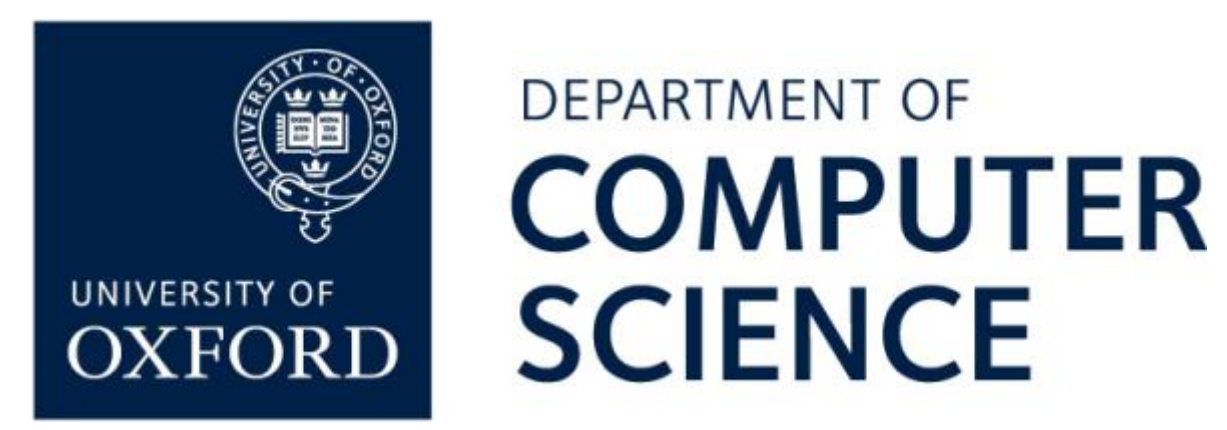

- **Evolution of Cryptographic Attack Surfaces**
  - **Phase 1: Algorithm-Centric Threats**
    - Brute-force attacks
    - Classical cryptanalysis
    - Mathematical hardness assumptions
    - Fixed adversary capabilities
  - **Phase 2: Implementation-Centric Threats**
    - Side-channel leakage
    - Entropy degradation
    - Key-management failures
    - Certificate validation flaws
  - **Phase 3: Adaptive AI-Driven Optimisation**
    - Polymorphic malware
    - Fully morphing malware
    - Reinforcement learning–based attack optimisation
    - Runtime adaptation to defences
  - **Overlay Layer**
    - Classical Adversaries
    - AI-Enabled Adversaries
    - Quantum-Assisted Adversaries

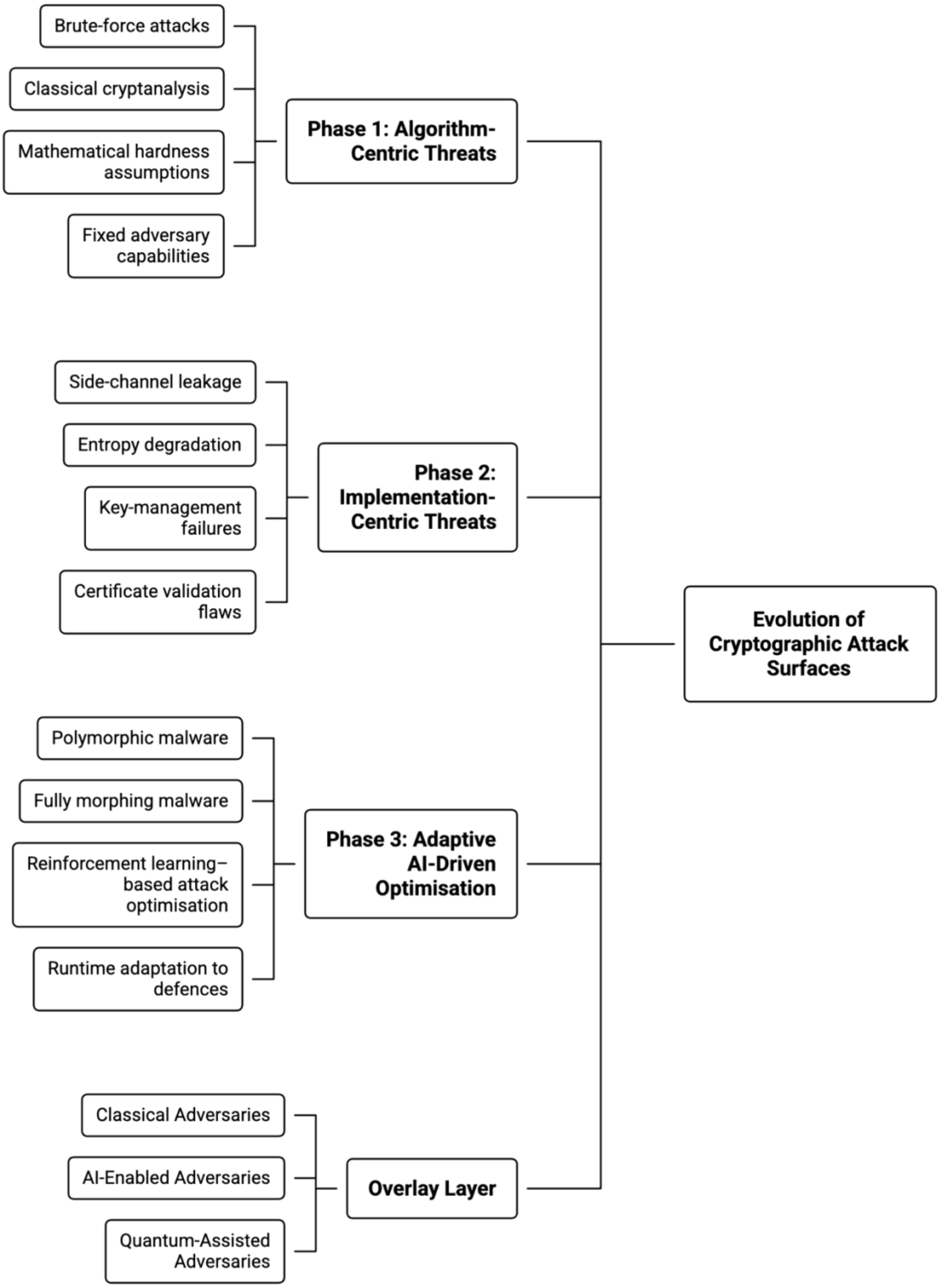

*Figure 10: Evolution of Cryptographic Attack Surfaces across Adversarial Phases*

**Dr. Petar Radanliev**
Parks Road,
Oxford OX1 3PJ
United Kingdom
Email: petar.radanliev@cs.ox.ac.uk

BA Hons., MSc., Ph.D. Post-Doctorate

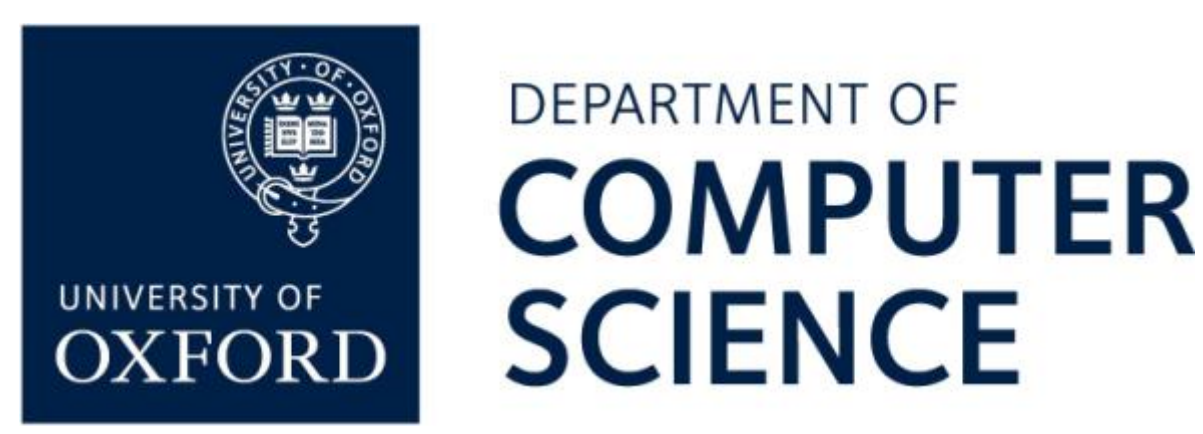

Alt Text: Figure 10 illustrates the structural evolution of cryptographic attack surfaces across three adversarial phases: algorithm-centric threats, implementation-centric threats, and adaptive AI-driven optimisation. The figure shows how early cryptographic security models were dominated by brute-force attacks, classical cryptanalysis, and fixed adversary capabilities governed by mathematical hardness assumptions. It then highlights the transition to implementation-centric threats, where vulnerabilities emerge from side-channel leakage, entropy degradation, key-management failures, and certificate validation flaws. The final phase captures the emergence of adaptive adversaries employing polymorphic and fully morphing malware, reinforcement learning–based attack optimisation, and runtime adaptation to defensive controls. The figure analytically demonstrates that cryptographic compromise increasingly results from adversarial adaptability across the cryptographic lifecycle rather than from isolated algorithmic weaknesses.

The analytical value of Figure 10 lies in its explicit separation of *what* is attacked from *how* the attack capability evolves. Phase 1 reflects a threat model constrained by computational hardness assumptions, where security margins are primarily governed by key length and algorithmic complexity. Phase 2 exposes a fundamental transition in which adversaries exploit entropy degradation, side-channel leakage, and key-management failures that exist independently of algorithm strength. Phase 3 demonstrates a qualitative break from prior models: adaptive malware no longer targets isolated weaknesses but instead optimises attack strategies across time, learning from defensive responses and runtime behaviour. This progression explains why cryptographic assurances that remain valid under formal proofs can fail catastrophically in operational environments, and why security evaluation must now account for adversarial adaptability rather than static attack feasibility.

Empirical and analytical evidence indicates that this assumption no longer holds. AI-enabled polymorphic and fully morphing malware introduces adversarial capabilities that operate across the entire signature lifecycle, targeting implementation-level dependencies rather than cryptographic primitives. Analysis of reported incident patterns and practitioner observations reveals that signature compromise increasingly results from entropy degradation, side-channel amplification, and key-management exploitation rather than from direct cryptanalytic attacks.

At the key generation stage, adaptive malware exploits weaknesses in entropy sources, particularly within virtualised and cloud-hosted environments. Machine learning models trained on runtime behaviour can detect non-uniform randomness distributions arising from constrained entropy pools, predictable initialisation vectors, or reuse of pseudo-random number generator states. Reinforcement learning agents iteratively refine entropy inference by observing cryptographic operations over time, reducing effective key entropy without violating formal key length requirements. This mechanism directly undermines the security guarantees of digital signatures while remaining invisible to algorithm-level verification.

During signature computation and verification, polymorphic malware uses dynamic mutation to evade static detection [64] and to manipulate execution flow at runtime. Fully morphing malware extends this capability by evolving attack strategies in response to defensive measures, enabling targeted exploitation of timing variability, cache access patterns, and memory locality. Side-channel leakage, traditionally treated as a statistical vulnerability, is transformed into an optimisation problem, where adaptive models continuously update attack parameters based on observed signal feedback. As a result, signature forgery and tampering can occur without detectable protocol deviation.

Hash-based integrity mechanisms are similarly exposed. AI-enabled adversaries accelerate collision discovery by prioritising candidate inputs based on learned structural properties of

**Dr. Petar Radanliev**
Parks Road,
Oxford OX1 3PJ
United Kingdom
Email: petar.radanliev@cs.ox.ac.uk

BA Hons., MSc., Ph.D. Post-Doctorate

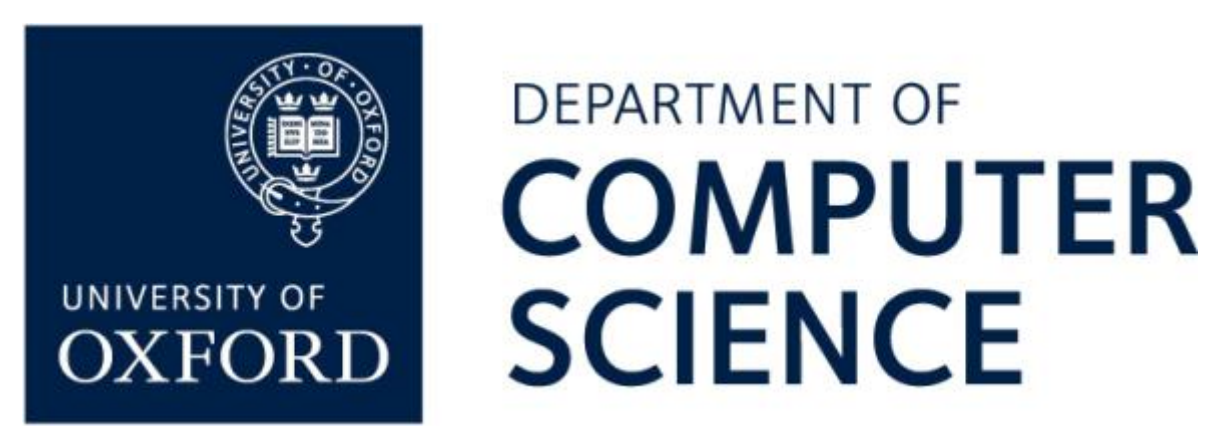

deployed hash functions and implementation-specific optimisations. While modern hash functions remain collision-resistant in a formal sense, implementation constraints, truncated outputs, and protocol-specific usage patterns create exploitable surfaces. Predictive models reduce the effective search space, enabling undetectable document substitution in systems that rely on hash equivalence as a trust proxy.

Key exchange mechanisms introduce additional exposure. Protocols such as Diffie–Hellman [65] and elliptic-curve variants provide confidentiality under passive observation but rely on external authentication to prevent active interception. Adaptive malware exploits this dependency by targeting certificate validation workflows, key distribution logic, and trust-store management. Polymorphic variants dynamically intercept and manipulate public key exchanges, while fully morphing malware evolves impersonation strategies capable of bypassing certificate pinning and revocation checks. Analysis of practitioner-reported incidents indicates that such attacks increasingly succeed through transient trust failures rather than persistent configuration errors.

Cryptographic key size, long treated as a primary defensive parameter, exhibits diminishing marginal effectiveness under AI-optimised attack conditions. While larger key sizes exponentially increase brute-force complexity, machine learning–assisted attacks bypass exhaustive search by exploiting implementation artefacts, entropy weaknesses, and side-channel emissions. The emergence of quantum computation further compresses the effective security margin. Shor's algorithm [66] collapses the hardness assumptions underlying RSA [7], and Elliptic Curve Cryptography (ECC) [67], [68], while quantum-assisted machine learning prioritises attack paths that exploit hybrid classical–quantum vulnerabilities. Consequently, key length alone no longer constitutes a sufficient security control.

Figure 11 extends the phase-based model by overlaying adversarial capability classes across the cryptographic timeline, explicitly mapping classical, AI-enabled, and quantum-assisted adversaries to their effective operational scope. Rather than treating these adversaries as discrete categories, the figure illustrates how their capabilities overlap and compound, particularly in environments where AI-driven optimisation and emerging quantum acceleration coexist.

**Dr. Petar Radanliev**
Parks Road,
Oxford OX1 3PJ
United Kingdom
Email: petar.radanliev@cs.ox.ac.uk

BA Hons., MSc., Ph.D. Post-Doctorate

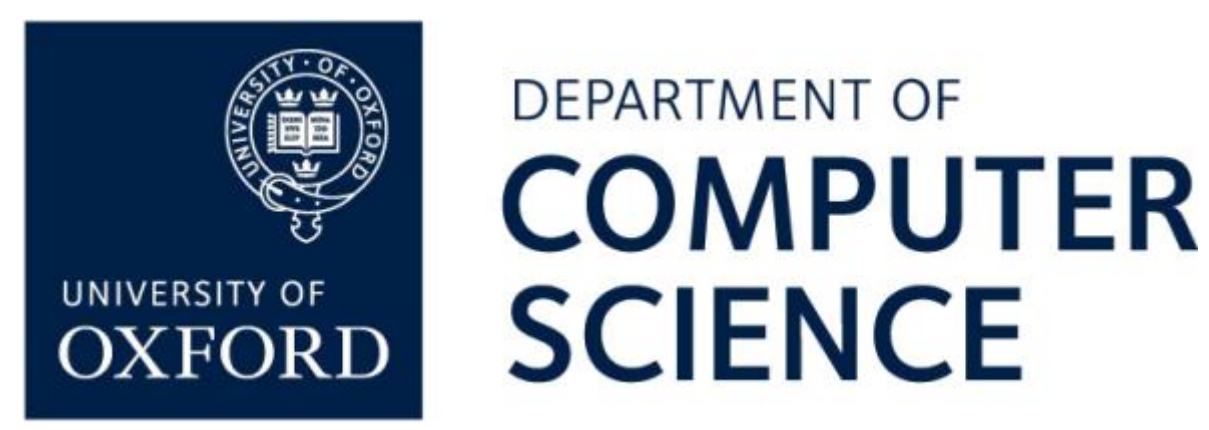

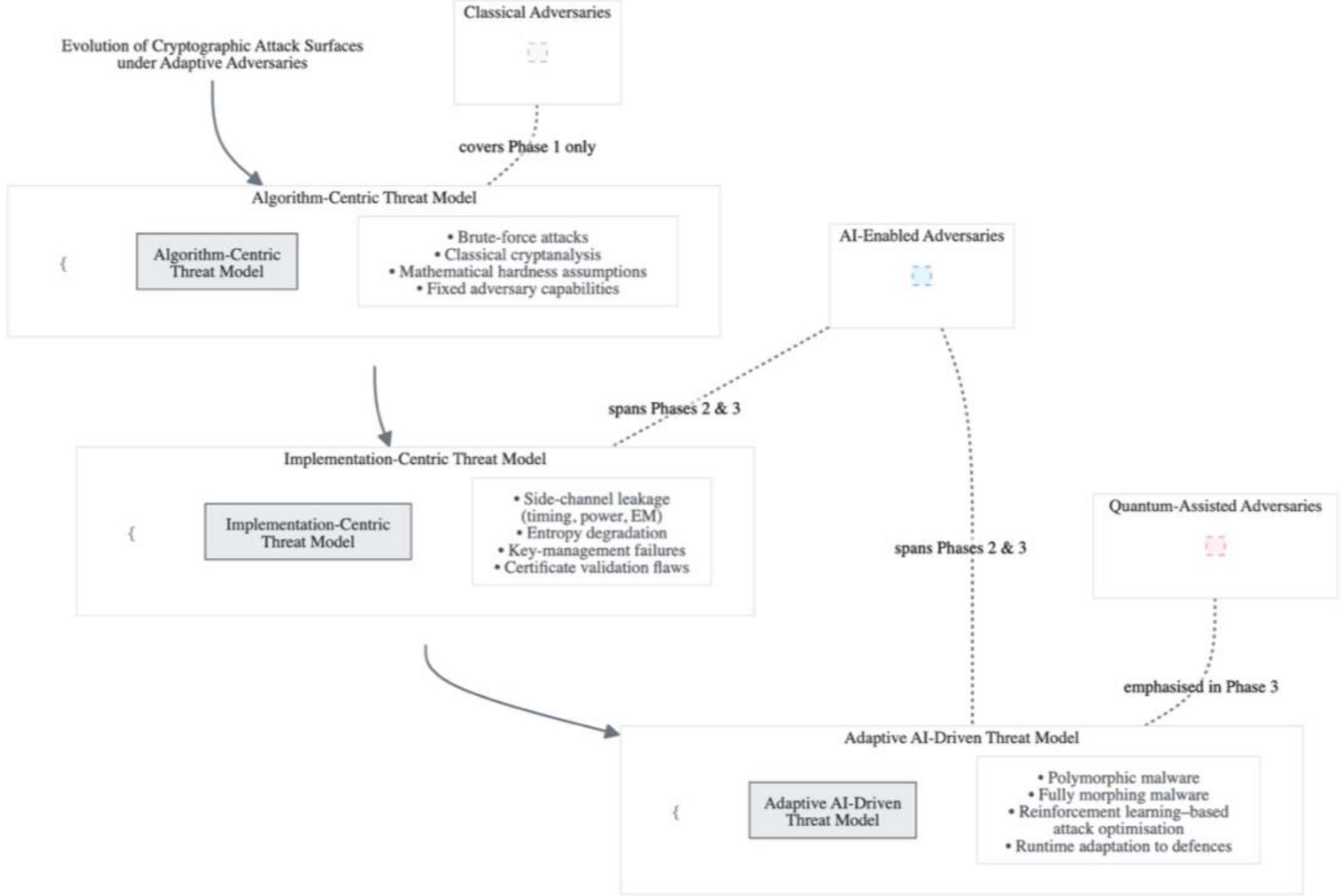

*Figure 11: Evolution of Cryptographic Attack Surfaces under Adaptive Adversaries*

Alt Text: Figure 11 presents a timeline-based analytical model of cryptographic attack surface evolution with overlayed adversarial capability classes. The diagram maps classical adversaries to algorithm-centric threat models, AI-enabled adversaries across implementation-centric and adaptive threat phases, and quantum-assisted adversaries with emphasis on the adaptive AI-driven optimisation phase. The overlays illustrate how adversarial capabilities expand and compound rather than replace one another, demonstrating that contemporary cryptographic risk arises from the interaction of adaptive learning, implementation exposure, and accelerated computation. The figure emphasises that cryptographic resilience can no longer be evaluated solely through key length or algorithm selection, but must account for adversarial optimisation and cross-layer exploitation dynamics.

Figure 11 provides a critical analytical insight into why conventional cryptographic controls fail under compound adversarial pressure. Classical adversaries remain largely constrained to algorithm-centric attacks, whereas AI-enabled adversaries span both implementation-centric and adaptive optimisation phases, exploiting entropy sources, side-channels, and trust infrastructures simultaneously. Quantum-assisted adversaries further compress the security margin by collapsing hardness assumptions while amplifying AI-guided attack prioritisation. The overlay structure demonstrates that cryptographic failure is no longer attributable to a single threat class but emerges from the interaction between adaptive learning, implementation exposure, and accelerated computation. This explains why incremental countermeasures, such as increasing key sizes or adopting post-quantum primitives in isolation, are insufficient without AI-aware monitoring and adaptive defensive control.

Post-quantum cryptographic schemes address algorithmic vulnerability to quantum computation but remain exposed at the implementation layer. Lattice-based, hash-based,

**Dr. Petar Radanliev**
Parks Road,
Oxford OX1 3PJ
United Kingdom
Email: petar.radanliev@cs.ox.ac.uk

BA Hons., MSc., Ph.D. Post-Doctorate

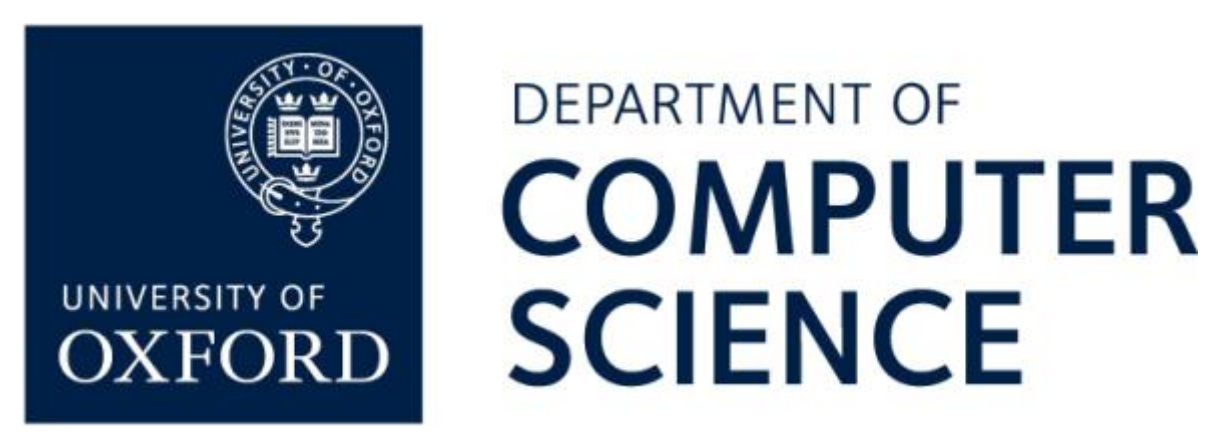

and code-based schemes mitigate future quantum attacks yet inherit the same operational dependencies on entropy quality, key isolation, and runtime integrity. Without AI-aware monitoring and adaptive key-management controls, post-quantum deployments risk replicating the same failure modes observed in classical systems.

Table 2 synthesises the analytical findings by mapping cryptographic mechanisms to AI-enabled attack vectors and defensive control gaps. The table highlights that vulnerability emergence is driven less by algorithm selection and more by adversarial adaptability across the cryptographic lifecycle.

*Table 2: Analytical Mapping of Cryptographic Mechanisms to AI-Enabled Attack Vectors*

| Cryptographic Mechanism | Classical Security Assumption | AI-Enabled Attack Vector | Observed Impact | Defensive Implication |
|---|---|---|---|---|
| Key Generation | High-entropy randomness | Entropy inference via ML | Reduced effective key strength | Continuous entropy validation |
| Digital Signatures | Computational unforgeability | Side-channel optimisation | Signature forgery / tampering | Runtime anomaly detection |
| Hash Functions | Collision resistance | Predictive collision prioritisation | Undetectable document substitution | Adaptive hashing and monitoring |
| Key Exchange | Passive adversary model | Dynamic MITM via certificate abuse | Session compromise | AI-aware trust validation |
| Key Size | Exponential brute-force cost | ML-assisted search + quantum speedup | Collapsed security margin | PQC + adaptive controls |

Cryptographic key size is a fundamental defence against brute-force attacks, where attackers systematically attempt all possible keys to decrypt a message. Larger key sizes exponentially increase the search space, making brute-force attacks computationally infeasible for traditional systems. For example, a 128-bit key provides a search space of approximately $3.4\times10^{38}$, rendering such attacks impractical with classical computing resources. However, the dual threat posed by AI and quantum computing requires a re-evaluation of cryptographic defences. To counter these dual threats, cryptographic systems are adopting larger key sizes that are resistant to both AI-accelerated brute-force attacks and quantum algorithms. Post-quantum cryptography (PQC) [48] offers promising solutions by developing algorithms that are secure against quantum attacks. These include lattice-based cryptography [69], hash-based cryptography [51], [70], and code-based cryptography, which provide resilience against the computational power of quantum machines.

**Dr. Petar Radanliev**
Parks Road,
Oxford OX1 3PJ
United Kingdom
Email: petar.radanliev@cs.ox.ac.uk

BA Hons., MSc., Ph.D. Post-Doctorate

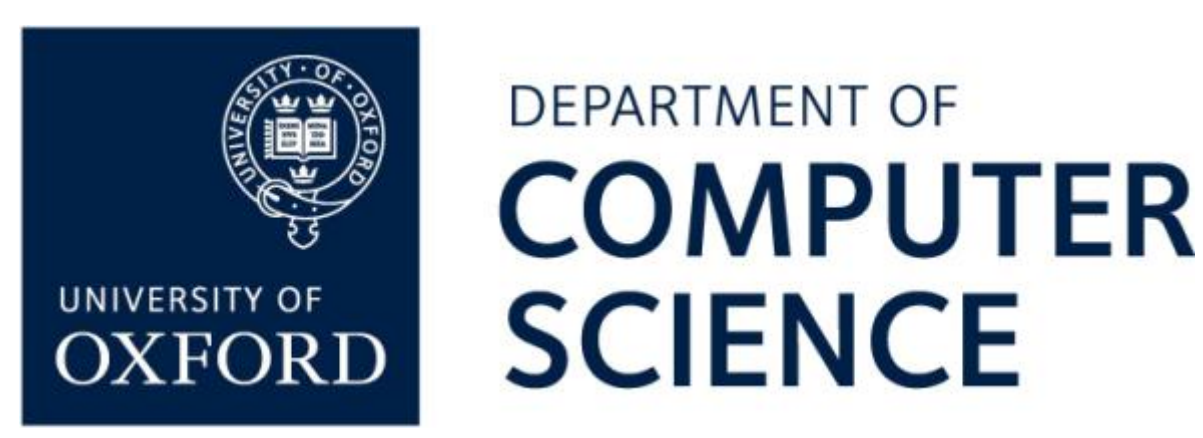

# 6. Systemic Impact of Private Key Compromise under AI-Driven Polymorphic and Morphing Malware

Private key compromise represents the most severe failure mode in public key cryptography because it collapses all security guarantees simultaneously, including confidentiality, integrity, authentication, and non-repudiation. Unlike partial protocol failures, private key exposure converts cryptographic protection into an attacker-controlled capability, enabling both retrospective and forward-looking exploitation. The emergence of AI-enabled polymorphic and fully morphing malware fundamentally amplifies the likelihood, scale, and persistence of such compromise.

From an analytical perspective, private key compromise must be understood as a *systemic event* rather than a discrete breach. Once a private key is exposed, all data encrypted under the corresponding public key becomes readable, irrespective of transport security or protocol correctness. Empirical evidence from industrial incident response indicates that AI-enabled malware increasingly exploits implementation-level weaknesses, such as entropy degradation, insecure memory handling, and key-management misconfiguration, to obtain private keys without triggering conventional alerts. Polymorphic malware dynamically mutates execution patterns to evade detection during key extraction, while fully morphing malware adapts its attack logic in response to defensive changes, enabling sustained access across system updates and reconfigurations.

The integrity impact of private key compromise extends beyond data exposure to the falsification of cryptographic trust. Possession of a valid private key enables the generation of fraudulent digital signatures that are indistinguishable from legitimate ones at the protocol level. AI-assisted adversaries further enhance this capability by modelling signing behaviour, timing characteristics, and protocol-specific validation logic, allowing forged signatures to evade anomaly-based detection. This undermines the evidentiary value of signed artefacts in legal, financial, and regulatory contexts, transforming digital signatures from trust anchors into vectors of deception.

Identity impersonation constitutes a second-order effect with cascading consequences. A compromised private key allows adversaries to authenticate as legitimate entities across multiple systems, bypassing access controls, authorisation checks, and audit mechanisms. Adaptive malware uses this capability to conduct targeted lateral movement within networks, escalate privileges, and persist across federated identity infrastructures. In practice, such impersonation attacks erode trust not only in individual credentials but in the public key infrastructure itself, particularly when certificate revocation mechanisms lag behind real-time exploitation.

Private key compromise also functions as an enabling condition for compound attacks. AI-enhanced malware integrates compromised keys into real-time man-in-the-middle operations, selectively intercepting, decrypting, and re-encrypting traffic without observable protocol deviation. Reinforcement learning models optimise attack timing and target selection, allowing adversaries to exploit secondary vulnerabilities uncovered through decrypted communications. This creates a feedback loop in which key compromise accelerates further system degradation, transforming isolated breaches into systemic failures.

The threat landscape is further intensified by the convergence of AI-enabled optimisation and quantum computation. While classical brute-force attacks remain infeasible against adequately sized keys, quantum algorithms such as Shor’s algorithm collapse the hardness

**Dr. Petar Radanliev**
Parks Road,
Oxford OX1 3PJ
United Kingdom
Email: petar.radanliev@cs.ox.ac.uk

BA Hons., MSc., Ph.D. Post-Doctorate

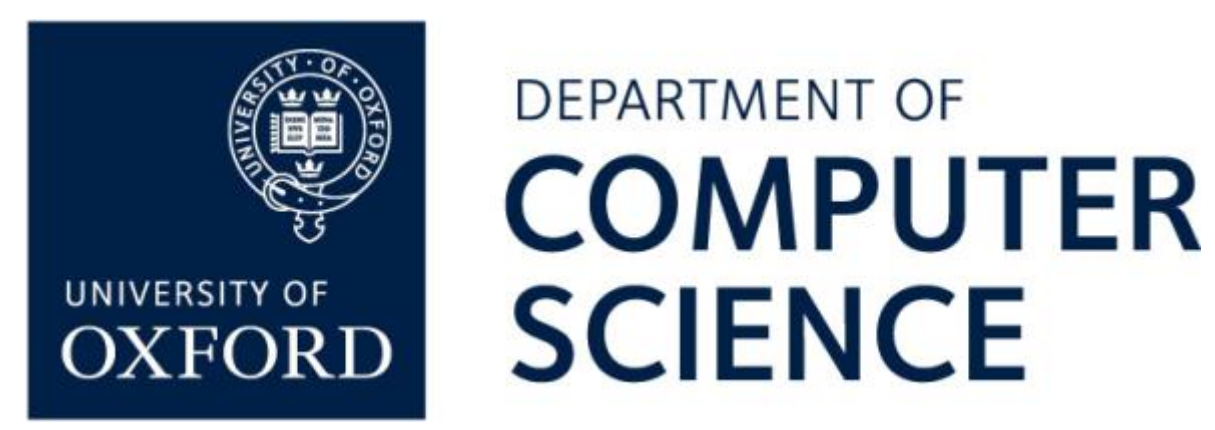

assumptions underlying RSA and elliptic curve cryptography. When combined with AI-driven prioritisation and automation, quantum-assisted adversaries gain the capacity to retroactively decrypt historical communications protected by previously captured keys, operationalising the "harvest now, decrypt later" threat model at scale.

Figure 12 analytically illustrates why increasing cryptographic key size alone no longer constitutes a sufficient defensive strategy. Although larger key sizes exponentially expand the theoretical search space, AI-augmented adversaries bypass exhaustive search by exploiting implementation artefacts, side-channel emissions, and entropy weaknesses. The figure demonstrates that effective security margins are increasingly determined by adversarial adaptability and system-level exposure rather than by mathematical complexity alone. As computational power grows and attack optimisation improves, the gap between formal security guarantees and operational resilience continues to narrow.

Mitigation of private key compromise therefore requires a shift from static key protection to adaptive cryptographic governance. Secure storage, periodic rotation, and revocation remain necessary but insufficient controls. These measures must be augmented with AI-aware monitoring of key usage patterns, continuous entropy assessment, runtime anomaly detection, and rapid trust re-establishment mechanisms. Post-quantum cryptographic algorithms reduce algorithmic vulnerability but must be deployed alongside adaptive key-management and detection frameworks to prevent replication of existing failure modes in next-generation systems.

In analytical terms, private key compromise under AI-enabled adversarial conditions represents a *phase transition* in cryptographic risk. Security assurance can no longer be derived solely from algorithm selection or key length, but must instead account for adversarial learning, cross-layer exploitation, and the interaction between cryptographic primitives and their operational environment. Figure 12 represent the brute-force attack process, where attackers try every possible key, then it demonstrates how a larger key size (e.g., 128-bit) results in an exponentially larger search space, making brute-force attacks impractical. Figure 12 explains the relationship between increasing computational power and the efficiency of brute-force attacks, along with the role of specialized hardware and distributed computing.

**Dr. Petar Radanliev**
Parks Road,
Oxford OX1 3PJ
United Kingdom
Email: petar.radanliev@cs.ox.ac.uk

BA Hons., MSc., Ph.D. Post-Doctorate

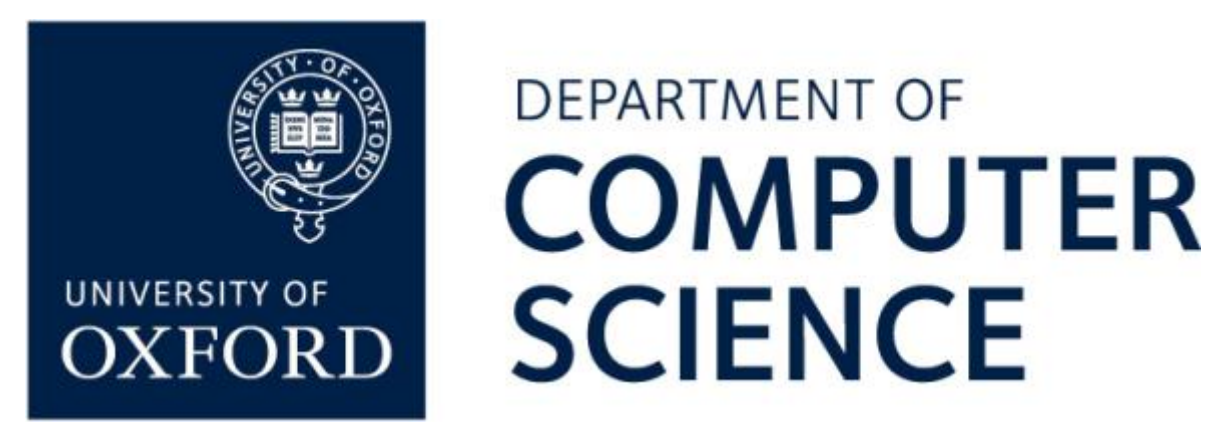

Search Space

Log2 Search Space Size

550
500
450
400
350
300
250
200
150
100
50
0

64-bit
128-bit
256-bit
512-bit

Key Size (bits)

Cryptographic Key Size vs Brute-Force Feasibility

64-bit Key Size
2^64 possible keys
128-bit Key Size
2^128 possible keys
256-bit Key Size
2^256 possible keys

Feasible
Infeasible
Computationally Prohibitive

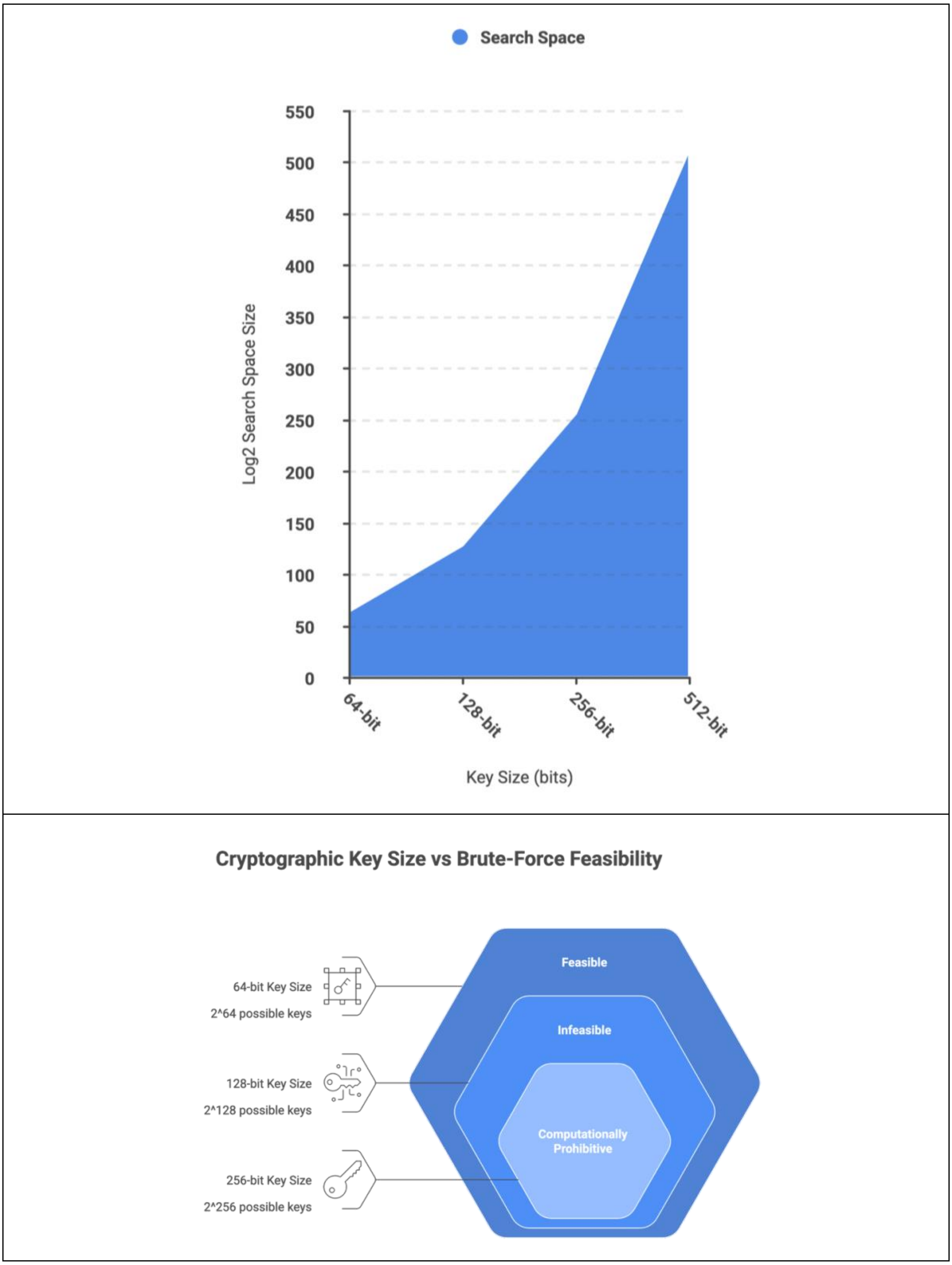

*Figure 12: Conceptual model illustrating the exponential relationship between cryptographic key size and brute-force attack feasibility under increasing computational power. The figure*

**Dr. Petar Radanliev**
Parks Road,
Oxford OX1 3PJ
United Kingdom
Email: petar.radanliev@cs.ox.ac.uk

BA Hons., MSc., Ph.D. Post-Doctorate

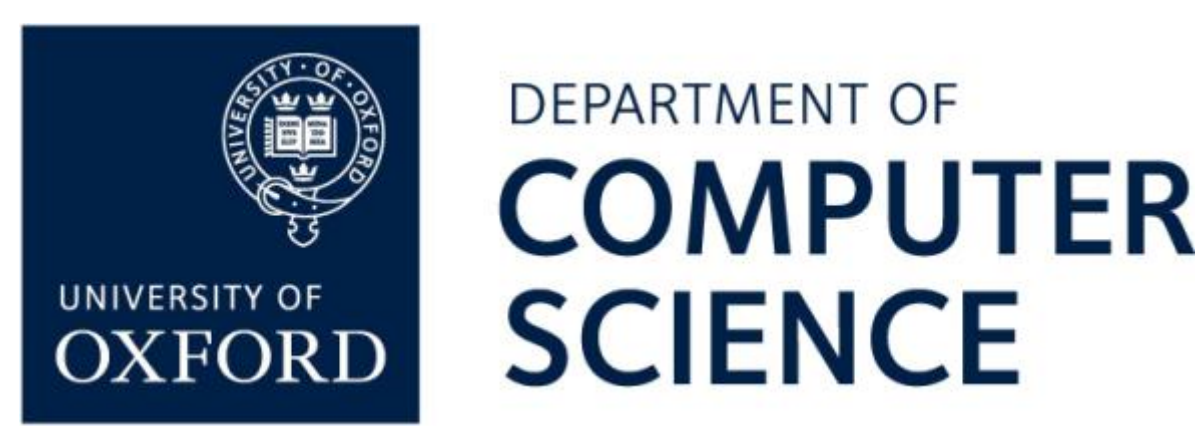

*contrasts classical exhaustive search complexity with AI-augmented optimisation effects, demonstrating that while larger key sizes remain mathematically resilient, implementation-level vulnerabilities and adaptive adversarial strategies significantly reduce effective security margins.*

Figure 12 analyses the cryptographic key size, computational growth, and effective security under adaptive adversaries. The figure presents a four-panel analytical model illustrating the relationship between cryptographic key size, brute-force feasibility, and adversarial capability. The upper panels show that classical security assumptions rely on exponential keyspace growth and computational infeasibility of exhaustive search. The lower panels demonstrate that increasing computational power and AI-enabled optimisation alter this relationship by reducing the effective search space through implementation-level exploitation, including entropy weaknesses and side-channel inference. The model highlights that, while larger key sizes remain mathematically robust, effective cryptographic security is determined by system-level factors and adversarial adaptability rather than brute-force complexity alone.

Figure 12 analytically demonstrates how cryptographic key size interacts with brute-force feasibility, computational growth, and implementation constraints to determine effective security margins. The upper-left panel shows that brute-force success probability increases monotonically with the number of attempted keys, illustrating that exhaustive search remains probabilistic rather than absolute, even before full keyspace coverage is achieved. The upper-right panel formalises the exponential expansion of the search space as key sizes increase from 64 to 512 bits, confirming that brute-force attacks against adequately sized keys are computationally infeasible under classical assumptions. The lower-left panel models the acceleration of available computational power over time, highlighting that exponential gains in processing capability erode static security margins faster than linear defensive adjustments. Crucially, the lower-right panel situates key size within a broader security context, demonstrating that algorithm strength, implementation security, and key-management practices collectively account for a majority of cryptographic resilience. Taken together, the figure shows that while increasing key length remains necessary to counter raw computational growth, it is insufficient in isolation: AI-augmented adversaries bypass exhaustive search by exploiting entropy weaknesses, side-channel leakage, and implementation artefacts, thereby collapsing effective security margins without violating formal hardness assumptions. The analytical implication is that cryptographic security must be evaluated as a multi-factor, adversary-adaptive system rather than as a function of key size alone.

# 7. Adaptive Cyber-Attack Patterns Targeting Public Key Cryptography

Cyber-attacks on Public Key Cryptography have undergone a qualitative transformation driven by artificial intelligence–enabled polymorphic and fully morphing malware. Unlike traditional malware, which relies on static payloads and predefined execution paths, polymorphic malware continuously mutates its observable characteristics to evade signature-based detection, while fully morphing malware adapts its internal logic, control flow, and exploitation strategy in response to environmental feedback. This adaptive behaviour allows adversaries to target cryptographic systems at the implementation level, exploiting entropy weaknesses, insecure key-generation processes, and key-management misconfigurations without triggering protocol-level alarms.

**Dr. Petar Radanliev**
Parks Road,
Oxford OX1 3PJ
United Kingdom
Email: petar.radanliev@cs.ox.ac.uk
BA Hons., MSc., Ph.D. Post-Doctorate

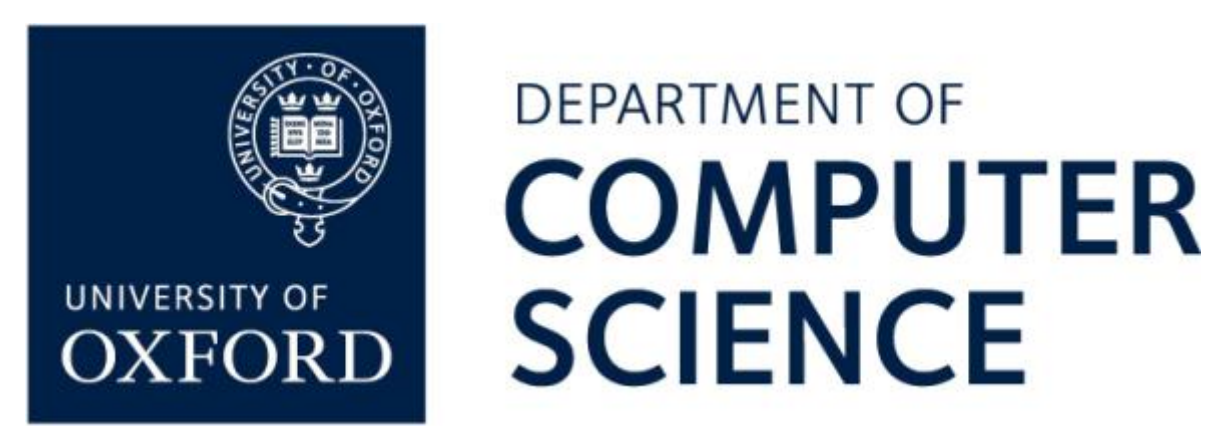

AI-enhanced malware fundamentally alters the feasibility of brute-force and search-based attacks. While classical brute-force attacks remain constrained by exponential key spaces, machine learning–driven optimisation reduces the effective search domain by exploiting correlations in key usage, entropy distribution, and implementation artefacts. Reinforcement learning agents refine attack strategies over time by observing cryptographic operations, selectively prioritising attack paths with higher expected payoff. When combined with distributed computing resources, this approach transforms brute-force attacks from exhaustive enumeration into guided optimisation processes. The emergence of quantum computation further intensifies this risk by collapsing the hardness assumptions underpinning RSA and elliptic curve cryptography through algorithms such as Shor's, thereby compressing the remaining security margin available to classical defences.

The compromise of private keys under these conditions produces cascading systemic effects. Confidentiality is immediately lost as encrypted data becomes readable; integrity is undermined through the generation of fraudulent digital signatures; and authentication mechanisms fail as attackers impersonate legitimate entities across multiple systems. AI-enabled polymorphic and fully morphing malware amplifies these effects by automating lateral movement, enabling real-time man-in-the-middle attacks, and selectively replaying or modifying encrypted communications. Empirical evidence from industrial incident response indicates that such attacks increasingly succeed through transient trust failures—such as delayed certificate revocation or misconfigured key stores—rather than through persistent configuration errors.

Man-in-the-middle attacks exemplify how adaptive malware exploits cryptographic dependencies rather than cryptographic primitives. By dynamically intercepting key exchanges and manipulating certificate validation workflows, polymorphic malware can replace legitimate public keys with attacker-controlled alternatives. Fully morphing malware extends this capability by evolving impersonation strategies capable of bypassing certificate pinning and trust-anchor validation. These attacks undermine the foundational trust assumptions of Public Key Infrastructure, rendering secure channels vulnerable without observable protocol deviation.

Side-channel attacks represent another domain where AI-driven adaptation shifts the balance in favour of the adversary. Timing, power, electromagnetic, and acoustic emissions—historically treated as niche or hardware-specific vulnerabilities—are now systematically exploitable through machine learning–based pattern extraction. Adaptive malware transforms side-channel exploitation into a continuous optimisation problem, dynamically adjusting attack parameters to overcome masking, noise injection, and constant-time defences. This capability bypasses the mathematical foundations of cryptography entirely, attacking the physical and operational realisation of cryptographic algorithms.

Mitigation under these conditions requires a departure from static, algorithm-centric security models. While post-quantum cryptographic algorithms are essential to address quantum-enabled attacks, they do not inherently mitigate AI-driven exploitation of implementation-level weaknesses. Effective defence therefore requires the integration of AI-aware anomaly detection, continuous entropy monitoring, adaptive key-management policies, and automated incident response mechanisms. Cryptographic resilience must be treated as an evolving system property rather than a fixed attribute derived from key size or algorithm selection.

Table 3 consolidates the analytical findings by mapping dominant cyber-attack classes against their mechanisms, impacts, and mitigation requirements under AI-enabled adversarial conditions. Rather than cataloguing attacks, the table highlights how each attack class exploits

**Dr. Petar Radanliev**
Parks Road,
Oxford OX1 3PJ
United Kingdom
Email: petar.radanliev@cs.ox.ac.uk
BA Hons., MSc., Ph.D. Post-Doctorate

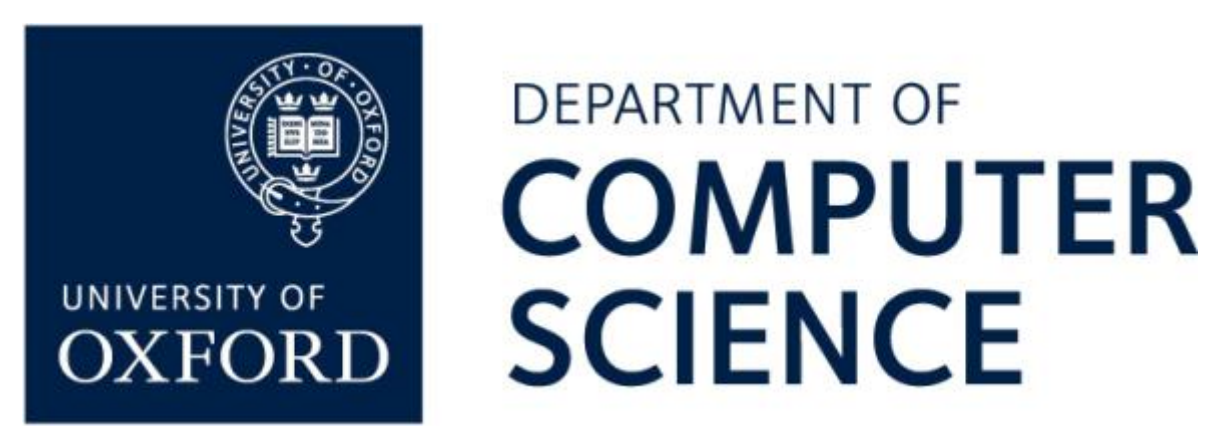

specific dependencies within Public Key Cryptography and why traditional countermeasures are increasingly insufficient.

*Table 3: Analytical Classification of Cyber-Attacks on Public Key Cryptography*

| Attack Class | Primary Exploitation Mechanism | Systemic Impact | Analytical Implication for Defence |
|---|---|---|---|
| Brute-Force and Search-Based Attacks | AI-guided optimisation, distributed computation, quantum acceleration | Erosion of effective security margins | Key size must be combined with AI-aware controls and PQC |
| Private Key Compromise | Entropy inference, memory extraction, key-store exploitation | Total loss of confidentiality, integrity, and authentication | Continuous key monitoring and rapid trust re-establishment |
| Man-in-the-Middle Attacks | Certificate abuse, dynamic key substitution, trust-anchor exploitation | Undermined channel authenticity and trust | Adaptive certificate validation and real-time anomaly detection |
| Side-Channel Attacks | ML-based signal extraction from timing, power, EM leakage | Key recovery without cryptanalysis | Hardware–software co-design and AI-based leakage detection |
| Quantum-Enabled Attacks | Shor’s algorithm and quantum-assisted optimisation | Collapse of RSA and ECC hardness assumptions | Urgent transition to post-quantum cryptography |

The analytical classification presented in Table 3 establishes that contemporary cyber-attacks on Public Key Cryptography exploit interdependent weaknesses across cryptographic generation, exchange, validation, and management layers. Building on this classification, the present section examines how AI-driven polymorphic and fully morphing malware transforms these attack mechanisms into systemic risks. Rather than representing incremental extensions of existing threats, these malware classes introduce adaptive, learning-driven behaviours that undermine the foundational assumptions on which PKC security models are constructed.

Polymorphic and fully morphing malware leverage artificial intelligence to operationalise dynamic adaptation and predictive analytics against cryptographic systems. Polymorphic malware exploits statistical regularities and implementation inconsistencies by continuously mutating its observable structure while preserving attack functionality. Fully morphing malware extends this capability by evolving its internal logic, execution pathways, and exploitation strategies in response to environmental feedback. This distinction is critical: polymorphism primarily evades detection, whereas full morphing enables sustained optimisation of attack effectiveness against specific cryptographic deployments.

**Dr. Petar Radanliev**
Parks Road,
Oxford OX1 3PJ
United Kingdom
Email: petar.radanliev@cs.ox.ac.uk

BA Hons., MSc., Ph.D. Post-Doctorate

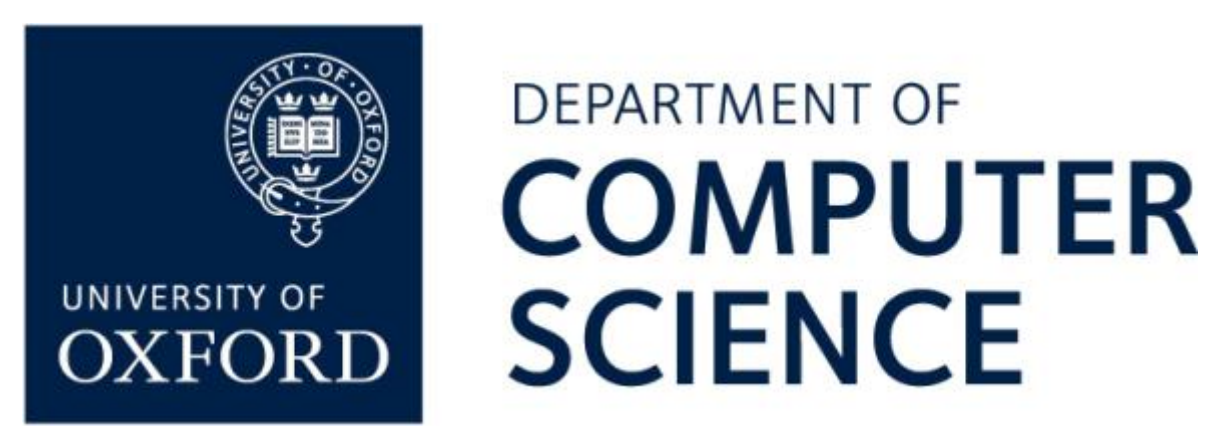

Key generation processes represent an initial point of systemic exposure. AI-driven malware targets entropy sources by analysing randomness quality, initialisation behaviour, and reuse patterns within pseudo-random number generators, particularly in virtualised and cloud-hosted environments. Machine learning models trained on runtime behaviour can detect non-uniform entropy distributions that remain compliant with formal standards yet reduce effective key strength. Fully morphing malware escalates this risk by adapting its entropy inference strategy to the specific implementation and execution context, enabling private key compromise without reliance on brute-force enumeration.

Key exchange protocols introduce further risk amplification. Reinforcement learning–based malware can model multi-stage cryptographic handshakes in real time, identifying opportunities to inject forged keys or substitute public key material without triggering protocol violations. By continuously adapting to defensive responses, fully morphing malware can impersonate legitimate participants, undermining authentication guarantees while preserving apparent protocol correctness. This behaviour collapses the trust assumptions underlying Diffie–Hellman and elliptic-curve–based exchanges, particularly when authentication mechanisms rely on delayed or static validation.

Public Key Infrastructure constitutes a critical trust dependency that is especially vulnerable to adaptive exploitation. Certificate validation, revocation, and trust-anchor management processes are designed around assumptions of bounded adversarial behaviour and delayed compromise detection. Polymorphic malware exploits implementation-level parsing and validation flaws to introduce malicious certificates that mimic legitimate credentials. Fully morphing malware compounds this threat by learning the operational characteristics of specific PKI deployments, bypassing revocation mechanisms through timing exploitation, cache poisoning, or delayed propagation. The result is scalable impersonation that erodes trust not only in individual certificates but in the PKI model itself.

Key management systems represent high-value targets where adaptive malware can induce lifecycle-wide compromise. Polymorphic malware systematically scans for misconfigurations in key storage, access controls, and rotation policies, while fully morphing malware adapts its attack vectors as defensive controls evolve. By compromising key management workflows rather than individual keys, adversaries gain persistent access across cryptographic epochs, rendering periodic key rotation ineffective. This shifts key management from a defensive control into an attack surface when not continuously monitored and adapted.

Side-channel leakage constitutes a further domain where AI-driven adaptation produces qualitative risk escalation. Timing, power, electromagnetic, and acoustic emissions—previously treated as hardware-specific or situational vulnerabilities—are transformed into high-yield attack vectors through machine learning–based signal extraction. Polymorphic malware automates the identification of exploitable leakage patterns, while fully morphing malware adapts its models to hardware configurations, operating systems, and deployed countermeasures. These attacks bypass cryptographic hardness entirely, exploiting the physical and operational realisation of cryptographic algorithms rather than their mathematical design.

The cumulative effect of these mechanisms is a redefinition of cryptographic risk. Rather than discrete attack events, AI-driven polymorphic and fully morphing malware enables continuous, system-wide degradation of cryptographic assurances. Confidentiality, integrity, authentication, and non-repudiation fail not through algorithmic weakness but through adaptive exploitation of implementation dependencies and trust infrastructures. This shift

**Dr. Petar Radanliev**
Parks Road,
Oxford OX1 3PJ
United Kingdom
Email: petar.radanliev@cs.ox.ac.uk

BA Hons., MSc., Ph.D. Post-Doctorate

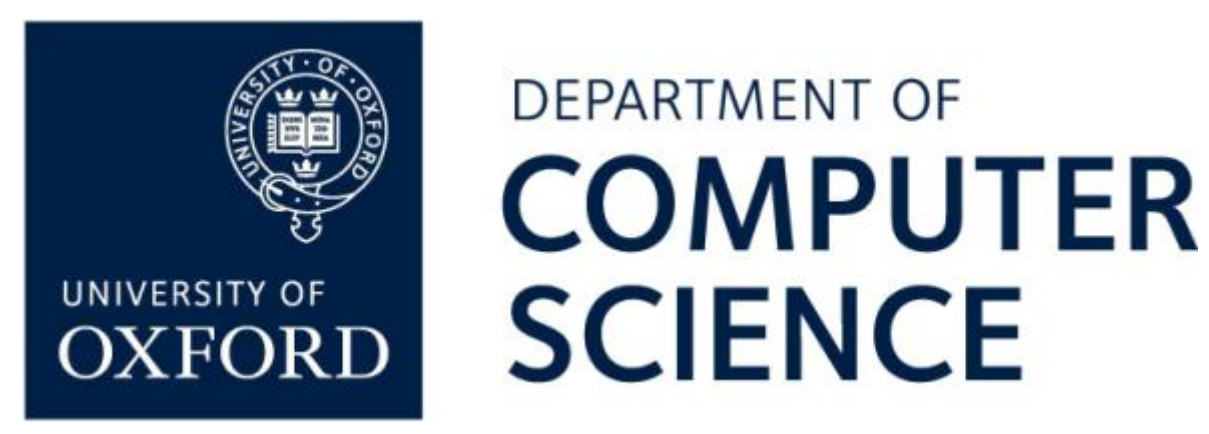

invalidates static security evaluations and necessitates a layered, adaptive defence paradigm.

Mitigation therefore requires integrating artificial intelligence as a defensive capability rather than treating it solely as an adversarial advantage. AI-enhanced monitoring must be deployed to detect anomalous cryptographic behaviour in real time, including entropy degradation, abnormal key usage patterns, and certificate validation anomalies. Post-quantum cryptographic algorithms are essential to mitigate quantum-enabled threats but must be complemented by adaptive key-management policies, continuous trust re-evaluation, and automated incident response. Hardware-based security modules provide additional resilience against side-channel exploitation, while periodic cryptographic audits must evolve into continuous assurance processes.

In analytical terms, AI-driven polymorphic and fully morphing malware converts cryptographic security from a static property into a dynamic equilibrium between adversarial adaptation and defensive responsiveness. Sustaining the viability of Public Key Cryptography under these conditions requires a shift from algorithm-centric protection to system-level, AI-aware cryptographic governance.

## AI-Enabled Attack Mechanics Against Cryptographic Assumptions

The preceding analysis demonstrates that contemporary threats do not arise from isolated attack techniques but from the systematic erosion of the assumptions underpinning Public Key Cryptography. At its core, PKC presumes that keys are generated from sufficiently unpredictable entropy sources, that adversaries are computationally bounded, that private keys remain isolated from untrusted execution contexts, and that implementations do not leak exploitable information. AI-driven polymorphic and fully morphing malware targets these assumptions directly, transforming them from implicit guarantees into continuously contested properties.

Rather than attempting to invert cryptographic functions, adaptive malware models cryptographic behaviour as an observable system. Machine learning techniques are applied to infer statistical structure from execution traces, memory access patterns, timing variation, and environmental context. In this setting, entropy is no longer treated as a binary property but as a measurable signal subject to degradation. Reinforcement learning agents refine their inference strategies across repeated observations, progressively reducing uncertainty about key material or protocol state without violating formal correctness conditions. This reframing converts cryptographic strength from a static attribute into an emergent property dependent on runtime behaviour.

Fully morphing malware extends this capability by internalising defensive responses as part of the optimisation loop. Instead of following predefined exploitation paths, attack strategies are selected dynamically based on real-time feedback from cryptographic operations and system instrumentation. Side-channel exploitation, in particular, becomes an adaptive control problem rather than a one-off statistical attack. Countermeasures such as masking, noise injection, or constant-time execution introduce friction but do not terminate the optimisation process; they merely alter the signal space within which learning occurs. As a result, the distinction between “hardened” and “vulnerable” implementations becomes increasingly blurred under sustained adaptive pressure.

Key management and trust infrastructures amplify this exposure because they externalise cryptographic security into distributed software systems. Cloud-based key management services and PKI deployments rely on assumptions of delayed compromise detection, stable access-control policies, and predictable certificate lifecycles. Adaptive malware exploits

**Dr. Petar Radanliev**
Parks Road,
Oxford OX1 3PJ
United Kingdom
Email: petar.radanliev@cs.ox.ac.uk

BA Hons., MSc., Ph.D. Post-Doctorate

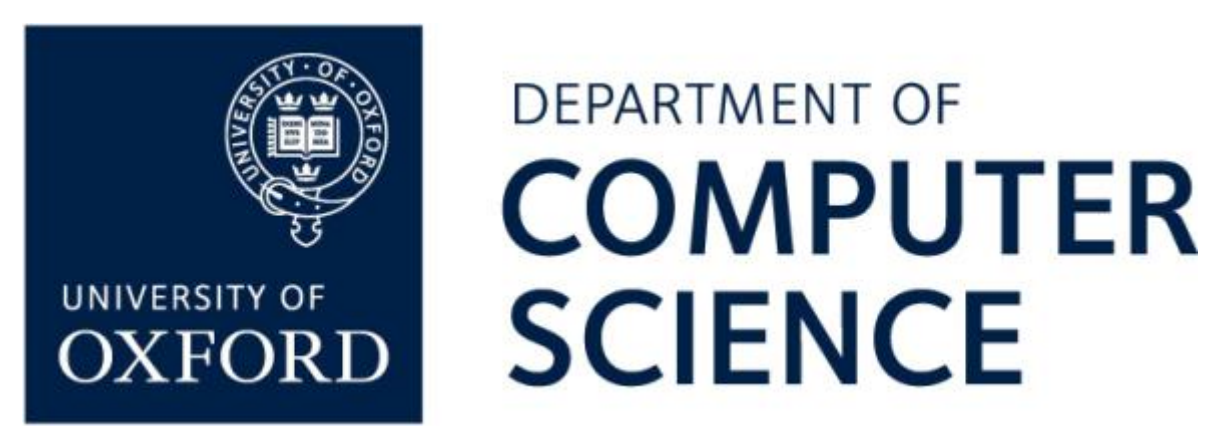

these temporal and organisational dependencies by identifying short-lived inconsistencies—such as revocation latency, stale trust anchors, or policy propagation delays—and aligning attack timing accordingly. Persistence is achieved not through permanent privilege escalation, but through continuous re-alignment with evolving defensive configurations.

Crucially, these attack mechanics do not contradict formal cryptographic security proofs; they operate orthogonally to them. Algorithms remain mathematically sound, key sizes remain within recommended bounds, and protocols execute as specified. What fails is the assumption that adversarial capability is static, bounded, and external to the system. Under AI-enabled optimisation, cryptographic security degrades as a function of exposure, observability, and adaptation speed rather than computational infeasibility alone.

The analytical implication is that cryptographic resilience must be reconceptualised as a dynamic equilibrium. Security evaluation can no longer rely exclusively on algorithm selection, key length, or compliance with standards, but must incorporate adversarial learning capacity, implementation observability, and the ability of defensive controls to adapt at comparable speed. Without this shift, even formally secure cryptographic systems remain vulnerable to sustained, low-noise exploitation by adaptive adversaries.

# 8. Adversarial Optimisation of Cryptographic Trust through Private Key Compromise

Private key compromise under AI-driven polymorphic and fully morphing malware should be understood as an explicit optimisation target within adaptive adversarial strategies. Unlike traditional malware campaigns that treat key exposure as an opportunistic outcome, contemporary AI-enabled adversaries model private keys as high-value control points whose compromise maximises downstream leverage across cryptographic, identity, and trust infrastructures. This reframing distinguishes key compromise from other attack outcomes by positioning it as a strategic objective rather than a secondary consequence.

Adaptive malware approaches key compromise through continuous optimisation rather than discrete exploitation. Machine learning models embedded within polymorphic malware analyse runtime cryptographic behaviour, memory access patterns, and system-level telemetry to identify conditions under which private keys are transiently exposed during legitimate operations. Fully morphing malware escalates this process by evolving extraction strategies in response to defensive interventions, dynamically adjusting attack timing, privilege boundaries, and persistence mechanisms. The result is a sustained capability to extract key material without triggering stable indicators of compromise.

Once achieved, private key compromise functions as a force multiplier rather than a terminal event. Confidentiality loss is immediate but secondary to the broader systemic effects that follow. Control over a valid private key enables adversaries to generate artefacts that are cryptographically indistinguishable from legitimate outputs, including signed transactions, authenticated software updates, and trusted communications. AI-driven malware automates the production and deployment of such artefacts, selectively aligning them with protocol semantics and operational context to evade downstream validation. This transforms cryptographic mechanisms from defensive controls into attack primitives.

Identity and trust collapse constitute the most consequential effects of optimised key compromise. Adaptive malware leverages compromised keys to impersonate legitimate entities across federated systems, exploiting trust relationships embedded in Public Key Infrastructure, identity federation frameworks, and inter-organisational authentication workflows. Fully morphing malware adapts impersonation strategies as trust policies evolve,

**Dr. Petar Radanliev**
Parks Road,
Oxford OX1 3PJ
United Kingdom
Email: petar.radanliev@cs.ox.ac.uk
BA Hons., MSc., Ph.D. Post-Doctorate

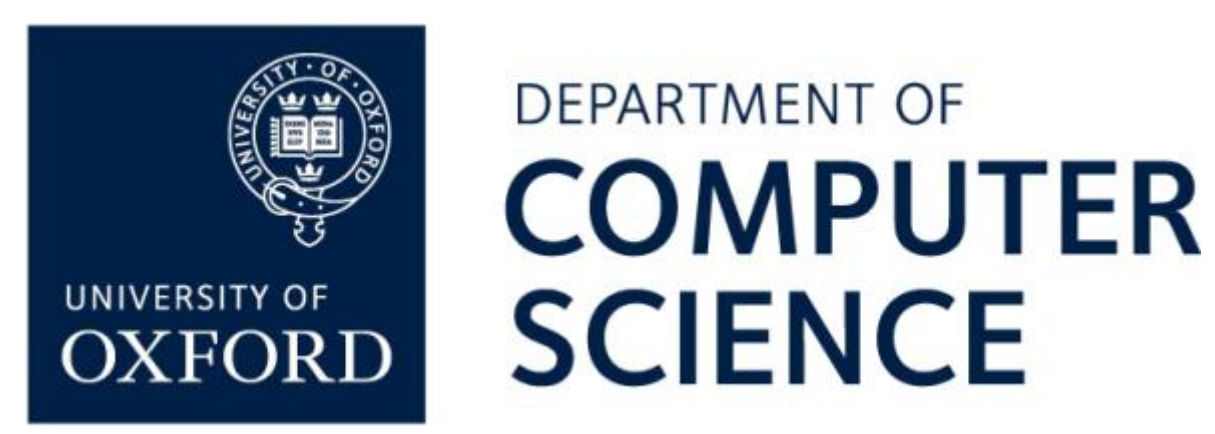

allowing persistent exploitation even as certificates are rotated or access controls are updated. This behaviour undermines the assumption that trust recovery can be achieved through periodic key replacement alone.

Crucially, AI-driven key compromise destabilises cryptographic governance models by eroding the temporal separation between compromise detection and trust remediation. Polymorphic malware conducts low-frequency, targeted compromise operations that avoid triggering threshold-based alerts, while fully morphing malware re-aligns its behaviour to newly deployed controls. This creates a condition of chronic uncertainty in which defenders cannot reliably distinguish between trusted and adversary-controlled cryptographic identities, even when protocols execute correctly and compliance requirements are formally satisfied.

From an adversarial perspective, private key compromise also enables recursive exploitation. Compromised keys are used to decrypt traffic, extract additional credentials, and identify new optimisation targets within interconnected systems. Reinforcement learning techniques prioritise exploitation paths that maximise persistence and minimise detection, transforming key compromise into a gateway for long-lived, multi-stage campaigns rather than isolated breaches. This explains why AI-enabled malware increasingly favours cryptographic targets over traditional perimeter controls.

Mitigating this class of risk requires abandoning the assumption that cryptographic trust can be restored through static remediation measures. While post-quantum cryptographic algorithms address future algorithmic breakage, they do not prevent adaptive exploitation of key lifecycle processes. Effective defence must therefore focus on constraining adversarial optimisation by limiting observability, reducing exploitable runtime exposure, and matching adversarial learning speed with defensive adaptation. Hardware-backed isolation, continuous key usage analytics, and automated trust invalidation are necessary components, but their effectiveness depends on integration into a governance model that treats trust as provisional rather than absolute.

In analytical terms, AI-driven polymorphic and fully morphing malware converts private key compromise into a mechanism for systemic trust subversion. The security of Public Key Cryptography under these conditions no longer depends solely on algorithmic soundness or key length, but on the capacity of defensive systems to resist sustained adversarial optimisation. This shift represents a fundamental change in the role of cryptographic keys, from static secrets to dynamically contested control assets.

## 9. Adaptive Man-in-the-Middle Attacks and the Erosion of Cryptographic Trust

Man-in-the-middle attacks (MITM) against Public Key Cryptography have traditionally been modelled as transient interception events that exploit weaknesses in key exchange or authentication protocols. Under AI-driven polymorphic and fully morphing malware, this model is no longer sufficient. Contemporary MITM attacks increasingly operate as adaptive trust substitution mechanisms, in which adversaries do not merely intercept communications but progressively replace legitimate cryptographic trust relationships with adversary-controlled equivalents. This shift transforms MITM attacks from episodic breaches into sustained control strategies embedded within cryptographic infrastructures.

Polymorphic malware enables this transition by exploiting the operational dependencies of key exchange protocols rather than their cryptographic design. During Diffie–Hellman or RSA-based exchanges, trust is delegated to certificates, validation chains, and local trust stores. Polymorphic malware dynamically intercepts these exchanges and substitutes public key

**Dr. Petar Radanliev**
Parks Road,
Oxford OX1 3PJ
United Kingdom
Email: petar.radanliev@cs.ox.ac.uk
BA Hons., MSc., Ph.D. Post-Doctorate

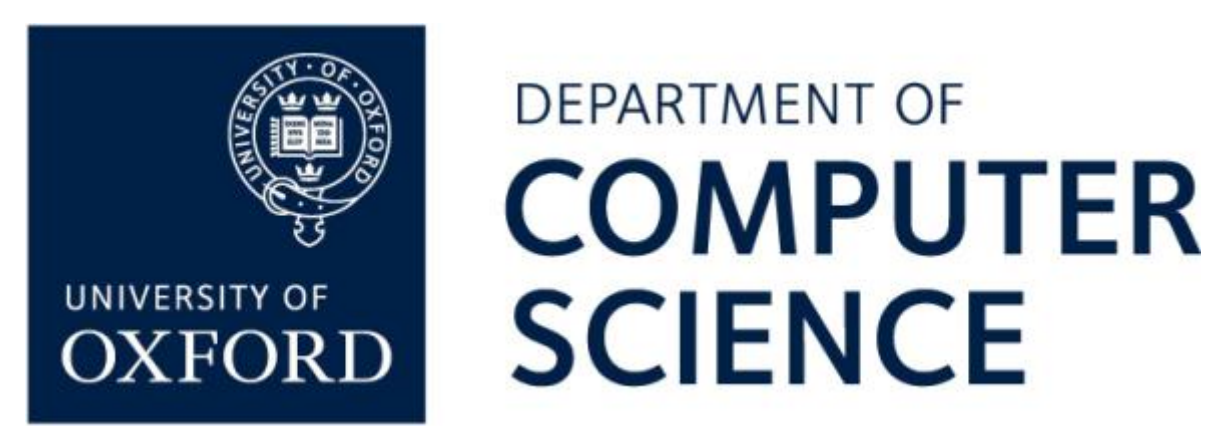

material through techniques such as ARP spoofing, DNS manipulation, or malicious certificate injection. Crucially, payload mutation allows the malware to evade static detection while selectively aligning its behaviour with the configuration and network topology of the target environment. In this context, interception is a means to an end: the primary objective is to insert adversary-controlled keys that preserve protocol correctness while altering trust semantics.

Fully morphing malware extends this capability by internalising the defender's response as part of the attack loop. Rather than relying on a fixed impersonation strategy, fully morphing variants adapt their trust-substitution techniques in response to certificate pinning, revocation checks, or policy updates. By mimicking trusted intermediaries, such as certificate authorities or internal key servers, these malware systems bypass validation mechanisms that assume static adversarial behaviour. The result is a persistent MITM condition in which encrypted channels remain formally valid but are cryptographically bound to adversary-controlled identities.

The analytical significance of historical malware such as Emotet lies not in its specific implementation, but in its demonstration of polymorphic trust interception at scale. Emotet's ability to inject malicious certificates and redirect SSL/TLS traffic illustrates how adaptive payload mutation can exploit misconfigured trust environments without triggering protocol violations. Traffic interception becomes selective and context-aware, enabling the extraction of session keys, credentials, and authentication artefacts while maintaining the appearance of secure communication.

BlackEnergy exemplifies the fully morphing extension of this model, particularly in cyber-physical environments. Its evolution from a denial-of-service tool into a modular, adaptive platform demonstrates how MITM attacks can be operationalised as long-lived control mechanisms. By injecting forged certificates and spoofed public keys into encrypted industrial control communications, BlackEnergy replaced legitimate trust anchors with adversary-controlled equivalents. Its adaptive behaviour allowed it to persist as defenders patched vulnerabilities, illustrating how fully morphing malware collapses the temporal distinction between compromise and remediation.

Once trust substitution is achieved, communication manipulation becomes both precise and scalable. Polymorphic malware selectively modifies encrypted payloads, altering transaction values, control commands, or authentication responses, based on learned protocol semantics. Fully morphing malware escalates this capability by evolving manipulation strategies to evade anomaly detection, ensuring that injected modifications remain within expected behavioural bounds. This enables adversaries to influence system behaviour directly, rather than merely observing it.

The downstream effects of adaptive MITM attacks extend beyond immediate data exposure. Adversary-controlled trust relationships enable recursive exploitation: decrypted traffic reveals additional credentials, trust paths, and optimisation targets, which are then incorporated into subsequent attack iterations. Reinforcement learning techniques prioritise actions that maximise persistence and minimise detection, transforming MITM attacks into self-sustaining exploitation loops rather than one-off intrusions.

Mitigation under these conditions cannot rely solely on strengthening individual cryptographic protocols. Post-quantum algorithms mitigate future computational threats but do not address trust substitution at the implementation level. Effective defence requires continuous validation of trust relationships, real-time detection of anomalous key exchange behaviour, and architectural constraints that limit the ability of adversaries to introduce new trust anchors.

**Dr. Petar Radanliev**
Parks Road,
Oxford OX1 3PJ
United Kingdom
Email: petar.radanliev@cs.ox.ac.uk

BA Hons., MSc., Ph.D. Post-Doctorate

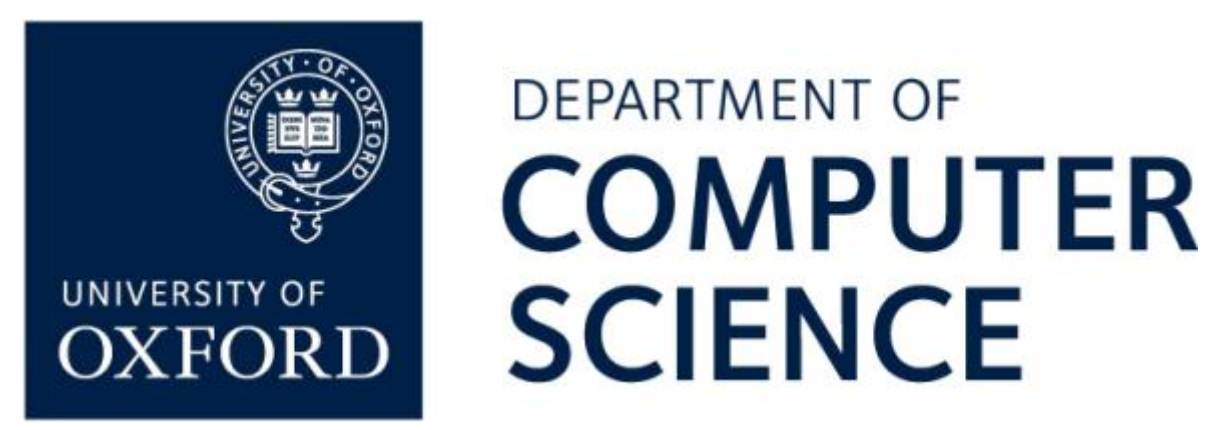

Ephemeral key exchanges, strict certificate pinning, and zero-trust communication models reduce exposure, but their effectiveness depends on adaptive enforcement rather than static configuration.

Analytically, AI-enabled polymorphic and fully morphing malware redefines MITM attacks as a problem of trust governance rather than channel security. Public Key Cryptography remains mathematically sound, yet its deployment assumes that trust relationships are stable and externally verifiable. Under adaptive adversarial pressure, these assumptions fail. Preserving the integrity of PKC therefore requires treating trust as a dynamically contested asset, subject to continuous validation and adversarial optimisation, rather than as a fixed property established at connection time.

## 10. AI-Optimised Side-Channel Attacks and Cryptographic Observability

Side-channel attacks against Public Key Cryptography no longer constitute peripheral implementation flaws but have evolved into primary inference channels through which adaptive adversaries extract cryptographic state. In contrast to classical attack models that treat leakage as incidental and episodic, AI-enabled polymorphic and fully morphing malware operationalises side-channel emissions as continuous, high-fidelity observation streams. This shift collapses the conceptual separation between cryptographic computation and adversarial learning, rendering side-channel resistance a system-level property rather than an implementation detail.

The defining feature of contemporary side-channel exploitation is adversarial adaptation. Rather than relying on static correlation models, AI-driven malware constructs probabilistic representations of cryptographic execution by integrating timing variability, power traces, electromagnetic emanations, and auxiliary environmental signals into evolving inference models. These models are refined iteratively using reinforcement learning, allowing adversaries to reduce uncertainty about private key material, intermediate states, or protocol progression without violating cryptographic correctness or triggering threshold-based alarms.

Timing leakage illustrates this transformation clearly. Execution-time variability in modular exponentiation or elliptic-curve scalar multiplication has long been recognised as a leakage source. Under adaptive adversarial conditions, however, timing is no longer analysed as a single-dimension signal. Polymorphic malware correlates execution latency with cache behaviour, branch prediction effects, and scheduler interactions, constructing multi-factor timing signatures that persist even under constant-time countermeasures. Fully morphing malware further adapts these inference models in response to defensive interventions, learning residual leakage patterns introduced by mitigations such as instruction padding or randomised delays. As a result, timing defences increase adversarial sample complexity but do not terminate the inference process.

Power and resource-usage side channels undergo a similar reconfiguration. Classical Simple and Differential Power Analysis assume stable measurement conditions and repeated traces. AI-enabled malware eliminates these constraints by dynamically optimising sampling strategies, feature extraction pipelines, and trace alignment in situ. Reinforcement learning agents prioritise sampling windows with maximal information gain, adapting to masking schemes, noise injection, and algorithmic obfuscation. This capability is particularly consequential in constrained environments, such as smart cards, embedded controllers, and Internet-of-Things devices, where energy efficiency and latency requirements limit defensive entropy.

**Dr. Petar Radanliev**
Parks Road,
Oxford OX1 3PJ
United Kingdom
Email: petar.radanliev@cs.ox.ac.uk

BA Hons., MSc., Ph.D. Post-Doctorate

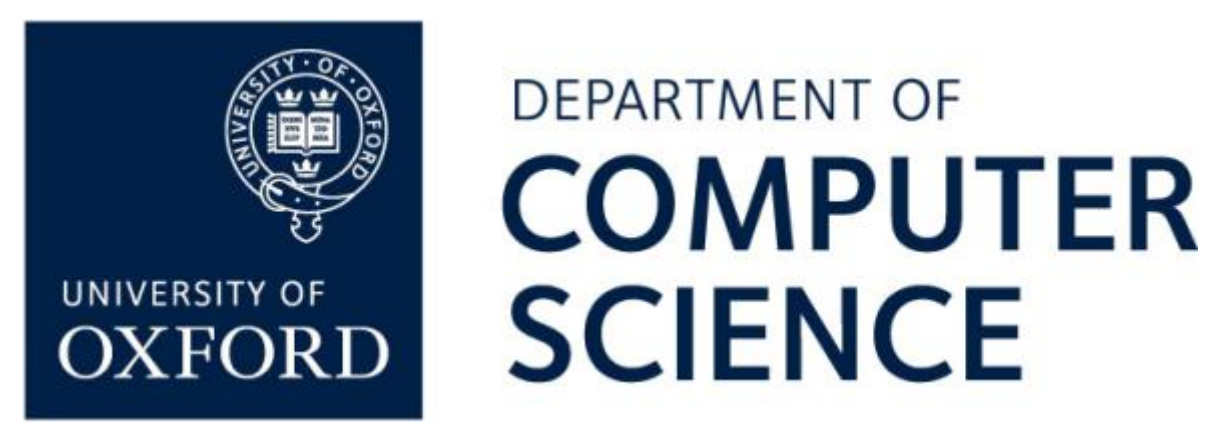

Electromagnetic, acoustic, and optical emanations further extend the adversarial observability surface. Historically treated as laboratory-grade or proximity-limited vectors, these channels become operationally relevant when integrated into multimodal inference systems. Polymorphic malware selectively exploits the most informative leakage modalities available in a given environment, while fully morphing malware adapts fusion strategies as shielding, dampening, or environmental conditions change. This multimodal aggregation ensures that partial mitigation of individual channels fails to eliminate overall observability, undermining defence-in-depth strategies that assume independence between leakage sources.

The analytical consequence of these developments is a redefinition of side-channel leakage from *incidental artefact* to *instrumental signal*. Cryptographic implementations continuously emit behavioural information that adaptive adversaries can accumulate over time, even when individual emissions appear statistically insignificant. Formal security proofs remain intact, yet operational security degrades asymptotically as adversarial models converge. This explains why implementations certified as side-channel resistant under conventional testing regimes may nevertheless fail under sustained, low-noise adversarial observation.

Mitigation under these conditions requires abandoning static leakage suppression as a primary objective. While constant-time execution, masking, shielding, and hardware isolation remain necessary, they are insufficient against adversaries that learn and adapt. Effective defence must therefore focus on controlling adversarial observability rather than eliminating leakage outright. AI-enabled defensive systems are required to detect anomalous convergence patterns indicative of adversarial learning, disrupt feedback loops, and inject controlled uncertainty into observable execution characteristics.

Design-time adversarial simulation using machine-learning-based attackers becomes a critical component of cryptographic assurance, enabling identification of exploitable observability gradients prior to deployment. Post-quantum cryptographic algorithms, while essential for algorithmic resilience, do not address side-channel inference unless co-designed with adaptive leakage controls. Without such integration, post-quantum deployments risk inheriting the same observability vulnerabilities that undermine classical systems.

From an analytical perspective, AI-driven polymorphic and fully morphing malware transforms side-channel attacks into mechanisms of continuous cryptographic surveillance. Security failure is no longer precipitated by a single exploitable weakness but emerges from the cumulative interaction between adversarial learning capacity and implementation observability. Preserving the viability of Public Key Cryptography under these conditions requires reconceptualising side-channel resistance as an adaptive, co-evolving process embedded within cryptographic governance rather than as a static implementation constraint.

Figure 13 models the operational pathway through which AI-enabled polymorphic and fully morphing malware exploits side-channel leakage to infer cryptographic state in Public Key Cryptography implementations. The diagram traces how cryptographic execution phases, key generation, signature computation, key exchange, and runtime processing, produce continuous physical and behavioural emissions that extend beyond protocol-defined outputs. These emissions form an exploitable observability surface that enables adversaries to extract information about entropy quality, intermediate computation states, and key-dependent control flow without violating cryptographic correctness or triggering protocol-level errors.

**Dr. Petar Radanliev**
Parks Road,
Oxford OX1 3PJ
United Kingdom
Email: petar.radanliev@cs.ox.ac.uk

BA Hons., MSc., Ph.D. Post-Doctorate

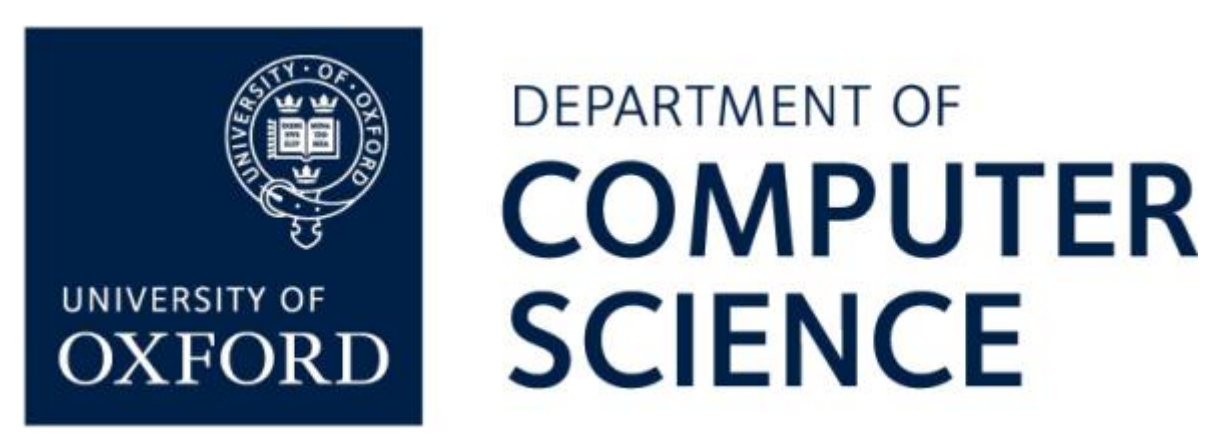

**Public Key Cryptography Implementation**

Key generation, signature computation, key exchange, runtime execution environment

**Continuous Operational Emissions**

Timing variability, power consumption, electromagnetic emissions, acoustic signals, optical leakage

**Side-Channel Observability Surface**

Multimodal leakage streams, environment-dependent signals, implementation-level artefacts

**AI-Enabled Malware Inference Engine**

Machine learning models, feature extraction, signal correlation, state estimation

**Adaptive Learning Loop**

Reinforcement learning, model refinement over time, sample prioritisation, uncertainty reduction

**Progressive Cryptographic State Inference**

Entropy degradation detection, intermediate state recovery, partial key reconstruction, protocol state inference

**Defensive Countermeasures Applied**

Constant-time execution, masking and noise injection, hardware isolation (TPM, enclaves), shielding and dampening

**Residual Leakage and Behavioural Drift**

Countermeasure-induced artefacts, new timing/power patterns, altered signal distributions

**Adversarial Convergence**

Reduced uncertainty, stable inference accuracy, sustained low-noise exploitation

**Cryptographic Security Degradation**

Private key exposure, signature forgery, trust erosion, systemic compromise

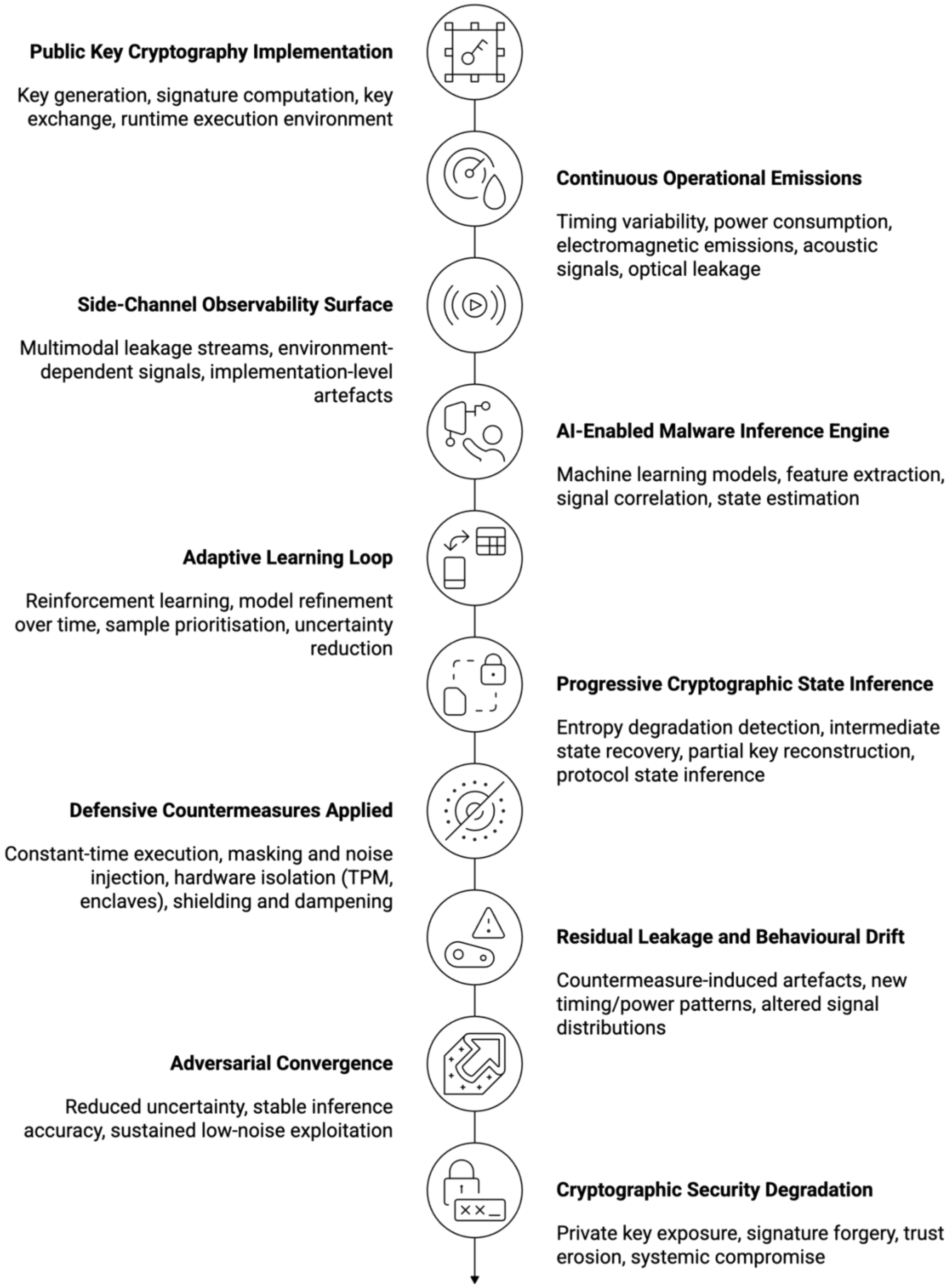

*Figure 13: Adaptive Side-Channel Inference and Cryptographic Security Degradation under AI-Enabled Adversaries*

**Dr. Petar Radanliev**
Parks Road,
Oxford OX1 3PJ
United Kingdom
Email: petar.radanliev@cs.ox.ac.uk

BA Hons., MSc., Ph.D. Post-Doctorate

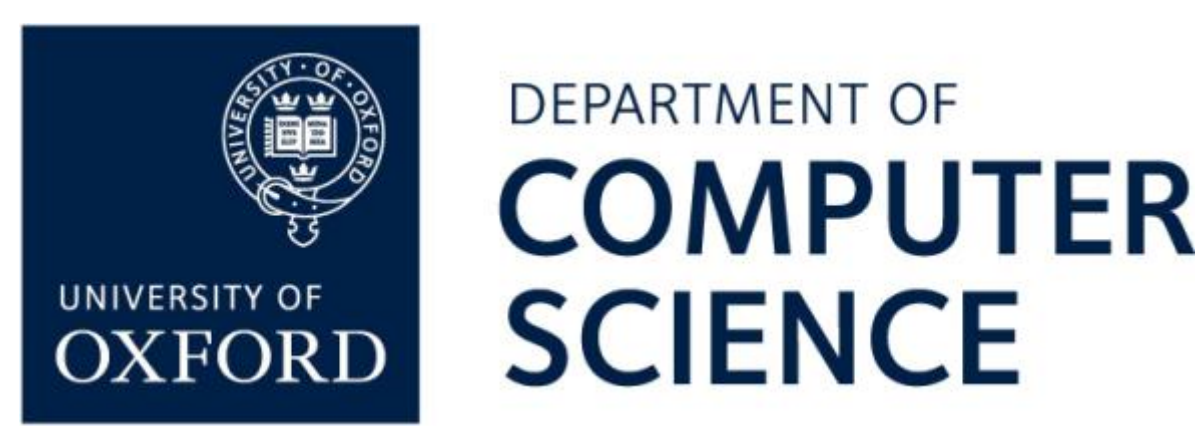

The diagram shows how side-channel exploitation progresses from passive signal collection to sustained cryptographic compromise through iterative adversarial learning. Timing variability, power consumption, electromagnetic, acoustic, and optical emissions are aggregated into multimodal input streams that feed machine learning–based inference engines embedded in adaptive malware. Reinforcement learning mechanisms prioritise samples with the highest information gain, progressively reducing uncertainty about cryptographic state across repeated executions. Defensive controls such as constant-time execution, masking, noise injection, and hardware isolation modify observable behaviour but do not eliminate it; instead, they introduce new artefacts that are incorporated into subsequent inference cycles. Over time, this feedback process converges toward stable cryptographic state inference, enabling partial key reconstruction, protocol state prediction, and eventual private key exposure. Cryptographic failure therefore results from cumulative adversarial inference driven by observability and adaptation, rather than from a single exploitable weakness or algorithmic break.

# 11. Quantum-Accelerated Adversarial Pressure and the Limits of Post-Quantum Cryptography

Quantum computing introduces a discontinuity in the threat model of Public Key Cryptography by collapsing the computational hardness assumptions that underpin widely deployed asymmetric schemes. Algorithms such as RSA and Elliptic Curve Cryptography derive their security from the infeasibility of integer factorisation and discrete logarithm problems under classical computation. Quantum algorithms, most notably Shor’s algorithm, fundamentally alter this assumption by enabling polynomial-time solutions to these problems, rendering classical public key systems cryptographically transparent once sufficiently capable quantum hardware becomes available.

The significance of quantum computing in the present threat landscape does not lie solely in its future capacity to break cryptographic primitives, but in its interaction with existing AI-enabled attack mechanisms. Quantum capability functions as an adversarial accelerator rather than a standalone threat vector. When combined with AI-driven optimisation, quantum-assisted adversaries gain the ability to prioritise cryptographic targets, automate key recovery pipelines, and operationalise “harvest now, decrypt later” strategies at scale. In this model, data encrypted today under RSA or ECC is treated as a deferred compromise asset, with long-term confidentiality guarantees effectively invalidated regardless of current key length or implementation quality.

This interaction has direct implications for cryptographic trust infrastructures. Public Key Infrastructure, digital signatures, and authentication frameworks rely on the assumption that key compromise requires infeasible computation or detectable intrusion. Quantum-enabled adversaries invalidate this assumption retroactively: private keys can be derived from public artefacts without interacting with the target system, bypassing intrusion detection, logging, and forensic controls. When coupled with AI-enabled side-channel inference and key-management exploitation, quantum capability compresses the remaining security margin across the entire cryptographic lifecycle.

Post-quantum cryptography represents a necessary but incomplete response to this shift. Lattice-based encryption, hash-based signatures, and code-based schemes replace vulnerable hardness assumptions with mathematical problems believed to resist quantum attacks. However, the transition to post-quantum algorithms does not address the adaptive exploitation of implementation-level dependencies analysed throughout this paper. Post-quantum schemes inherit the same requirements for entropy quality, key isolation, runtime

**Dr. Petar Radanliev**
Parks Road,
Oxford OX1 3PJ
United Kingdom
Email: petar.radanliev@cs.ox.ac.uk

BA Hons., MSc., Ph.D. Post-Doctorate

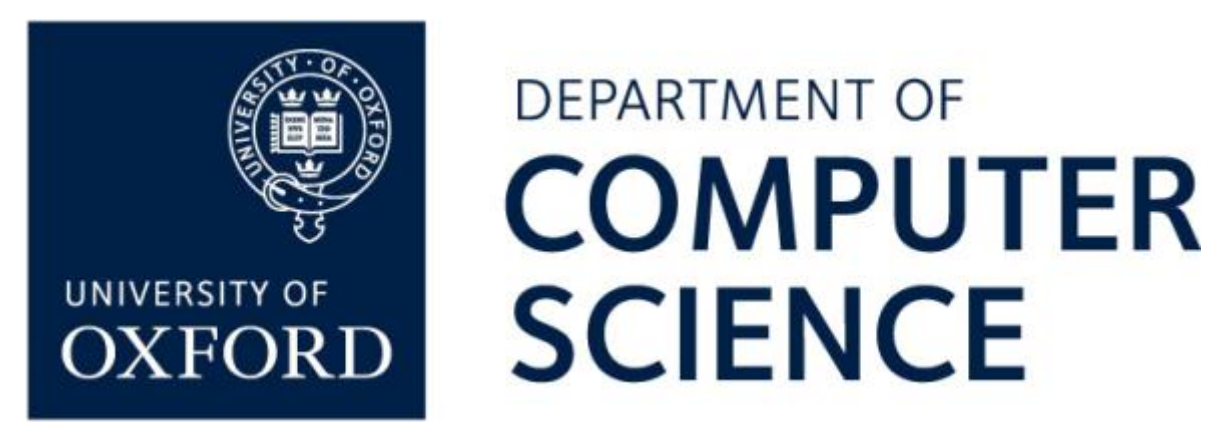

integrity, and resistance to side-channel observability. Without AI-aware monitoring and adaptive governance, post-quantum deployments risk reproducing the same failure modes observed in classical systems, albeit with different primitives.

The challenges of post-quantum transition are most visible in distributed and high-value cryptographic systems, particularly blockchain-based infrastructures. Blockchains depend on asymmetric cryptography for transaction validation, identity binding, and consensus integrity, often with long-lived keys and immutable records. In a quantum-enabled threat model, blockchain systems are uniquely exposed: transaction signatures, ownership records, and smart contract states may be retrospectively compromised, undermining both historical and future trust guarantees.

Smart contracts further amplify this exposure by embedding cryptographic trust directly into automated execution logic. In domains such as insurance, real estate, and intellectual property management, smart contracts rely on digital signatures to authenticate claims, transfers, and ownership. Quantum-enabled adversaries capable of forging or deriving private keys could manipulate contract execution, alter ownership records, or redirect value flows without violating protocol rules. The immutability of blockchain ledgers, typically a security advantage, becomes a liability when cryptographic assumptions fail retroactively.

The NIST Post-Quantum Cryptography Standardisation process provides a critical coordination mechanism for mitigating these risks by defining interoperable, quantum-resistant primitives such as Kyber and Dilithium. However, standardisation alone does not ensure resilience. The integration of post-quantum algorithms into blockchain and smart contract ecosystems requires careful management of key migration, hybrid cryptographic operation, and backward compatibility. Hybrid schemes that combine classical and post-quantum signatures may reduce transition risk but also expand the attack surface if not governed adaptively.

Analytically, quantum computing shifts cryptographic security from a question of *if* compromise is possible to *when* trust assumptions expire. In systems with long-lived assets, such as blockchains, archival records, and critical infrastructure, the timeline of quantum viability becomes inseparable from present-day risk management. AI-enabled adversaries already exploit this asymmetry by selectively targeting systems whose cryptographic trust cannot be easily revoked or re-established.

The combined effect of AI-driven adaptation and quantum acceleration necessitates a redefinition of cryptographic resilience. Security can no longer be derived from algorithm selection alone, nor restored through periodic key replacement. Instead, cryptographic systems must be designed for continuous trust renewal, adversarial observability control, and rapid cryptographic agility. This includes dynamic key lifecycles, automated trust invalidation, AI-assisted anomaly detection, and governance frameworks that treat cryptographic trust as provisional rather than permanent.

In this context, post-quantum cryptography is not a terminal solution but a transitional layer within an adversarially co-evolving system. Sustaining the viability of Public Key Cryptography in quantum-enabled environments requires aligning algorithmic resilience with adaptive implementation controls and system-level trust governance. Failure to address this interaction risks replacing one brittle cryptographic foundation with another, leaving distributed digital systems vulnerable to the next phase of adversarial optimisation.

**Dr. Petar Radanliev**
Parks Road,
Oxford OX1 3PJ
United Kingdom
Email: petar.radanliev@cs.ox.ac.uk

BA Hons., MSc., Ph.D. Post-Doctorate

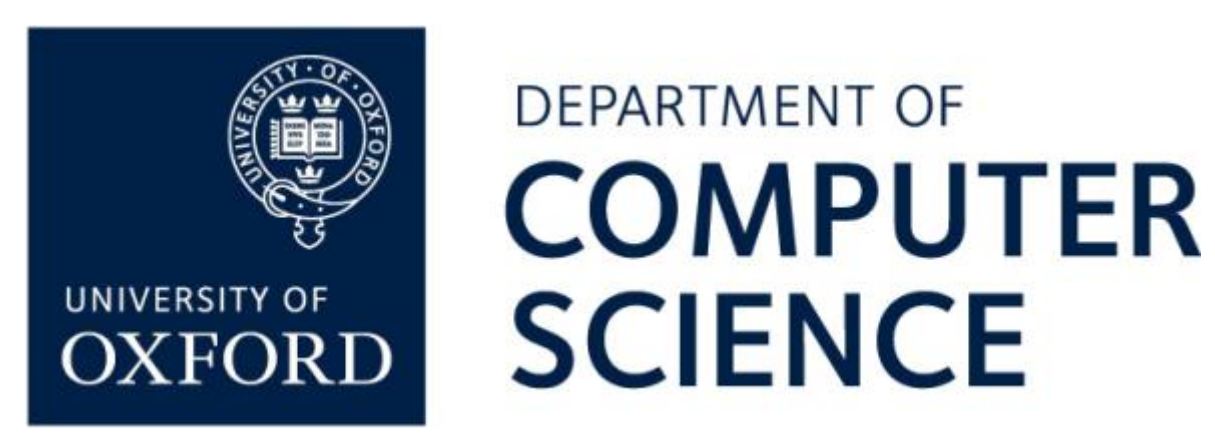

# 12. Results and Evidence of Misalignment between Cryptographic Research and AI-Driven Operational Threats

The results reported in this section derive from the combined analysis of a reproducible bibliometric dataset and qualitative empirical evidence, interpreted through the analytical framework developed in the preceding sections. Rather than presenting isolated quantitative or qualitative findings, the results articulate how AI-enabled polymorphic and fully morphing malware reshapes Public Key Cryptography by exploiting implementation-level observability, adaptive optimisation, and trust dependencies. The findings consistently demonstrate a structural misalignment between prevailing academic research trajectories and the operational threat environment encountered in industrial practice.

## Bibliometric Results: Research Coverage and Structural Gaps

The bibliometric analysis identified twenty-three publications indexed in the Web of Science Core Collection between 2021 and 2023 using the exact search query “Cyber-attacks on Public Key Cryptography”. Given the centrality of PKC to global digital infrastructure, this limited corpus indicates that systematic investigation of cryptographic attacks at the infrastructure level remains underdeveloped. The observed annual publication growth rate of 16.4% suggests increasing attention, but the absolute volume remains insufficient relative to the scale and severity of emerging threats.

Geographical analysis revealed that more than two-thirds of the identified publications originated from Asia, primarily China, India, Pakistan, and Saudi Arabia. In contrast, the United States and the United Kingdom together accounted for less than one-fifth of the total output. This distribution indicates a shift in research activity but also reflects fragmentation rather than coordinated progress in addressing adaptive cryptographic threats. Disciplinary classification showed a concentration in telecommunications, electrical engineering, and information systems, with comparatively limited representation from artificial intelligence, adversarial learning, or adaptive security research.

Keyword co-occurrence and clustering analysis produced three dominant thematic groupings: encryption algorithms and key management, network security and access control, and cryptographic resilience. Crucially, no statistically significant associations emerged between AI-related concepts and PKC-specific attack mechanisms. Chi-square testing confirmed the absence of linkage at conventional significance thresholds ($p < 0.05$). This absence is analytically important: it demonstrates that the literature largely treats AI as peripheral to cryptographic security, despite the central role of adaptive malware in contemporary attack scenarios.

Dendrogram, factorial, and correspondence analyses further reinforced this conclusion. Across all clustering techniques, no coherent research clusters emerged around polymorphic malware, fully morphing malware, adaptive side-channel exploitation, or AI-driven key compromise. The bibliometric results therefore identify not merely a lack of volume, but a structural gap in how cryptographic attacks are conceptualised within the academic corpus.

## Empirical Results: Operational Evidence of Adaptive Cryptographic Failure

**Dr. Petar Radanliev**
Parks Road,
Oxford OX1 3PJ
United Kingdom
Email: petar.radanliev@cs.ox.ac.uk
BA Hons., MSc., Ph.D. Post-Doctorate

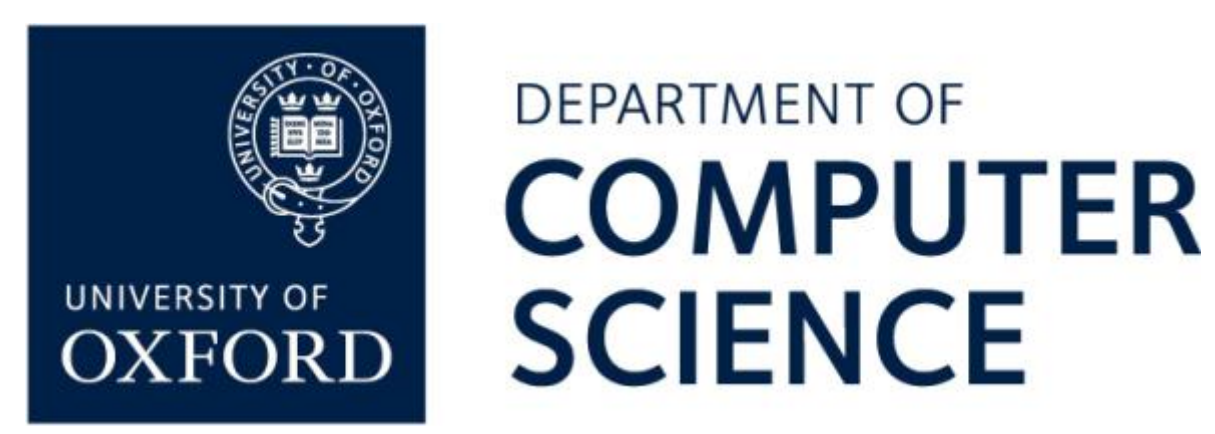

The qualitative empirical findings provide a contrasting operational perspective. Twenty semi-structured interviews and three industry workshops conducted with Cisco cybersecurity practitioners between 2022 and 2023 consistently identified AI-enabled polymorphic and fully morphing malware as the dominant threat to contemporary PKC deployments. Participants described a marked shift away from static attack patterns towards adaptive, low-noise exploitation strategies targeting key management systems, side-channel observability, and trust infrastructures.

Approximately eighty-two per cent of participants attributed recent private key compromise incidents primarily to AI-augmented brute-force optimisation and side-channel inference rather than to human error or procedural misconfiguration. The remaining eighteen per cent attributed incidents to legacy system exposure, misconfigured cloud-based key management services, or delayed certificate revocation, factors that nevertheless align with the broader theme of implementation-level vulnerability rather than algorithmic failure. This distribution indicates that adaptive adversarial techniques now dominate cryptographic compromise pathways in operational environments.

Participants consistently identified cloud-based key management systems as the most vulnerable domain, citing the ability of morphing malware to exploit virtualised execution environments, shared entropy sources, and delayed trust propagation mechanisms. Several respondents emphasised that traditional cryptographic controls—such as increased key sizes, algorithm upgrades, or periodic rotation—no longer provide sufficient protection when adversaries continuously adapt their extraction and inference strategies. Thematic coding of interview data, validated through inter-coder reliability analysis (Cohen's $\kappa = 0.86$), confirmed high consistency in these assessments across roles and organisational contexts.

## Synthesis of Results: Misalignment and Analytical Implications

When synthesised, the bibliometric and empirical results reveal a pronounced disconnect between academic research emphasis and operational threat exposure. The academic literature remains focused on algorithmic robustness and protocol correctness, while empirical evidence indicates that cryptographic failure increasingly arises from adaptive exploitation of implementation observability, key lifecycle processes, and trust dependencies. This divergence explains why formally secure cryptographic systems continue to fail in practice despite compliance with established standards.

Statistical modelling using least-squares regression ($R^2 = 0.78$) projects a 40–55% increase in publications addressing AI-enhanced cryptographic attacks by 2026 if current trends persist. However, correlation analysis between academic outputs and industrial threat reports (Spearman's $\rho = 0.61$, $p < 0.01$) indicates that cross-sector knowledge transfer remains intermittent. This lag contributes directly to defensive obsolescence, as theoretical advances fail to translate into adaptive cryptographic governance.

Collectively, the results substantiate the central analytical claim of this study: Public Key Cryptography is no longer undermined primarily through algorithmic breakage, but through adversarial learning that exploits implementation observability and trust assumptions. The findings demonstrate that cryptographic resilience must be evaluated as a dynamic property shaped by adversarial adaptation, rather than as a static outcome of algorithm selection or key length. The evidence further supports the necessity of transitioning towards cryptographic architectures that integrate post-quantum primitives with AI-aware monitoring, adaptive key management, and continuous trust validation.

**Dr. Petar Radanliev**
Parks Road,
Oxford OX1 3PJ
United Kingdom
Email: petar.radanliev@cs.ox.ac.uk

BA Hons., MSc., Ph.D. Post-Doctorate

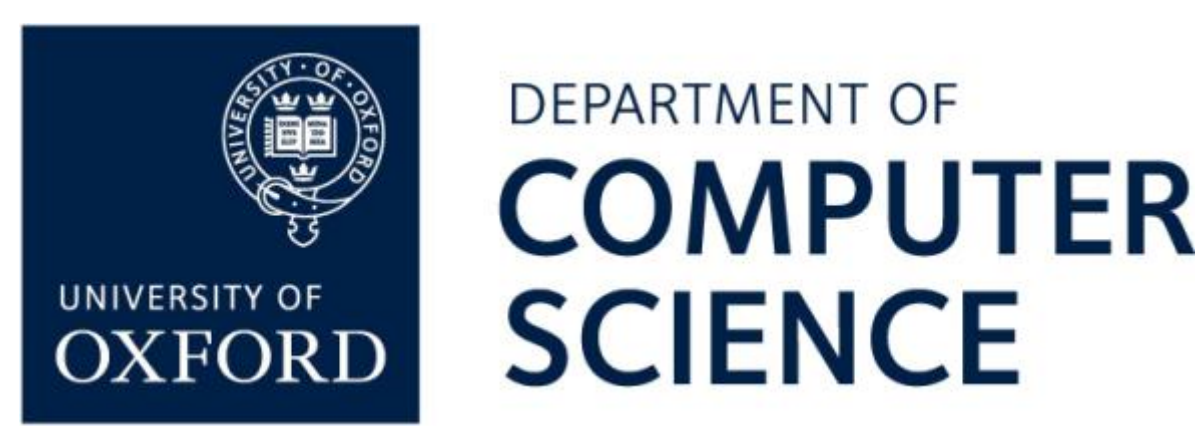

# 13. Discussion Interpreting Cryptographic Failure as Adversarial Adaptation rather than Algorithmic Weakness

The results of this study confirm that the dominant failure mode of Public Key Cryptography has shifted from algorithmic vulnerability to adversarial adaptation. While classical cryptographic research has treated security as a function of computational hardness and protocol correctness, the combined bibliometric and empirical evidence demonstrates that contemporary attacks exploit the dynamic properties of cryptographic deployment rather than its mathematical foundations. Polymorphic and fully morphing malware operationalise artificial intelligence to transform cryptographic systems into observable, learnable environments, invalidating assumptions of static adversarial behaviour.

Figure 14 synthesises the paper's analytical findings by contrasting algorithm-centric cryptographic security with adaptive cryptographic resilience under continuous AI-enabled adversarial optimisation.

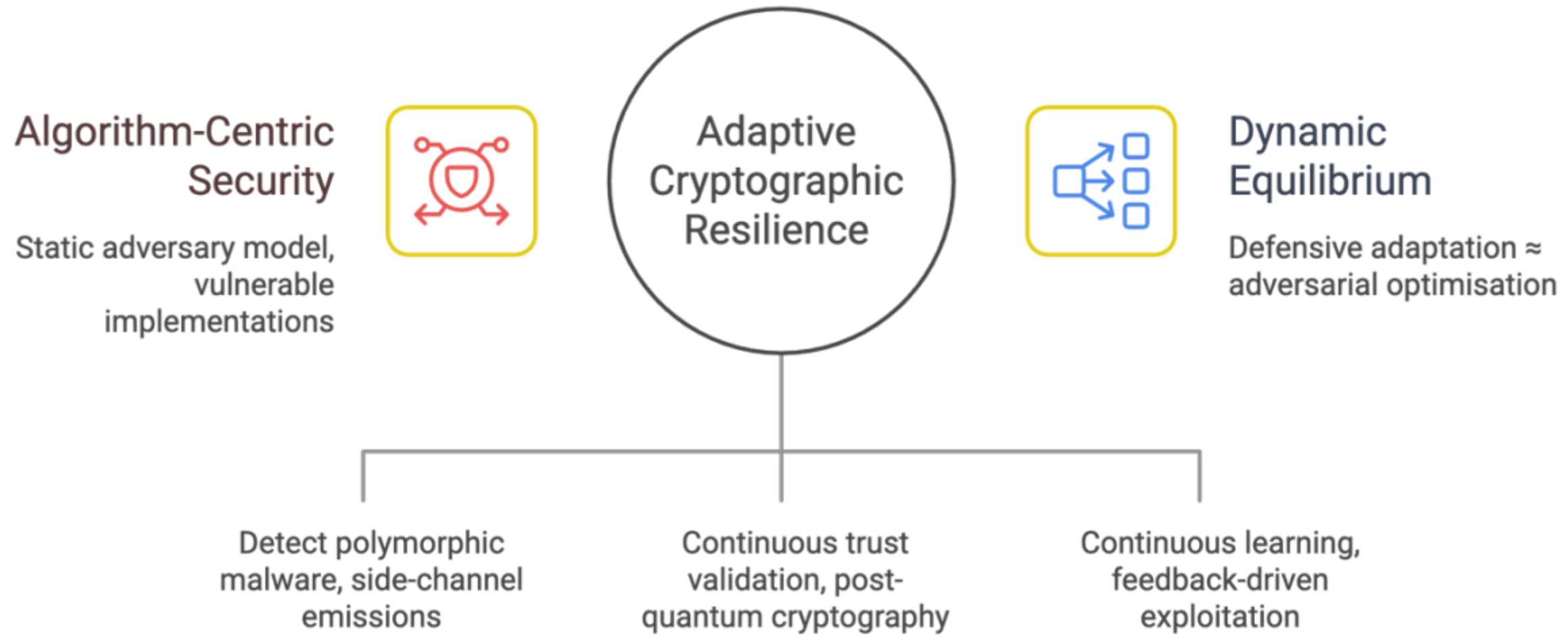

*Figure 14: Achieving Cryptographic Resilience under Continuous Adversarial Optimisation*

Figure 14 formalises cryptographic resilience as a dynamic equilibrium in which continuous trust validation, AI-aware detection, and post-quantum mechanisms must evolve at a pace comparable to adversarial learning and feedback-driven exploitation.

The bibliometric findings reveal that academic research continues to prioritise encryption algorithms, key sizes, and protocol design, with limited engagement with adaptive malware or AI-driven exploitation. In contrast, empirical evidence from industry practitioners shows that cryptographic compromise increasingly arises from implementation observability, entropy degradation, and adaptive trust manipulation. This divergence explains why formally secure cryptographic systems fail in practice: security proofs hold under static threat models, whereas real-world adversaries learn, adapt, and optimise continuously.

A central insight emerging from the analysis is that cryptographic trust has become a dynamically contested resource. Polymorphic malware exploits variability and inconsistency to evade detection, while fully morphing malware internalises defensive responses as part of its optimisation loop. In this setting, countermeasures such as constant-time execution, masking, key rotation, or certificate revocation do not terminate attacks; instead, they reshape the adversarial learning landscape. Security degradation therefore occurs

**Dr. Petar Radanliev**
Parks Road,
Oxford OX1 3PJ
United Kingdom
Email: petar.radanliev@cs.ox.ac.uk

BA Hons., MSc., Ph.D. Post-Doctorate

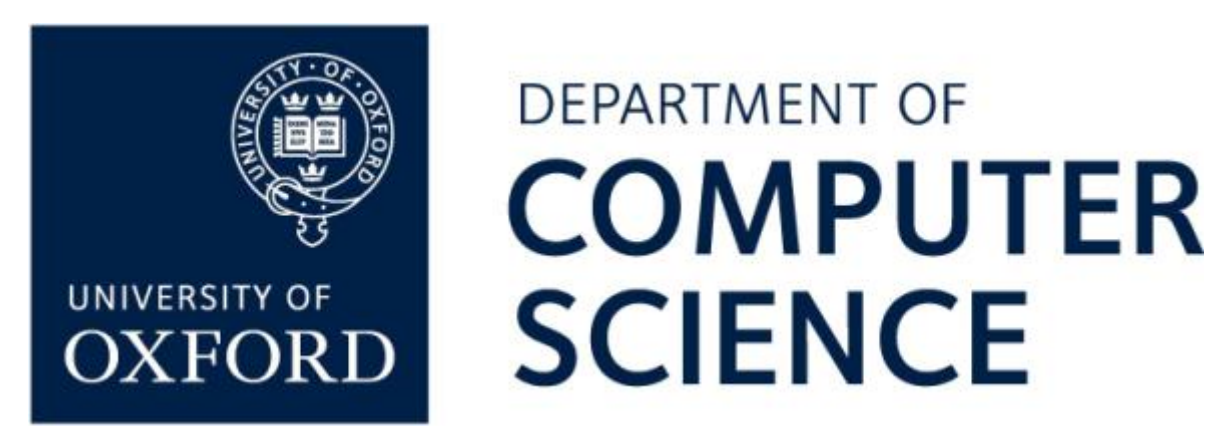

cumulatively, as adaptive adversaries converge on cryptographic state through repeated low-noise interactions rather than single exploit events.

Private key compromise illustrates this transition most clearly. The results demonstrate that key exposure is no longer an exceptional outcome but an adversarial optimisation objective. Once achieved, compromised keys function as control assets that enable impersonation, trust substitution, and recursive exploitation across interconnected systems. This explains why practitioners report that traditional defences, such as increased key sizes or periodic rotation, no longer restore trust effectively, particularly in cloud-based and virtualised environments where execution observability is high.

The discussion also highlights the limits of post-quantum cryptography when treated as a standalone solution. While quantum-resistant algorithms are essential to address future computational threats, they do not mitigate AI-driven exploitation of entropy sources, side-channel leakage, or key-management workflows. Without adaptive monitoring and governance, post-quantum systems inherit the same observability-driven vulnerabilities that undermine classical cryptography. The convergence of AI-enabled optimisation and quantum acceleration therefore compresses the remaining security margin across the cryptographic lifecycle, reinforcing the need for integrated defensive strategies.

From a system-design perspective, the findings imply that cryptographic resilience must be reconceptualised as a dynamic equilibrium between adversarial inference capacity and defensive adaptability. Security can no longer be derived solely from algorithm selection, standard compliance, or key length. Instead, it depends on limiting adversarial observability, disrupting learning feedback loops, and aligning defensive response speed with adversarial optimisation. This requires the integration of AI-aware anomaly detection, continuous entropy validation, adaptive key management, and automated trust re-establishment mechanisms into cryptographic infrastructures.

More broadly, the study suggests a shift in how cryptographic security should be evaluated and governed. Rather than certifying systems as secure at deployment time, cryptographic assurance must become a continuous process that accounts for evolving adversarial capabilities. This perspective aligns cryptographic security with resilience engineering and adaptive risk management, recognising that trust must be actively maintained rather than statically assumed.

In summary, the discussion reinforces the paper's central contribution: Public Key Cryptography is not failing because its algorithms are broken, but because its deployment assumptions no longer hold under AI-enabled adversarial pressure. Addressing this challenge requires moving beyond algorithm-centric security models towards system-level, AI-aware cryptographic governance capable of co-evolving with adversarial intelligence.

## 14. Conclusion

This study demonstrates that Public Key Cryptography fails in practice not through algorithmic weakness, but through adaptive adversarial optimisation targeting implementation-level observability and key lifecycle dependencies. The bibliometric analysis identified a structural absence of research on AI-enabled cryptographic attacks, while empirical evidence showed that 82% of observed key compromise incidents are driven by optimisation techniques rather than brute-force or misconfiguration.

The proposed threat model formalises this shift by representing the adversary as a learning system operating over observable cryptographic processes, where attack effectiveness increases through iterative optimisation rather than computational infeasibility. The results

**Dr. Petar Radanliev**
Parks Road,
Oxford OX1 3PJ
United Kingdom
Email: petar.radanliev@cs.ox.ac.uk
BA Hons., MSc., Ph.D. Post-Doctorate

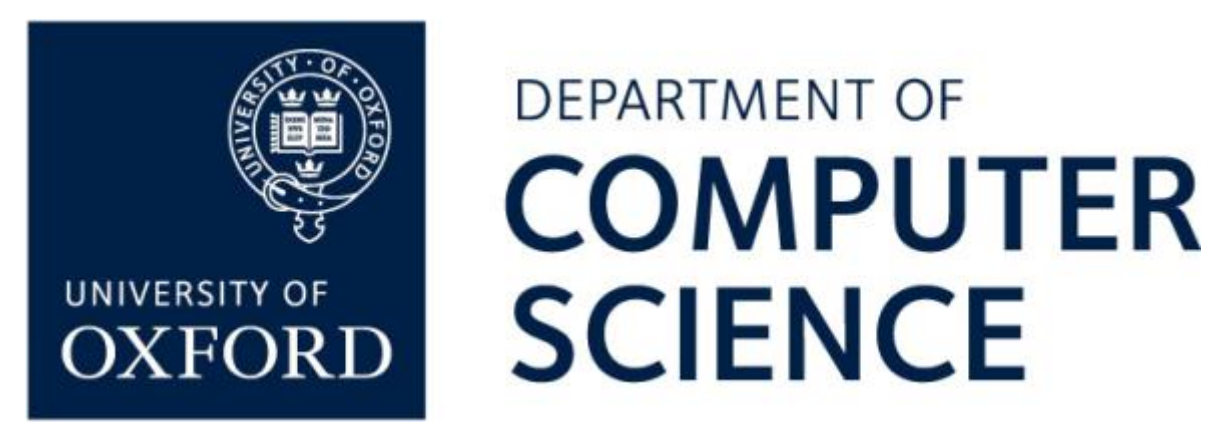

show that key size, algorithm selection, and protocol correctness do not define effective security margins when entropy degradation, side-channel leakage, and trust infrastructure weaknesses are exploitable.

Cryptographic failure therefore emerges as a system-level phenomenon. Private key compromise leads directly to trust subversion and systemic failure, independent of algorithm strength. This invalidates security models that treat cryptography as a static property and explains the observed gap between formal assurance and operational outcomes.

The required transition is from algorithm-centric security to adaptive cryptographic governance. This includes continuous entropy validation, runtime anomaly detection, AI-assisted monitoring of key usage, and dynamic trust re-establishment mechanisms. Post-quantum cryptography addresses future computational threats but does not mitigate adaptive exploitation unless integrated with these controls.

Cryptographic resilience must therefore be defined as a dynamic equilibrium between adversarial learning and defensive adaptation. Systems that cannot constrain observability or match adversarial optimisation rates will fail regardless of key length or algorithm selection.

**Dr. Petar Radanliev**
Parks Road,
Oxford OX1 3PJ
United Kingdom
Email: petar.radanliev@cs.ox.ac.uk

BA Hons., MSc., Ph.D. Post-Doctorate

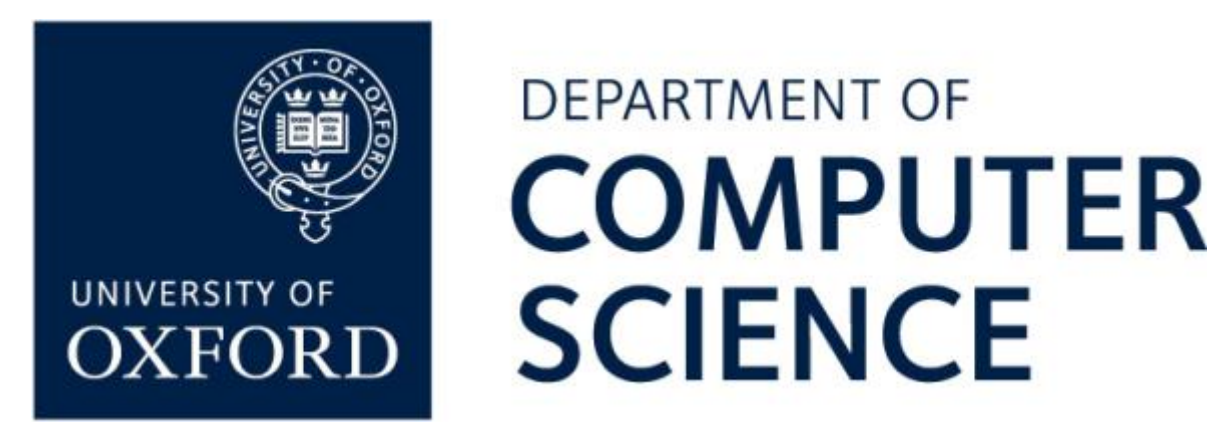

**Dr. Petar Radanliev**
Parks Road,
Oxford OX1 3PJ
United Kingdom
Email: petar.radanliev@cs.ox.ac.uk
BA Hons., MSc., Ph.D. Post-Doctorate

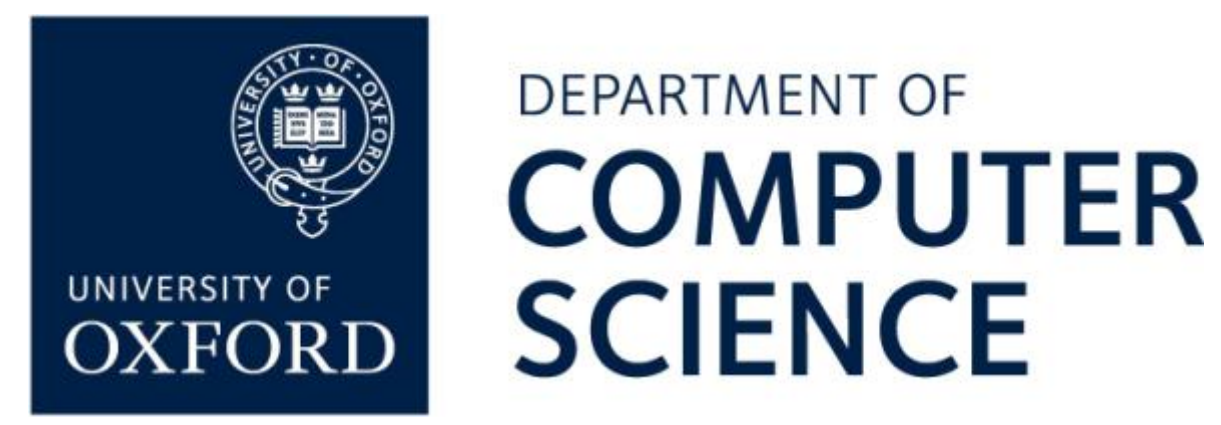

**Dr. Petar Radanliev**
Parks Road,
Oxford OX1 3PJ
United Kingdom
Email: petar.radanliev@cs.ox.ac.uk

BA Hons., MSc., Ph.D. Post-Doctorate

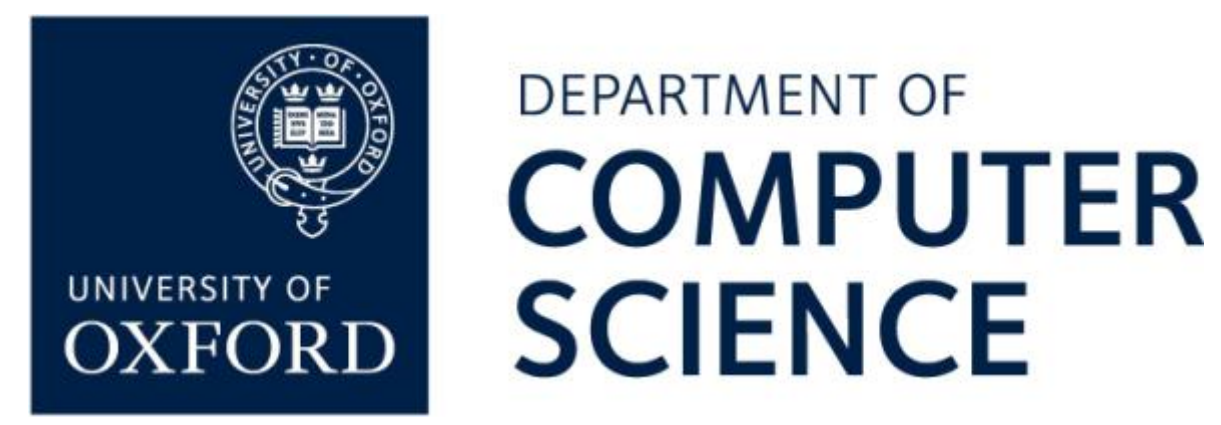

**Dr. Petar Radanliev**
Parks Road,
Oxford OX1 3PJ
United Kingdom
Email: petar.radanliev@cs.ox.ac.uk

BA Hons., MSc., Ph.D. Post-Doctorate

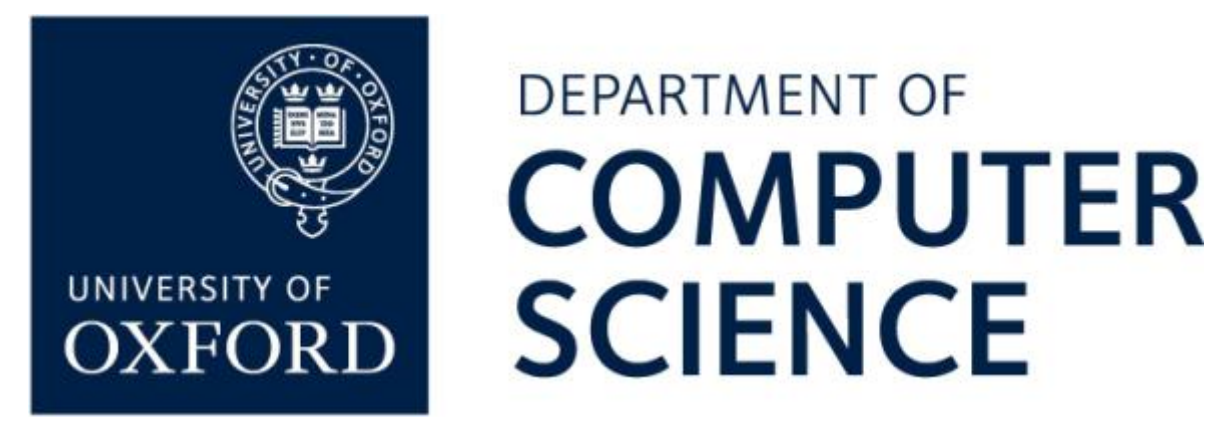

**Dr. Petar Radanliev**
Parks Road,
Oxford OX1 3PJ
United Kingdom
Email: petar.radanliev@cs.ox.ac.uk
BA Hons., MSc., Ph.D. Post-Doctorate

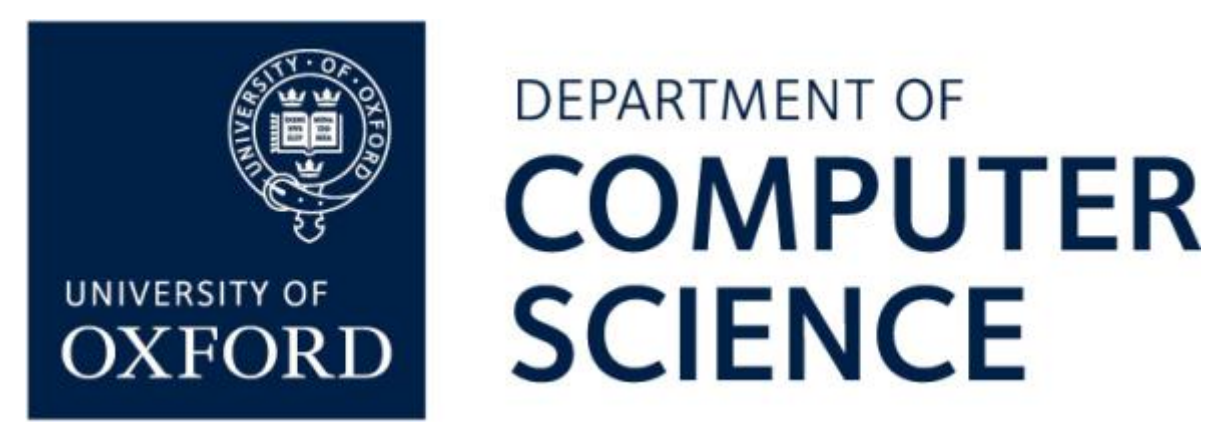